\documentclass[journal]{IEEEtran}
\IEEEoverridecommandlockouts
\usepackage{amsfonts,amsmath,amssymb}
\usepackage{cite}
\usepackage{graphicx}
\usepackage{url}
\usepackage{epstopdf}
\usepackage{bm}

\usepackage{xcolor}

 \newtheorem{lemma}{Lemma}
 \newtheorem{proposition}{Proposition}
 \newtheorem{corollary}{Corollary}
 \newtheorem{definition}{Definition}
  \newtheorem{remark}{Remark}

 \newtheorem{subsidiary theorem}{Subsidiary Theorem}

\ifCLASSINFOpdf
\else
\fi
\usepackage{lineno}

\begin{document}
%

\title{On the Subtleties of $q$-PAM Linear Physical-Layer Network Coding }

\author{Long~Shi,~\IEEEmembership{Member,~IEEE}, ~Soung~Chang~Liew,~\IEEEmembership{Fellow,~IEEE},~
        and ~Lu~Lu,~\IEEEmembership{Member,~IEEE}}
%
\maketitle

\begin{abstract}

This paper investigates various subtleties of applying linear physical-layer network coding (PNC) with $q$-level pulse amplitude modulation ($q$-PAM) in two-way relay channels (TWRC). A critical issue is how the PNC system performs when the received powers from the two users at the relay are imbalanced. In particular, how would the PNC system perform under slight power imbalance that is inevitable in practice, even when power control is applied? To answer these questions, this paper presents a comprehensive analysis of $q$-PAM PNC. Our contributions are as follows: 1) We give a systematic way to obtain the analytical relationship between the \emph{minimum distance of the signal constellation induced by the superimposed signals of the two users} (a key performance determining factor) and the \emph{channel-gain ratio of the two users}, for all $q$. In particular, we show how the minimum distance changes in a piecewise linear fashion as the channel-gain ratio varies. 2) We show that the performance of $q$-PAM PNC is highly sensitive to imbalanced received powers from the two users at the relay, even when the power imbalance is slight (e.g., the residual power imbalance in a power-controlled system). This sensitivity problem is exacerbated as $q$ increases, calling into question the robustness of high-order modulated PNC. 3) We propose an asynchronized PNC system in which the symbol arrival times of the two users at the relay are deliberately made to be asynchronous. We show that such asynchronized PNC, when operated with a belief propagation (BP) decoder, can remove the sensitivity problem, allowing a robust high-order modulated PNC system to be built.

\end{abstract}


\IEEEpeerreviewmaketitle

\section{Introduction}

The past decade has witnessed a  boost of research on exploiting interference  to increase  throughput in wireless networks. Among the interference-exploitation schemes, physical-layer network coding (PNC)  can  potentially double the throughput in a two-way relay network (TWRN) \cite{pnc}.

In TWRN,  two users exchange information via a relay.   When operated with PNC, time in the TWRN is divided into two slots \cite{pnc}. In the first timeslot, the two users transmit simultaneously to the relay. The relay then maps the overlapped received signals to a network-coded message. This process is referred to as PNC mapping \cite{pnc}. In the second timeslot,  the relay broadcasts the network-coded message to the two users. Each of the users then extracts the information from the other user by subtracting its own message from the network-coded message.
Compared with   traditional relaying, PNC boosts the throughput of TWRN  by reducing  the number of time slots  required  for the  information exchange to the minimum \cite{pnc,liew}.

 The principle of PNC was first studied assuming binary XOR PNC mapping on BPSK modulated signals \cite{pnc, pop}, although it was known that in general higher-order PNC mappings were also possible \cite{liew}. Subsequently, \cite{koi,choi,  chang, latin, yanglet,yangtwc} provided detailed studies of PNC systems with higher-order PNC mappings on higher-order modulated signals to improve the throughput in the high SNR regime.
In particular, \cite{yanglet} and \cite{yangtwc} proposed a linear PNC mapping  scheme for  $q$-PAM modulated signals  that minimizes PNC mapping errors caused by noise.

  Our work here assumes largely the same system model and builds on top of the results in \cite{yanglet} and \cite{yangtwc}. Although  the  linear PNC design proposed in \cite{yanglet} and \cite{yangtwc} can optimize the error performance, an important assumption there was that the channel gains between the users and the relay  have  irrational values  (in so far as the proofs of the  working mechanism is concerned). In real  implementations of communication systems, however,  channel gains  are invariably  represented by  rational values since processors have finite resolution. Our current paper shows that, fortunately, the irrationality assumption is not a fundamental requirement. In fact, the PNC error performance changes in a continuous fashion as the channel gains vary, and there is no abrupt breakdown at rational channel gains. This is  positive news in that the linear PNC mapping as proposed in \cite{yanglet} and \cite{yangtwc} remains intact even for rational channel gain representations.

On the negative side, however, we find that the performance of the $q$-PAM linear PNC scheme is very sensitive to channel gains, i.e., a little change in channel gains can result in drastically  different performance. To provide the context, we remark that it is widely believed that PNC systems would have good performance when the powers of the two users are balanced \cite{liew}. However, with the $q$-PAM linear PNC scheme, even a slight deviation from perfect power balance will cause a drastic performance degradation. How to overcome this super sensitivity to the received powers is of paramount importance to a practical PNC system.

In this paper, we propose an  asynchronized $q$-PAM linear PNC system that has robust performance under channel variations.  By deliberately introducing symbol misalignment between the received signals of the two users at the relay, and with an appropriate PNC decoding scheme that makes use of the belief propagation (BP) algorithm, the sensitivity to  channel gains can be overcome.

Overall, our current work contributes to the fundamental understanding of   $q$-PAM linear PNC mapping scheme operated in the high SNR regime (high code rate regime). In particular, we offer insights to its fundamental weaknesses and show how these weaknesses can potentially be overcome.

The remainder of this paper is organized as follows. Section II overviews prior related work.   Section III describes the key idea of  the $q$-PAM   linear PNC design  and the   motivations behind our work. Section IV focuses on the specific case of $7$-PAM PNC to bring out the various issues and subtleties in general $q$-PAM PNC. Sections V and  VI detail our   general analytical results for the   $q$-PAM linear PNC designs. Section VII  shows how  an asynchronized  $q$-PAM linear PNC design can overcome the  sensitivity problem.  Simulation results and discussions are given in Section VIII. The relationships among propositions, lemmas, theorems, and corollaries in this paper are presented in \emph{Appendix I}.

\section{Related Work }

\subsection{Linear PNC Mapping}

In  TWRN   operated with linear PNC, the relay computes a network-coded packet  as a linear combination of the received packets  from users. Linear PNC mappings have been extensively investigated in  \cite{choi, chang, zhang, yang3, naz1, yanglet,to,yangtwc,nam,wang,chen,wub, hern, yuan,naz2, zhanginf}. Typically, the linear PNC mapping consists of a weighted sum of the packets of the two users over a finite field.

Our current paper is related to such linear PNC. More specifically, our results are applicable to nonchannel-coded linear PNC systems  as well as  end-to-end channel-coded linear PNC systems, as elaborated below.

In  nonchannel-coded PNC TWRN \cite{zhanginf, pnc,liew,yanglet,yangtwc}, users $A$ and $B$ send their source packets, $\mathbf{w}_A=(w_A[n])_{n=1,2,\ldots, N}$ and $\mathbf{w}_B=(w_B[n])_{n=1,2,\ldots, N}$  respectively, to the relay simultaneously. Each packet consists of  $N$ symbols. The relay $R$ then maps the overlapped received signals to a linear network-coded packet $\mathbf{w}_N=\alpha\mathbf{w}_A+\beta\mathbf{w}_B=(\alpha{w}_A[n]+\beta{w}_B[n])_{n=1,2,\ldots, N}$ and broadcasts $\mathbf{w}_N$ to users $A$ and $B$. In particular, the network coding is performed on a symbol-by-symbol pairwise basis.

In end-to-end channel-coded PNC TWRN \cite{liew}, the source packets of users $A$ and $B$ are $\mathbf{s}_A=(s_A[m])_{m=1,2,\ldots, M}$ and $\mathbf{s}_B=(s_B[m])_{m=1,2,\ldots, M}$,  respectively. They perform channel coding on the source packets to obtain channel-coded packets $\mathbf{w}_A=(w_A[n])_{n=1,2,\ldots, N}$ and $\mathbf{w}_B=(w_B[n])_{n=1,2,\ldots, N}$, where $N>M$.  The relay is oblivious of the channel coding, and it performs symbol-by-symbol PNC mapping in exactly the same way as in the nonchannel-coded PNC system. Errors are only corrected at the receiver ends of users $A$ and $B$ \cite{liew}.

Channel-coded PNC TWRN can also operate in a link-by-link manner (as opposed to end-to-end). There has been a large body of work on link-by-link channel-coded PNC \cite{liew, zhang,yang3,nam,naz2,hern, naz1, chen,to,wub}. Our current paper is not directly related to link-by-link channel-coded PNC, although there is relevance. In link-by-link channel-coded PNC, the relay is aware of the channel coding performed by users $A$ and $B$ and knows their codebooks. With this knowledge, the relay can exploit the correlations among $w_A[n]$ for different $n$, and $w_B[n]$ for different $n$, induced by channel coding to reduce errors when computing $\mathbf{w}_N=(\alpha{w}_A[n]+\beta{w}_B[n])_{n=1,2,\ldots, N}$. In general, link-by-link channel-coded PNC has better performance than end-to-end channel-coded PNC, at the expense of higher complexity at the relay.

%


\subsection{Nonlinear PNC Mapping}

 Nonlinear PNC was studied in \cite{koi, latin}. In nonlinear PNC, the network-coded (NC) symbol   is not a linear weighted sum of the symbols transmitted by the end nodes.  A closest-neighbor clustering algorithm based on the exclusive law  was proposed in \cite{koi}.    By the nonlinear mapping in \cite{koi},  the number of NC symbols induced by the superimposed constellation at the relay is not the same as  the cardinality of the symbol constellation of each end node. For example, if QPSK is used at the end nodes ($4$ symbols in each user constellation), $5$-QAM is used at the relay for NC symbols ($5$ NC symbols in the NC constellation).

 Latin square based PNC mapping was proposed in \cite{latin} that can satisfy the exclusive law.  The rows correspond to the symbols transmitted by one end node, and the columns correspond to the symbols transmitted by the other end node. An entry  in the Latin square corresponds to the NC symbol induced by  row and column of the entry. The constraint of Latin square, i.e., an NC symbol appears once and only once in each row and each column,   ensures the exclusivity required for network decoding based on self information (e.g., knowledge of an NC symbol and the row in which it appears yield the knowledge of the column of the entry).

  Investigations on nonlinear PNC mappings typically assume the use of low-order modulated signals because exhaustive search or near-exhaustive search is required to identify an appropriate nonlinear PNC mapping. Although Latin square based PNC mapping is applicable for $ M$-ary constellation such as $M$-PSK, the computational  complexity of table construction becomes prohibitively  high as $M$ increases.   By contrast, as will be seen later, there are simple ways to find the appropriate linear PNC mapping  for the linear system under study here and \cite{yanglet,yangtwc}. The linear scheme is much more scalable in terms of the network coding operation  when high-order $q$-PAM modulations are used.

  Another advantage of linear PNC mapping is that it can be integrated with linear channel coding in a natural way in link-by-link channel-coded PNC systems. Specifically, suppose that the channel codes used by the two end nodes are the same, and they are linear in that each coded symbol of a user (i.e., $w_A[n]$ and $w_B[n], n=1,2,\ldots,N$) is a linear weighted sum of some of its source symbols. Then, at the relay,  symbol-by-symbol mappings of the sequence of channel-coded symbol pairs of the two users to a sequence of NC symbols (i.e., $w_N[n]=\alpha w_A[n] + \beta w_B[n],  n=1,2,\ldots,N$) yield a valid codeword (i.e., $(w_N[n])_{n=1,2,\ldots,N}$ is a valid codeword) as per the codebook used by the end users. This allows us to perform PNC mapping followed by channel decoding to find an NC source symbol sequence that network-codes the two source symbol sequences of the two users \cite{liew}  (i.e., the source symbols corresponding to the codeword $(w_N[n])_{n=1,2,...,N}$ is $(s_N[n] = \alpha s_A[n] + \beta s_B[n])_{n=1,2,\ldots,N}$). The facility for the integration of channel decoding and network coding as such at the relay is not available if nonlinear PNC is used.

\section{$q$-PAM Linear PNC design}\label{syn}

In this section, we first introduce the system model of  $q$-PAM linear PNC. Then we briefly revisit the $q$-PAM linear PNC system and its design criterion as proposed in \cite{yanglet,yangtwc}. After that, we point out some outstanding issues and problems with the $q$-PAM linear PNC design that motivate the investigations of our current paper.

 \subsection{System Model}

\begin{figure}[t]
 \centering
        \includegraphics[width=0.6\columnwidth]{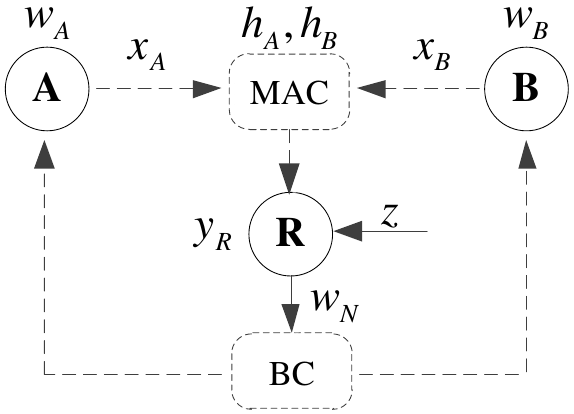}
       \caption{System model of a TWRN.}
        \label{fig:twr}
\end{figure}

 Fig. \ref{fig:twr} shows the TWRN under study.  Two nodes   $A$ and $B$ communicate with each other via a relay $R$. We assume that all nodes operate in the half-duplex mode. We further assume  each node ($A$, $B$, or $R$) has single antenna and there is no direct link between nodes $A$ and $B$. The  packets from nodes $A$ and $B$ are denoted by $(w_A[n])_{n=1,2,\ldots, N}$ and $(w_B[n])_{n=1,2,\ldots, N}$, respectively, where $N$ is the number of symbols in a packet. We assume that $w_i[n] \in GF(q)$, i.e., $w_i[n]\in \{0, 1,   \ldots, q-1\}$ for $i\in \{A,B\}$, where $q$ is   prime.


 Information exchange between nodes $A$ and $B$ is realized in two phases (i.e.,  multiple-access channel (MAC) phase and broadcast channel (BC) phase). In the MAC phase, nodes $A$ and $B$ modulate $w_A[n]$ and $w_B[n]$ into modulated symbols   $x_A[n]$ and $x_B[n]$ for transmission to  $R$. The bijective mapping  from $w_i[n]$ to $x_i[n]$ is effected via $q$-level pulse amplitude modulation ($q$-PAM), i.e., $x_i[n]=\frac{1}{\mu}(w_i[n]-\frac{q-1}{2})$, where $\mu$ is a power normalization factor such that $E(x^2_i[n])=1$.

 Nodes  $A$ and $B$ transmit  $x_A[n]$ and $x_B[n]$ simultaneously. At the relay node, if the symbol arrival times of nodes $A$ and $B$ are synchronized,  the received signal  at baseband is given by
 \begin{align}\label{gyr1}
 \nonumber y_R(t) = &\sum^N_{n=1}\big\{ h_A \sqrt{P} x_A[n] p(t-nT) + \\
  &h_B \sqrt{P} x_B[n]p(t-nT) \big\} +z(t),
    \end{align}
where $p(\cdot)$ is the pulse shaping function for the baseband signal, $T$ is the duration of a symbol,  $h_i$ is the   real channel coefficient between  node $i$ and the relay $R$ for $i\in \{A,B\}$,   $P$ is the transmit power,  and $z(t)$ is an additive white Gaussian noise (AWGN) with zero mean and variance of $\sigma^2=N_0/2$. In addition, we assume that $h_A$ and $h_B$ can be perfectly estimated at $R$ in the MAC phase.

After match filtering and sampling on the received signal $y_R(t)$, the corresponding digital   samples  are given by
 \begin{align}\label{syr2}
   y_R[n] =     h_A \sqrt{P} x_A[n]  +  h_B \sqrt{P} x_B[n]+z[n], n=1,2, \ldots, N.
    \end{align}

Since we only consider symbol-by-symbol PNC mappings  (i.e., the PNC mapping of a pair of overlapping symbols is performed independently of the PNC mapping of other pairs), for simplicity, we henceforth omit the sample index $n$. Then, \eqref{syr2} can be written as
 \begin{align}\label{yr1}
  \nonumber &y_R  =       h_A \sqrt{P} x_A   +  h_B \sqrt{P} x_B    +z\\
   &=   \frac{\sqrt{P}}{\mu}( {h_A w_A + h_B w_B} )+z- \frac{\sqrt{P}(q-1)}{2\mu}(h_A+h_B),
    \end{align}
where the second equality holds due to the $q$-PAM modulation.

 \subsection{General Idea of   $q$-PAM Linear PNC}

This paper focuses on the MAC phase where the linear PNC mapping proposed in   \cite{yanglet} and \cite{yangtwc} is   adopted  at relay $R$.

We define the pair of symbols transmitted by the two nodes, denoted by $(w_A, w_B)$, as a  {\em joint symbol}. Here,  $(w_{A}, w_{B})$ is a \emph{valid} joint symbol if and only if $0 \leq w_A, w_B \leq q-1$.

For a given joint symbol $(w_A, w_B)$,  a linear  network-coded (NC)  symbol is a linear combination of  $w_A$ and $w_B$  given by
 \begin{eqnarray}\label{yr3}
w_N^{(\alpha, \beta)} = f^{(\alpha, \beta)}_N(w_A, w_B)\triangleq\alpha\otimes w_A \oplus \beta \otimes w_B,
    \end{eqnarray}
where $\alpha, \beta \in GF(q)\setminus\{0\}$, i.e., $\alpha, \beta\in \{1,2,\ldots, q-1\}$, $a \oplus b \triangleq {\rm mod}(a+b,q)$, and  $a \otimes b \triangleq {\rm mod}(ab,q)$. That is, $\oplus$ and $\otimes$ denote addition and multiplication in the finite field $GF(q)$ respectively.  We refer to $w_N^{(\alpha, \beta)}$ as  an {\em NC symbol}  and $(\alpha, \beta)$ as NC coefficients.

 In the BC phase, relay $R$   broadcasts an estimated version of  ${w}^{(\alpha, \beta)}_N$  to nodes $A$ and $B$. Ideally, suppose that the estimation at relay $R$ is perfect. Upon receiving the NC symbol ${w}^{(\alpha, \beta)}_N$, node  $A$  recovers the signal $w_B$ as follow
 \begin{align}\label{Abc}
\beta^{-1}\otimes(w^{(\alpha, \beta)}_N\ominus \alpha\otimes w_A)= \beta^{-1} \otimes \beta\otimes w_B =w_B,
    \end{align}
 where $\beta^{-1}$ denotes the multiplicative inverse of $\beta$, and $\ominus$ denotes subtraction,  in $GF(q)$. Node $B$ can recover $w_A$ in a similar manner.

   To form a valid NC symbol, neither $\alpha$ nor $\beta$ can be $0$ (i.e., each of them must have a multiplicative inverse in $GF(q)$). To see why, suppose that  if $\alpha\neq 0$ and $\beta = 0$, then the NC symbol will contain no information about $w_B$. Thus, by \eqref{Abc}, node $A$ cannot recover $w_B$.

 Given a pair of $(h_A, h_B)$, we further define  a \emph{superimposed symbol} associated with the joint symbol $(w_A, w_B)$ as
 \begin{eqnarray}\label{ws}
w_S=f_S(w_A, w_B)\triangleq h_A w_A + h_B w_B,
  \end{eqnarray}
  where ``$+$'' is a summation in real field.

 Overall,  we see from \eqref{yr3} and \eqref{ws} that for a given joint symbol  $(w_A, w_B)$, there is an associated NC symbol and an associated superimposed symbol given by the mapping functions $f^{(\alpha, \beta)}_N: \mathcal{W}_{(A,B)}\rightarrow \mathcal{W}^{(\alpha, \beta)}_{N}$ and $f_S: \mathcal{W}_{(A,B)}\rightarrow  \mathcal{W}_{S}$, respectively.
 \begin{figure}[t]
 \centering
        \includegraphics[width=0.5\columnwidth]{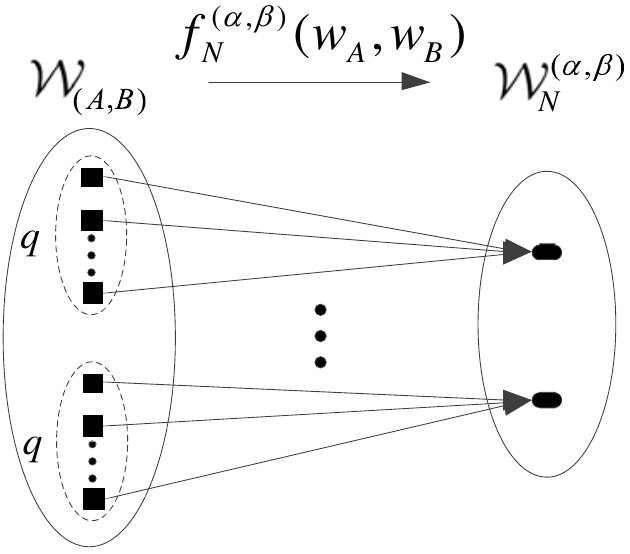}
       \caption{$f^{(\alpha, \beta)}_N(w_A, w_B): \mathcal{W}_{(A,B)}\rightarrow  \mathcal{W}^{(\alpha, \beta)}_{N}$ given a pair of $(\alpha, \beta)$.}
        \label{fig:wabwn}
\end{figure}

The domain of $f^{(\alpha, \beta)}_N$ and $f_S$ is
\begin{align}\label{setallab}
 \mathcal{W}_{(A,B)}=  GF(q)\times GF(q).
\end{align}

The range of $f^{(\alpha, \beta)}_N$ for a given pair ${(\alpha, \beta)}$ is
   \begin{align}\label{wn}
\nonumber \mathcal{W}^{(\alpha, \beta)}_{N} = &\big\{ w^{(\alpha, \beta)}_{N} \in GF(q) | \exists (w_A, w_B) \in \mathcal{W}_{(A, B)}: \\& w_N^{(\alpha, \beta)} = f_N^{(\alpha, \beta)} (w_A, w_B)  \big\}.
    \end{align}
Given that $\alpha$ and $\beta$ are both non-zero elements of $GF(q)$, it can be verified that $\mathcal{W}^{(\alpha, \beta)}_{N}=GF(q)$.  In particular, $f^{(\alpha, \beta)}_N: \mathcal{W}_{(A,B)}\rightarrow \mathcal{W}^{(\alpha, \beta)}_{N}$ is a $q$-to-$1$ mapping  as illustrated  in Fig. \ref{fig:wabwn}. Based on the $q$-to-$1$ mapping,  $\mathcal{W}_{(A,B)}$ can be partitioned into $q$ subsets. Each of the subsets is induced by a specific $w_N^{(\alpha, \beta)}$, expressed as follows:
      \begin{align}\label{setwab}
\nonumber \mathcal{W}_{(A,B)}(w^{(\alpha, \beta)}_{N}) =& \big\{(w_{A}, w_{B})\in \mathcal{W}_{(A, B)}| \\ & w^{(\alpha, \beta)}_{N}=f^{(\alpha, \beta)}_N(w_A, w_B)\big\}.
    \end{align}

The difference between any two different joint symbols $(w_A, w_B)$ and $(w'_A, w'_B)$ is given by
  \begin{align}\label{del}
(\delta_A, \delta_B)\triangleq (w_A-w'_A, w_B-w'_B),
\end{align}
where $ \delta_A, \delta_B \in \{-(q-1), \ldots, q-1\} $ and $(\delta_A, \delta_B)\neq (0,0)$.  Given this $(\delta_A, \delta_B)$,  we define the associated ``mod $q$ difference'' as follows:
  \begin{align}\label{delq}
(\delta^{(q)}_A, \delta^{(q)}_B)\triangleq \big({\rm mod}(\delta_A, q), {\rm mod} (\delta_B, q)\big),
\end{align}
where $\delta^{(q)}_A, \delta^{(q)}_B\in GF(q)$ and $(\delta^{(q)}_A, \delta^{(q)}_B)\neq (0,0)$.

For a given joint symbol, \emph{Proposition \ref{pro:clu}} below specifies the joint symbols that will be clustered with the given joint symbol under a specific $(\alpha, \beta)$. \emph{Proposition \ref{pro:wab12}}, on the other hand, specifies the different possible $(\alpha, \beta)$ that can be used to cluster  two specific joint symbols together\footnote{The results in \emph{Propositions 1}, \emph{2}, and \emph{3} can alternatively be expressed in terms of cosets in group theory. For simplicity, we choose not to do so in the main body of this paper. For interested readers who are familiar with cosets, let us consider \emph{Proposition 1}.  Define a subgroup of $GF^2(q)$ as follows:
   \begin{eqnarray}\label{coset}
\nonumber H =\{ (\delta^{(q)}_A, \delta^{(q)}_B)\in GF^2(q)|  (\delta^{(q)}_A, \delta^{(q)}_B)=\nu \otimes(\ominus\beta, \alpha), \nu\in GF(q)\}.
    \end{eqnarray}
 The coset induced by $(w_A, w_B)$ over $H$, $(w_A, w_B)\oplus H$, is a group of joint symbols mapped to the same NC symbol by $(\alpha, \beta)$. Readers familiar with cosets should be able to extrapolate from the above on how \emph{Propositions 1}, \emph{2}, and \emph{3} can be framed in terms of cosets.}.

\begin{proposition}\label{pro:clu}

 Consider an arbitrary joint symbol $(w_A, w_B)\in \mathcal{W}_{(A,B)}$. Suppose that a given pair of $(\alpha, \beta)$, $\alpha\neq0$, $\beta\neq0$,  maps   $(w_A, w_B)$ to the NC symbol $w^{(\alpha, \beta)}_N$. Then,  $(\alpha, \beta)$ maps altogether $q$  joint symbols   (one of which is $(w_A, w_B)$) to the same $w^{(\alpha, \beta)}_N$ as expressed below:
     \begin{align}\label{cluab}
 (w'_{A}, w'_{B})  = (w_A, w_B)\oplus (\delta^{(q)}_A, \delta^{(q)}_B),
    \end{align}
 where $(\delta^{(q)}_A, \delta^{(q)}_B)= \nu \otimes (\ominus\beta, \alpha)$, $\nu=0,1, \ldots,q-1$ (note that $\ominus\beta$ is the additive inverse of $\beta$ in $GF(q)$, i.e., $\ominus\beta = q - \beta$). Furthermore, two distinct $(w_A, w_B)$ and $(w'_A, w'_B)$  mapped to the same NC symbol must satisfy $w_A\neq w'_A$ and $w_B\neq w'_B$.

\end{proposition}

\begin{IEEEproof}[Proof of Proposition \ref{pro:clu}]
We first prove the last statement in the proposition.  The NC coefficients  $(\alpha, \beta)$ cluster two distinct  $(w'_A, w'_B)$ and $(w_A, w_B)$ to the same NC symbol  if and only if
  \begin{align}\label{cluab1}
\alpha \otimes w'_A\oplus \beta\otimes w'_B=\alpha \otimes w_{A}\oplus \beta\otimes w_B.
    \end{align}
Equivalently, we can rewrite \eqref{cluab1} as
\begin{align}\label{cluab2}
\alpha \otimes (w'_A \ominus w_{A})\oplus \beta\otimes (w'_B\ominus w_B)=0,
    \end{align}
where $a \ominus b={\rm mod}(a-b, q)$.

Since $(w'_A, w'_B)$ is distinct from $(w_A, w_B)$, we cannot have $w_A=w'_A$ and $w_B=w'_B$ at the same time. According to \eqref{cluab2}, we cannot have $w_A=w'_A$ and  $w_B\neq w'_B$, or $w_A\neq w'_A$ and $w_B=w'_B$ either, since   $\alpha\neq 0$ and $\beta\neq 0$ for linear PNC mapping (see \eqref{yr3}). Thus, $w'_A \neq w_A$ and $w'_B\neq w_B$.

For the first statement of the proposition, we note that  $(w'_A\ominus w_{A}, w'_B\ominus w_B) = (\ominus\beta, \alpha)$   is a solution to \eqref{cluab2}. Thus, the other $q-2$ different solutions to \eqref{cluab2} (in $GF(q)$) are $\nu\otimes(\ominus\beta, \alpha)$, $\nu=2, \ldots, q-1$. Altogether, this gives $q-1$ other joint symbols that are mapped to the same NC symbol as $(w_A, w_B)$.

\end{IEEEproof}

\begin{proposition}\label{pro:twop}

Consider a pair of joint symbols $(w_{A,1}, w_{B,1})$ and $(w_{A,2}, w_{B,2})$ with $(\delta^{12}_A, \delta^{12}_B)\triangleq (w_{A,1}-w_{A,2}, w_{B,1}-w_{B,2})$. Suppose that a given pair of $(\alpha, \beta)$, $\alpha\neq 0$, $\beta\neq 0$, clusters $(w_{A,1}, w_{B,1})$ and $(w_{A,2}, w_{B,2})$ together. Then, $(\alpha, \beta)$   also  clusters  another pair of joint symbols $(w_{A,3}, w_{B,3})$ and $(w_{A,4}, w_{B,4})$ with $(\delta^{34}_A, \delta^{34}_B)\triangleq (w_{A,3}-w_{A,4}, w_{B,3}-w_{B,4})$ together  if and only if
  \begin{align}\label{clutwop}
{\rm mod}((\delta^{34}_A, \delta^{34}_B),q)=\nu \otimes {\rm mod}((\delta^{12}_A, \delta^{12}_B), q),
    \end{align}
where $\nu\in \{1, \ldots, q-1\}$.

\end{proposition}

Remark: note that the NC symbol of $(w_{A,1}, w_{B,1})$ and $(w_{A,2}, w_{B,2})$ and the NC symbol of $(w_{A,3}, w_{B,3})$ and $(w_{A,4}, w_{B,4})$ could be different even if \eqref{clutwop} is satisfied.

\begin{IEEEproof}[Proof of Proposition \ref{pro:twop}]

Since $(\alpha, \beta)$ clusters $(w_{A,1}, w_{B,1})$ and $(w_{A,2}, w_{B,2})$ together, we have
 \begin{align}\label{clutwop1}
\alpha \otimes {\rm mod}(\delta^{12}_A, q) \oplus \beta\otimes {\rm mod}(\delta^{12}_B, q)=0.
    \end{align}

We first prove the ``if'' part. If ${\rm mod}((\delta^{34}_A, \delta^{34}_B),q)=\nu \otimes  {\rm mod}((\delta^{12}_A, \delta^{12}_B), q)$ where $\nu\in \{1, \ldots, q-1\}$, then multiplying \eqref{clutwop1} by $\nu$ in $GF(q)$ and substitution from \eqref{clutwop} gives us (by the commutative and associative laws of multiplication, and the distributive law, in $GF(q)$ arithmetic)
\begin{align}\label{clutwop4}
\nonumber & \alpha \otimes \nu \otimes  {\rm mod}(\delta^{12}_A, q) \oplus \beta\otimes \nu \otimes  {\rm mod}(\delta^{12}_B, q)\\
=&\alpha \otimes {\rm mod}(\delta^{34}_A, q) \oplus \beta \otimes {\rm mod}(\delta^{34}_B, q)=0
    \end{align}

Thus,  $(w_{A,3}, w_{B,3})$ and $(w_{A,4}, w_{B,4})$  are clustered together by the same $(\alpha, \beta)$.

We next  prove the ``only if'' part. Since $(\alpha, \beta)$  clusters  $(w_{A,3}, w_{B,3})$ and $(w_{A,4}, w_{B,4})$ together, we have
 \begin{align}\label{clutwop2}
\alpha \otimes {\rm mod}(\delta^{34}_A,q) \oplus \beta\otimes {\rm mod}(\delta^{34}_B, q)=0.
    \end{align}

From \eqref{clutwop1} and \eqref{clutwop2}, we have
 \begin{align}\label{clutwop3}
 {\rm mod}\bigg(\left[\begin{array}{cc}
   \delta^{12}_A & \delta^{12}_B\\
     \delta^{34}_A & \delta^{34}_B
  \end{array}
  \right] \left[\begin{array}{c}
   \alpha\\
     \beta
  \end{array}
  \right], q\bigg)=\left[\begin{array}{c}
   0\\
   0
  \end{array}
  \right].
    \end{align}
    Since $\alpha\neq 0$ and $\beta\neq 0$,  $(\delta^{12}_A, \delta^{12}_B)$ and $(\delta^{34}_A, \delta^{34}_B)$ must be linearly dependent in $GF(q)$. Thus, ${\rm mod}((\delta^{34}_A, \delta^{34}_B), q)=\nu \otimes  {\rm mod}((\delta^{12}_A, \delta^{12}_B), q)$,  $\nu\in \{1, \ldots, q-1\}$.

\end{IEEEproof}

\begin{proposition}\label{pro:wab12}
Consider two distinct joint symbols $(w_A, w_B)$ and $(w'_A, w'_B)$ such that $\delta_A=w_A-w'_A\neq 0$ and $\delta_B=w_B-w'_B\neq 0$. There exist $q-1$ pairs of $(\alpha, \beta)$ that can cluster $(w_A, w_B)$ and $(w'_A, w'_B)$ to the same NC symbol given by $(\alpha, \beta)=\nu\otimes(\ominus\delta^{(q)}_B, \delta^{(q)}_A)$, $\nu=1,\ldots, q-1$.

\end{proposition}

\begin{IEEEproof}[Proof of Proposition \ref{pro:wab12}]
  We first note that $-(q-1)\leq \delta_A, \delta_B \leq q-1$, $\delta_A\neq 0$ and $\delta_B\neq 0$ implies $\delta^{(q)}_A\neq 0$ and $\delta^{(q)}_B\neq 0$. Define an NC coefficient pair $(\alpha, \beta)=(\ominus\delta^{(q)}_B, \delta^{(q)}_A)$. Then, $(\alpha, \beta)$ maps $(w_A, w_B)$ and $(w'_A, w'_B)$ to the NC symbols given by
  \begin{align}\label{p2wn1}
w_N&=\ominus\delta^{(q)}_B \otimes w_A\oplus \delta^{(q)}_A\otimes w_B,\\
  w'_N&=\ominus\delta^{(q)}_B \otimes w'_{A}\oplus \delta^{(q)}_A\otimes w'_B,
    \end{align}
    respectively. The difference of the two NC symbols is
      \begin{align}\label{p2wn2}
\nonumber w_N\ominus w'_N&=\ominus\delta^{(q)}_B \otimes (w_A\ominus w'_{A}) \oplus \delta^{(q)}_A\otimes (w_B\ominus w'_{B})\\
\nonumber&={\rm mod} \big((q-\delta_B)   (w_A- w'_{A}) + \delta_A (w_B- w'_{B}), q\big)\\
&={\rm mod}(-\delta_B\delta_A+\delta_A\delta_B, q)=0.
    \end{align}

Thus, the above $(\alpha, \beta)$ maps $(w_A, w_B)$ and $(w'_A, w'_B)$ to the same NC symbol. That is, $\alpha\otimes (w_A\ominus w'_{A}) \oplus \beta \otimes (w_B\ominus w'_{B})=0$.  Now, multiplying this equation by any $\nu \in$ $GF(q)\setminus\{0\}$,  we see that all the $q-1$ NC coefficient pairs $\nu\otimes (\alpha, \beta)=\nu\otimes(\ominus\delta^{(q)}_B, \delta^{(q)}_A)$, $\nu=1,\ldots, q-1$  can map $(w_A, w_B)$ and $(w'_A, w'_B)$ to the same NC symbol.

 \end{IEEEproof}

\begin{remark}\label{rem:wn}
  In the proof of \emph{Proposition \ref{pro:wab12}}, the NC symbol to which $(w_A, w_B)$ and $(w'_A, w'_B)$ are mapped depends on $\nu$. For a given $\nu$, $(w_A, w_B)$ and $(w'_A, w'_B)$ are always mapped to the same NC symbol.
\end{remark}
\rightline{$\blacksquare$}

\begin{figure}[t]
 \centering
        \includegraphics[width=0.9\columnwidth]{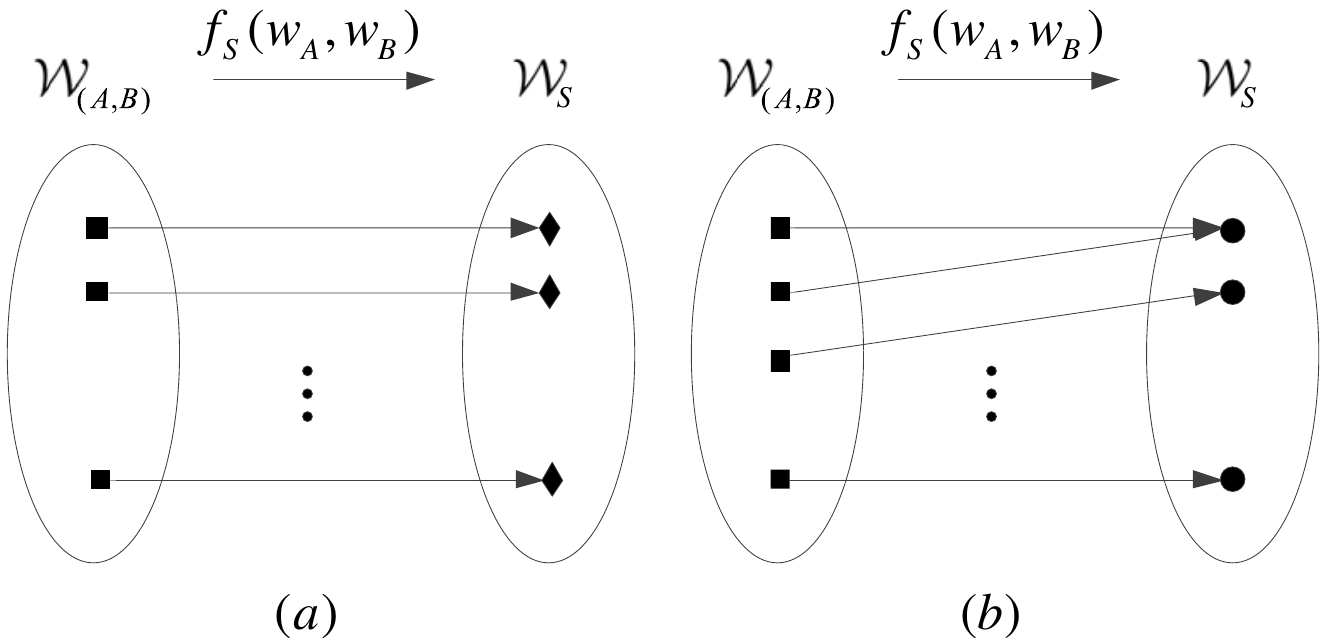}
       \caption{$f_S(w_A, w_B): \mathcal{W}_{(A,B)}\rightarrow  \mathcal{W}_{S}$ given a pair of $(h_A, h_B)$. (a) Bijective mapping for irrational and certain rational values of $h_A/h_B$, e.g., $h_A/h_B=q$ ; (b)   Non-injective  mapping for other rational values of $h_A/h_B$, e.g., $h_A/h_B=1$.}
        \label{fig:wabws}
\end{figure}

Returning to the definition of superimposed symbols in \eqref{ws}, the  range of $f_S$ for a given pair $(h_A, h_B)$ is
      \begin{align}\label{setalls}
 \mathcal{W}_{S}=\big\{w_{S} \in \mathbb{R} |\exists(w_{A}, w_{B})\in  \mathcal{W}_{(A,B)}: w_S=f_S(w_A, w_B)\big\}.
    \end{align}
 Note that while $\mathcal{W}_{(A, B)} = GF^2(q)$, $\mathcal{W}_{S} \subset \mathbb{R}$ (since $h_A$ and $h_B$ are real).  The mapping  $f_S: \mathcal{W}_{(A,B)}\rightarrow  \mathcal{W}_{S}$ is illustrated in Fig. \ref{fig:wabws}.

 Since each $(w_A, w_B)$ corresponds to a $w_S$, we may have a maximum of  $q^2$   elements in  $\mathcal{W}_S$. For irrational and certain rational $h_A/h_B$, the  mapping from $\mathcal{W}_{(A,B)}$ to $\mathcal{W}_{S}$ is bijective (see Fig. \ref{fig:wabws} (a)),  in which case there are  $q^2$ elements in $\mathcal{W}_S$. However, for $h_A/h_B$ of other  rational values, some distinct joint symbols $(w_A, w_B)$ may be mapped to the same   $w_S$ (see Fig. \ref{fig:wabws} (b)), in which case there are fewer than $q^2$ elements in $\mathcal{W}_S$.

As with $\mathcal{W}_{(A, B)}$, we can also partition $\mathcal{W}_{S}$ into $q$ subsets, each induced by one specific $w_N^{(\alpha, \beta)}$:
  \begin{align}\label{setws}
\nonumber &\mathcal{W}_S (w_N^{(\alpha, \beta)}) =
\big\{w_S \in  \mathcal{W}_S | \exists (w_A, w_B) \in  \mathcal{W}_{(A, B)}:\\ &w_N^{(\alpha, \beta)} = f_N^{(\alpha,\beta)}(w_A, w_B) ~{\rm and}~ w_S = f_S(w_A, w_B)\big\}.
  \end{align}

  The general problem in $q$-PAM linear PNC is to find a pair of NC coefficients $(\alpha,\beta)$ that   minimizes detection errors. To do so,  since the received signal $y_R$ in (3) (after subtracting the constant term $\sqrt{P}(q-1)(h_A + h_B)/(2\mu)$) depends on the superimposed symbol $w_S$, we focus on the constellation of $\mathcal{W}_S$  and its partition as in \eqref{setws}.   Two distance metrics associated with the constellation  and its partition can be defined \cite{yangtwc}:
\begin{itemize}
  \item  \emph{Minimum symbol distance} $l_{\min}$
\begin{align}\label{lmin}
l_{\min}\triangleq
\mathop{\arg\min}\limits_{\substack{ (w_{A}, w_B) \neq (w'_{A}, w'_B), \\ (w_A, w_B), (w'_A, w'_B) \in \mathcal{W}_{(A,B)}, \\ w_{S}=f_S(w_A, w_B), w'_{S}=f_S(w'_A, w'_B) }} |w_{S}-w'_{S}|.
\end{align}

\item  \emph{Minimum set distance} $d^{(\alpha, \beta)}_{\min}$
\begin{align}\label{dmin}
d^{(\alpha, \beta)}_{\min}\triangleq
\mathop{\arg\min}\limits_{\substack{(w_{A}, w_B) \neq (w'_{A}, w'_B), \\  (w_A, w_B), (w'_A, w'_B) \in \mathcal{W}_{(A,B)}, \\ w_{S}=f_S(w_A, w_B), w'_{S}=f_S(w'_A, w'_B),\\ f^{(\alpha, \beta)}_N(w_A, w_B) \neq f^{(\alpha, \beta)}_N(w'_A, w'_B)}}  |w_S-w'_S|.
\end{align}
\end{itemize}
We remark that $l_{\min}$  is  the minimum distance between two superimposed symbols in the superimposed  constellation, and it depends on $(h_A, h_B)$ only;  while $d^{(\alpha, \beta)}_{\min}$ is the minimum distance between superimposed symbols belonging to different partitions in \eqref{setws} (i.e., two superimposed symbols associated with two joint symbols mapped to different NC symbols), and it depends on both $(\alpha, \beta)$ and $(h_A, h_B)$.

The interpretations of the above distances are as follows. If our goal is to detect the joint symbol $(w_A, w_B)$, then maximizing $l_{\min}$ will minimize the symbol error probability in the high SNR regime. On the other hand, if our goal is to detect the NC symbol $w_N^{(\alpha, \beta)}$, such as in the case of PNC, then maximizing $d_{\min}$ will minimize the symbol error probability in the high SNR regime. Thus, for PNC, given a pair of channel coefficients $(h_A, h_B)$, our problem is to find a pair of NC coefficients $(\alpha_{opt}, \beta_{opt})$ such that $d^{(\alpha_{opt},\beta_{opt})}_{\min}\geq d^{(\alpha,\beta)}_{\min}$ for all $(\alpha, \beta) \neq (\alpha_{opt},\beta_{opt})$.

\subsection{Decision Rules of NC Symbol Detection}

For given $(\alpha, \beta)$, upon receiving $y_R$ in \eqref{yr1},   there are two possible decision rules that can be employed by relay $R$ to compute the NC symbols, as described below:

1) \emph{Maximum Likelihood (ML) Rule }

The ML decision rule is an optimal decision rule. For a given pair of NC coefficients  $(\alpha, \beta)$, the NC symbol computed by this rule has the lowest symbol error rate (SER).

 For this rule, relay $R$ first computes the likelihood functions for all  $w^{(\alpha, \beta)}_N \in GF(q)$:
 \begin{align}\label{map}
 &Pr(w^{(\alpha, \beta)}_N |y_R)=\sum_{{\substack{(w_A, w_B)\in \mathcal{W}_{(A,B)}(w^{(\alpha, \beta)}_N): \\ w^{(\alpha, \beta)}_N=\alpha\otimes w_A\oplus \beta\otimes  w_B}}}Pr\big( (w_A, w_B)|y_R\big),
\end{align}
where $Pr\big((w_A, w_B)|y_R\big)$ is the likelihood function of   $(w_A, w_B)$ given by
\begin{align}
\nonumber &Pr\big((w_A, w_B)|y_R\big)=\frac{Pr\big(y_R|(w_A, w_B)\big)}{q^2 Pr(y_R)}\\
\nonumber& =\frac{1}{q^2Pr(y_R)\sqrt{2\pi \sigma^2}} \exp\bigg\{ -\frac{\big(y_R-(h_A \sqrt{P} x_A  +  h_B \sqrt{P} x_B)\big)^2}{2\sigma^2}\bigg\}\\
&\propto\exp\bigg\{ -\frac{\big(y_R-(h_A \sqrt{P} x_A  +  h_B \sqrt{P} x_B)\big)^2}{2\sigma^2}\bigg\}.
\end{align}

Then, relay $R$ chooses the NC symbol that corresponds to the maximum likelihood function:
\begin{align}\label{map1}
\hat{w}^{(\alpha, \beta)}_N =\mathop{\arg\max}\limits_{w^{(\alpha, \beta)}_N  \in GF(q)}Pr( w^{(\alpha, \beta)}_N|y_R).
\end{align}

2) \emph{Minimum Distance  (MD) Rule }

The MD decision rule is near optimal, especially in the high SNR regime. For this rule, for a given pair $(\alpha, \beta)$, relay $R$ first estimates $(w_A, w_B)$ according to the minimum distance (MD) rule. Specifically, the estimate is given by
  \begin{align}\label{ml}
\nonumber(\hat{w}_A, \hat{w}_B)=&
\mathop{\arg\min}\limits_{(w_A, w_B) \in \mathcal{W}_{(A,B)}} \| \frac{\mu}{\sqrt{P}} y_R+ \frac{(q-1)}{2}(h_A+h_B)\\& -  ( {h_A w_A + h_B w_B} )\|.
    \end{align}

Then, $(\hat{w}_A, \hat{w}_B)$ is  mapped  to a $q$-PAM NC symbol  $\hat{w}^{(\alpha, \beta)}_N$ according to \eqref{yr3}, i.e., $\hat{w}^{(\alpha, \beta)}_N =\alpha\otimes \hat{w}_A \oplus \beta\otimes\hat{w}_B$.

Note that in the high SNR regime, we expect the SER of the MD rule to approach the SER of the ML rule. To see this, we note that for the ML rule, if we only retain the dominant term  $Pr((w_A, w_B)|y_R)$ in the summation $\sum_{{(w_A, w_B)\in \mathcal{W}_{(A,B)}(w^{(\alpha, \beta)}_N)}}Pr\big( (w_A, w_B)|y_R\big)$ in \eqref{map}, and then replace \eqref{map1} by the equivalent
\begin{align}
\hat{w}^{(\alpha, \beta)}_N =\mathop{\arg\max}\limits_{w^{(\alpha, \beta)}_N  \in GF(q)}\log Pr( w^{(\alpha, \beta)}_N|y_R),
\end{align}
we will then have the MD rule. In the high SNR regime, we expect the dominant term  $Pr((w_A, w_B)|y_R)$ in the summation $\sum_{(w_A, w_B)\in \mathcal{W}_{(A,B)}(w^{(\alpha, \beta)}_N) }Pr\big( (w_A, w_B)|y_R\big)$   to be much larger than the other terms. In particular, we can view the MD rule as the result of applying the log-max approximation $\log \sum\exp(z_i)\approx \max_i z_i$  on the ML rule.

In the high SNR regime, the decoding error probability of the MD rule is given by \cite{yangtwc}
\begin{align}\label{ser}
Pr^{(\alpha, \beta)}_e \lessapprox \frac{1}{q^2} \mathcal{A}^{(\alpha, \beta)}_{\min} Q\big(\sqrt{\frac{\rho}{2  \mu^2}d^{(\alpha, \beta)}_{\min}}\big),
\end{align}
where $\mathcal{A}^{(\alpha, \beta)}_{\min}$ denotes a total multiplicity with respect to the minimum set distance event, $Q(x)$ denotes the $Q$-function, and $\rho=P/N_0$.

%

 \subsection{Optimal $(\alpha, \beta)$ }

 In \emph{Part C}, we discussed the decision rules for NC symbol detection for a given pair $(\alpha, \beta)$. This part, on the other hand, focuses on  the optimal $(\alpha, \beta)$.

 We note the following subtleties when defining the optimal $(\alpha, \beta)$:



\begin{itemize}
  \item  In general, the SER-optimal $(\alpha, \beta)$ is  the $(\alpha, \beta)$ that yields the lowest SER under the ML rule.  MD with any $(\alpha, \beta)$ will not yield SER better than ML under this $(\alpha, \beta)$.

  \item In the high SNR regime, the SER-optimal $(\alpha, \beta)$ is the same as the $d_{\min}$-optimal $(\alpha, \beta)$  (i.e., the optimal $(\alpha, \beta)$ under the MD rule) given by
 \begin{align}\label{optab}
(\alpha_{opt}, \beta_{opt})=
\mathop{\arg\max}\limits_{\alpha, \beta \in  \{1, \ldots, q-1\}} d^{(\alpha,\beta)}_{\min}.
    \end{align}
for most  channel gain ratio $h_A/h_B$ (this will be verified by simulations in \emph{Part A}, \emph{Section \ref{sec:sim}}). In this case, $d^{(\alpha_{opt}, \beta_{opt})}_{\min}$ dominates over $\mathcal{A}^{(\alpha, \beta)}_{\min}$ in \eqref{ser}.

  \item The SER-optimal $(\alpha, \beta)$, however, is not the same as $d_{\min}$-optimal $(\alpha, \beta)$ in \eqref{optab} for certain $h_A/h_B$. For these $h_A/h_B$, $d_{\min}$ is not much larger than $l_{\min}$ for all $(\alpha, \beta)$. As a consequence, $d^{(\alpha, \beta)}_{\min}$ may not dominate over $\mathcal{A}^{(\alpha, \beta)}_{\min}$ in \eqref{ser}. \emph{Part A, Section \ref{sec:sim}} will elaborate on this case.

\end{itemize}

Henceforth, except when stated otherwise, the notation $(\alpha_{opt}, \beta_{opt})$ refers to the $d_{\min}$-optimal $(\alpha, \beta)$ given by (33). Also, unless stated otherwise, by optimal $(\alpha, \beta)$, we mean the $d_{\min}$-optimal $(\alpha_{opt}, \beta_{opt})$. Much of the rest of this paper focuses on $(\alpha_{opt}, \beta_{opt})$. As will be seen, knowing   $(\alpha_{opt}, \beta_{opt})$ also helps us to find the SER-optimal $(\alpha, \beta)$ even when they are not the same---a systematic way to to identify the SER-optimal $(\alpha, \beta)$ from the analysis of $(\alpha_{opt}, \beta_{opt})$ will be provided in \emph{Section \ref{sec:sim}}.



Let us now dig deeper into $(\alpha_{opt}, \beta_{opt})$ in \eqref{optab}. Consider all $q^2(q^2 - 1)/2$ pairs of joint symbols: $\big\{\{(w_A, w_B), (w'_A, w'_B)\}|(w_A, w_B), (w'_A, w'_B) \in \mathcal{W}_{(A,B)},  (w_A, w_B) \neq (w'_A, w'_B)\big\}$. For each pair $\{(w_A, w_B), (w'_A, w'_B)\}$, the distance between the two superimposed symbols induced by the two joint symbols,  $(w_A, w_B)$ and $(w'_A, w'_B)$, is given by $ |[\delta_A, \delta_B] [h_A, h_B]^T|$. Define $(\Delta_A, \Delta_B)$ as the difference of two joint symbols whose superimposed symbols are  separated by $l_{\min}$ as follows \cite{yangtwc}:
 \begin{align}\label{th1deta}
 [\Delta_A, \Delta_B]=\mathop{\arg\min}\limits_{\substack{{\delta_A,\delta_B\in \{1-q, \ldots, q-1\},}\\{|\delta_A|+|\delta_B|\neq 0}}}  |[\delta_A, \delta_B][h_A, h_B]^T|.
\end{align}

In \cite[\emph{Theorem 1}]{yangtwc}, $(\alpha, \beta)$ is said to be optimal if it can cluster the superimposed symbols separated by $l_{\min}$ to the same NC symbol:
\begin{align}\label{th1yang}
\mod\big((\Delta_A, \Delta_B)(\alpha_{opt}, \beta_{opt})^T, q\big)=0.
\end{align}

Furthermore, for positive real values of $h_A$ and $h_B$ (as explained in \emph{Appendix II}, this assumption does not cause loss of generality),  a possible  $(\alpha_{opt}, \beta_{opt})$ pair is given by \cite[\emph{Corollary 1}]{yangtwc}
 \begin{align}\label{optabnz}
 (\alpha_{opt}, \beta_{opt})=\mathop{\arg\min}\limits_{\alpha, \beta \in \{1, \ldots, q-1\}}  |\frac{\alpha}{h_A}-\frac{\beta}{h_B}|.
    \end{align}
By \emph{Proposition \ref{pro:wab12}}, other isomorphic  solutions are given by $(\kappa\otimes \alpha_{opt}, \kappa\otimes \beta_{opt})$ with $\kappa \in \{2, \ldots, q-1\}$. The groupings of the joint symbols into the NC symbols are the same in all these isomorphic solutions. Just that the labels of the NC symbols are different.

In \cite{yangtwc}, $h_A/h_B$ was assumed to be   \emph{irrational}. With this assumption,  $(\Delta_A, \Delta_B)$ in \eqref{th1deta} is then unique, except for signs\footnote{ That is,  $(\Delta_A, \Delta_B)$  is a solution to \eqref{th1deta}, so is   $(-\Delta_A, -\Delta_B)$. Furthermore, the same $(\alpha_{opt}, \beta_{opt})$ apply for  both $(\Delta_A, \Delta_B)$ and $(-\Delta_A, -\Delta_B)$ in \eqref{th1yang}. Thus, assuming irrational $h_A/h_B$,  $(\alpha_{opt}, \beta_{opt})$ is unique.}. In particular, with the uniqueness of $\pm(\Delta_A, \Delta_B)$   induced by irrational $h_A/h_B$, any two superimposed symbols separated by $l_{\min}$ can always be mapped to the same NC symbol by $(\alpha_{opt}, \beta_{opt})$ given in \eqref{th1yang}. Therefore, $d^{(\alpha_{opt}, \beta_{opt})}_{\min}$ is always larger than $l_{\min}$.

Our current paper considers arbitrary $h_A/h_B$,   rational or irrational. Note that the $(\alpha_{opt}, \beta_{opt})$ in \eqref{optab} is optimal for both rational and irrational  $h_A/h_B$. However, we will show that  $d_{\min}^{(\alpha_{opt}, \beta_{opt})}$ is equal to $l_{\min}$ for   certain rational values of  $h_A/h_B$. Furthermore, $d_{\min}^{(\alpha_{opt}, \beta_{opt})}$ can be very close to $l_{\min}$ for some irrational values of $h_A/h_B$.

 \subsection{Issues not Addressed in \cite{yangtwc}}

In this paper, we focus on the following issues that were not addressed in \cite{yangtwc} but are crucial for practical implementations of  $q$-PAM linear PNC.
\begin{itemize}
  \item  [1)]  An important underlying assumption of \cite{yangtwc} is that $h_A/h_B$ is irrational. In real implementation, $h_A/h_B$ is  rational, since  the digital processors  have   finite resolution. In this paper, we analyze the performance of linear PNC systems for arbitrary $h_A/h_B$. We find that the statement of \emph{Theorem 1} in \cite{yangtwc} is correct for both rational and irrational  $h_A/h_B$\footnote{Even for irrational $h_A/h_B$, the proof of \emph{Theorem 1} in \cite{yangtwc} was still not complete, as elaborated below. Suppose that among all the superimposed symbol pairs, there are $J$ unique distances of different values arranged in ascending order: $l_{\min}=l_1<l_2<\ldots<l_{J}$. The proof of \emph{Theorem 1} in \cite{yangtwc} assumed that $d^{({\alpha, \beta})}_{\min}$ could be maximized by clustering superimposed symbol pairs  with distance $l_{\min}$ (please see \eqref{lmin}). It is not clear, however, that among the solutions that cluster superimposed symbol pairs with distance $l_{\min}$, whether there are some solutions that are better in that they also cluster superimposed symbol pairs with distance $l_2, l_3, \ldots$ up to $l_j$. If so, we want to choose a solution that maximizes $j$ so that $d^{(\alpha_{opt}, \beta_{opt})}_{\min} = l_{j+1}$. In order that focusing on clustering the symbol pair with $l_{\min}$ is enough to maximize $d^{(\alpha_{opt}, \beta_{opt})}_{\min}$,  we first need to show that clustering the symbol pair with $l_{\min}$ implies that we cannot cluster the symbol pairs with $l_2$ at the same time. This was the missing part in the proof in \cite{yangtwc}. This paper will provide this missing part later.}.   In addition, we find that for certain (but not all) rational values of $h_A/h_B$,     $d^{(\alpha_{opt}, \beta_{opt})}_{\min}=l_{\min}$. Yet for certain other rational $h_A/h_B$,  $l_{\min} =0$ but $d^{(\alpha_{opt}, \beta_{opt})}_{\min}$ is much larger than $0$.


 \item  [2)]  With the new analysis in this paper, we find that $d^{(\alpha_{opt}, \beta_{opt})}_{\min}$ varies in a continuous piecewise linear manner as the value of $h_A/h_B$ changes. This means that when  $h_A/h_B$ is irrational, $d^{(\alpha_{opt}, \beta_{opt})}_{\min}$ may still be very close to $l_{\min}$. This is the case, for example, if the irrational $h_A/h_B$ is very close to the aforementioned rational $h_A/h_B$ in 1)  for which $d^{(\alpha_{opt}, \beta_{opt})}_{\min} = l_{\min}$.

     \item [3)] For most values of $h_A/h_B$, $d^{(\alpha, \beta)}_{\min}$ dominates the SER as expressed in  \eqref{ser} in the high SNR regime, as stated in \cite{yangtwc}. However,  for certain values of $h_A/h_B$ where $d^{(\alpha, \beta)}_{\min}$ is near $l_{\min}$, the consideration of $\mathcal{A}^{(\alpha, \beta)}_{\min}$  becomes important.
         In this case, maximizing $d^{(\alpha, \beta)}_{\min}$ may not lead to an SER-optimal solution. In this paper, we give a systematic way to find $h_A/h_B$ where considering $d^{(\alpha, \beta)}_{\min}$ alone is not enough for SER-optimality.

 \item  [4)] As will be shown in this paper, we find that for high-order $q$-PAM where $q$ is large, a slight deviation of $h_A/h_B$ can cause a drastic drop in $d^{(\alpha_{opt}, \beta_{opt})}_{\min}$. This sensitivity may result in systems that are not robust. In particular, the SER  performance may degrade drastically with a slight change in channel coefficients. In this paper, we propose an asynchronized linear PNC design to overcome this sensitivity problem to stabilize the SER performance.

\end{itemize}

Details of  1) are given in \emph{Section V},   2)   in \emph{Section VI},   3) in \emph{Part A, Section VIII}, and   4) in  \emph{Section VII}. Before a rigorous treatment, we first illustrate with a specific example in the next section.

\section{An Example of   $7$-PAM Linear PNC}\label{sec:7pam}

In \emph{Part A} of this section, we consider  $7$-PAM linear PNC  to illustrate various issues and subtleties. In \emph{Part B}, we pose a series of questions to be answered for an in-depth understanding of  general $q$-PAM linear PNC.

\subsection{An Example }

Throughout this paper, we assume without loss of generality that both $h_A$ and $h_B$ are positive and that $h_A\geq h_B$. As explained in \emph{Appendix II}, there is no loss of generality in assuming positive $h_A$ and $h_B$.  Consider  a normalized version of \eqref{yr1} as follows:
\begin{align}\label{yrs}
    \frac{y_R}{h_B} =  \eta \sqrt{P} x_A +  \sqrt{P} x_B +\frac{z}{h_B},
    \end{align}
where $\eta=h_A/h_B$ with $\eta\geq 1$ and $x_i, i\in \{A, B\}$ is a $7$-PAM modulated symbol. For simplicity, let us assume normalization such  that $h_B=1$ and thereby $\eta=h_A$.  Accordingly, we have the scaled version  of    superimposed symbol   $w_S=\eta w_A+w_B$.

\begin{figure}[t]
 \centering
        \includegraphics[width=0.9\columnwidth]{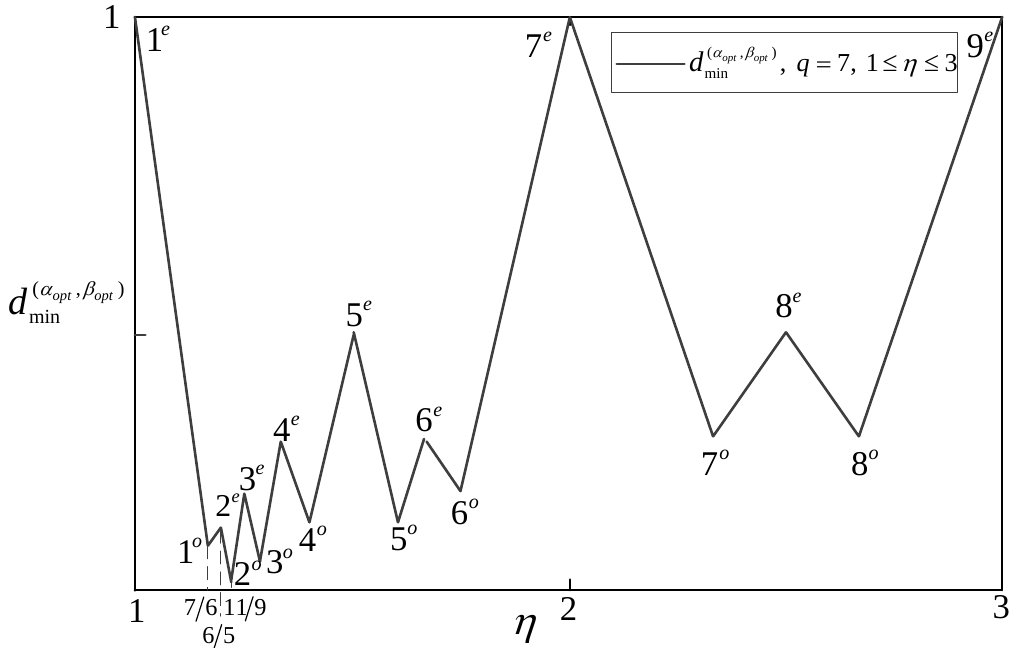}
       \caption{$d^{(\alpha_{opt}, \beta_{opt})}_{\min}$ versus $\eta\in [1,3]$ for $7$-PAM linear PNC.  $\{1^o, 2^o, \ldots, 8^o\}$ and $\{1^e, 2^e, \ldots, 9^e\}$ denote the positions of the odd turning points and the  even turning points respectively.}
        \label{fig:P713}
\end{figure}
\begin{figure}[t]
 \centering
        \includegraphics[width=0.9\columnwidth]{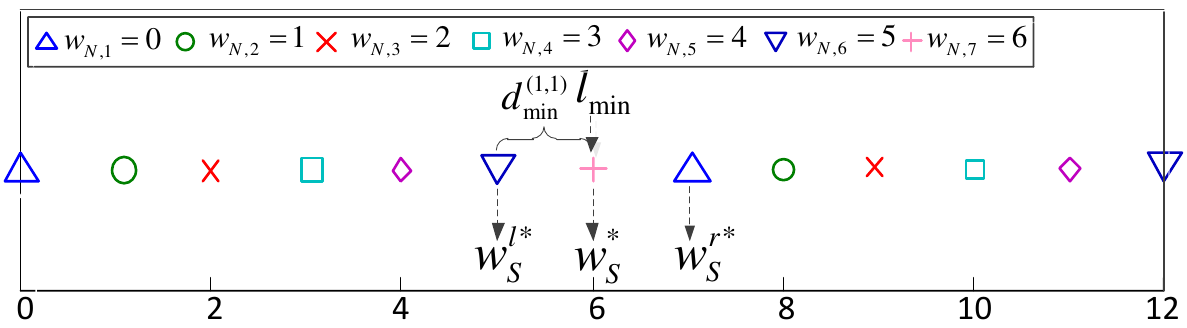}
       \caption{Clustered constellation of $\mathcal{W}_S$ for $7$-PAM linear PNC,  when $\eta=1$ and $(\alpha_{opt}, \beta_{opt})=(1,1)$. }
        \label{fig:P7h1cn}
\end{figure}
\begin{figure}[t]
 \centering
        \includegraphics[width=1\columnwidth]{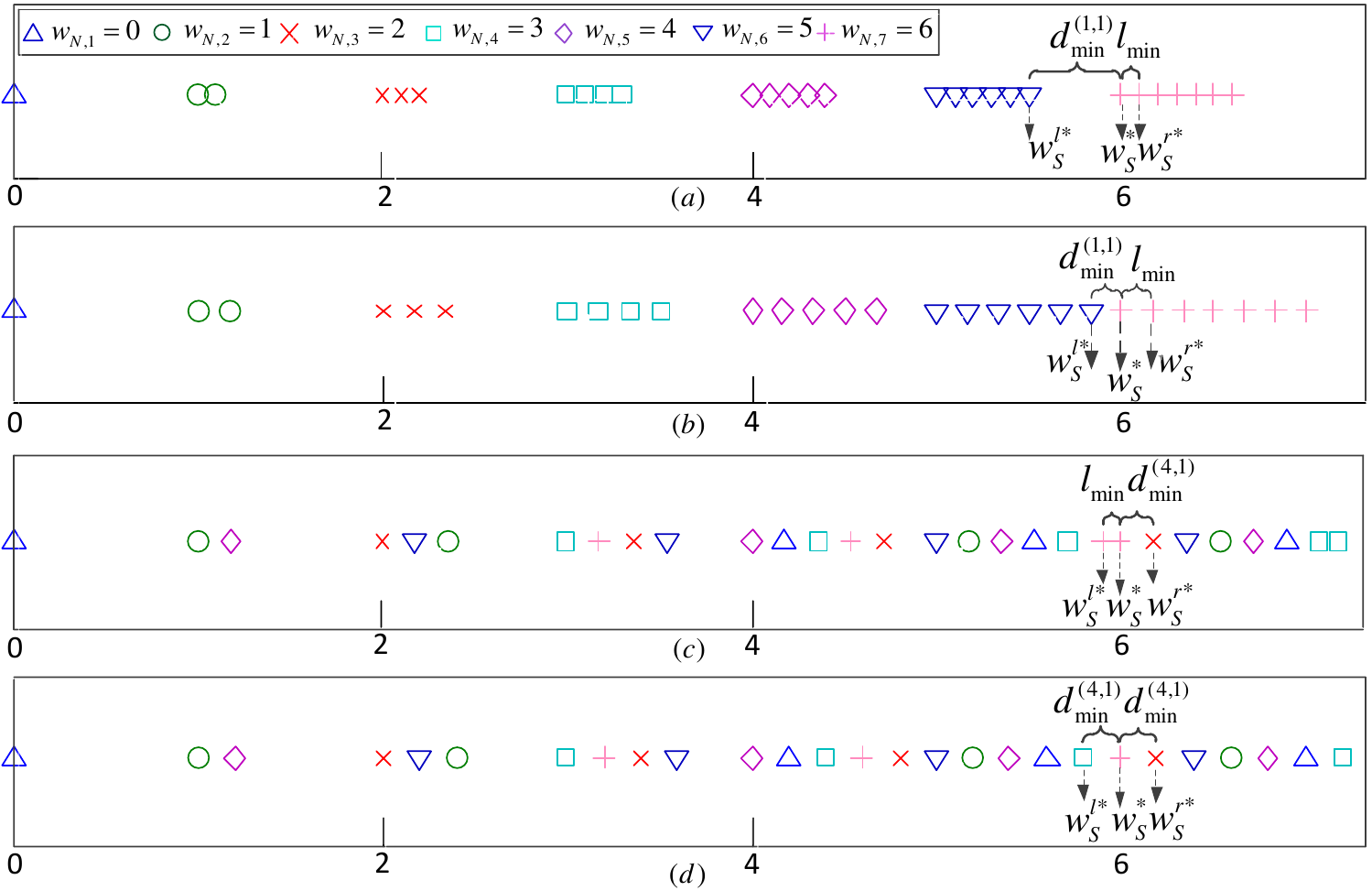}
       \caption{Clustered constellations of $\mathcal{W}_S$ for $7$-PAM linear PNC, when (a) $\eta=1.1$ and $(\alpha_{opt}, \beta_{opt})=(1,1)$; (b)  $\eta=7/6$ and $(\alpha_{opt}, \beta_{opt})=(1,1)$; (c) $\eta=1.18$ and $(\alpha_{opt}, \beta_{opt})=(4,1)$; (d)  $\eta=6/5$ and $(\alpha_{opt}, \beta_{opt})=(4,1)$. Constellations beyond $7$ on the real line are omitted to avoid cluttering.}
        \label{fig:comb}
\end{figure}

In the following, we analyze  how $d^{(\alpha_{opt}, \beta_{opt})}_{\min}$ varies with the gradual increase of $\eta$  starting with $\eta=1$. A goal of this paper is to derive a systematic method to compute the $d^{(\alpha_{opt}, \beta_{opt})}_{\min}$ versus $\eta$ curve for general $q$-PAM PNC. Fig. \ref{fig:P713} shows $d^{(\alpha_{opt}, \beta_{opt})}_{\min}$ for  $1\leq \eta \leq 3$.   Figs. \ref{fig:P7h1cn} and   \ref{fig:comb} illustrate how the superimposed symbols in $\mathcal{W}_S$ move   as $\eta$ increases. In Figs. \ref{fig:P7h1cn} and   \ref{fig:comb},  we use different shapes to denote  different sets of clustered superimposed symbols and each set corresponds to the same NC symbol. Figs. \ref{fig:P7h1cn} and   \ref{fig:comb} assumes the $(\alpha_{opt}, \beta_{opt})$ as per \eqref{optab}.  As we gradually increase $\eta$, the constellation, and the corresponding $(\alpha_{opt}, \beta_{opt})$,  evolves accordingly. In particular,  as $\eta$ increases,  the overlapped symbols in  Fig. \ref{fig:P7h1cn}  separate  and move to the right,  as illustrated in Fig. \ref{fig:comb}.

 In this paper, we introduce the concept of a reference symbol,  which plays  a critical role in aiding our understanding of the interplay between $\eta$ and $d_{\min}^{(\alpha_{opt}, \beta_{opt})}$.
 For general $q$-PAM, we define $(w^*_A, w^*_B)=(0,q-1)$ as a \emph{reference joint symbol}, and the corresponding $w^*_S=q-1$ as a \emph{reference superimposed symbol}.
%

Two reasons stand behind our choice of $(w^{*}_A, w^{*}_B)=(0,q-1)$ as the reference symbol. First, $w^{*}_S$ is  invariant (a static point on the constellation) as $\eta$ increases (see Figs. \ref{fig:P7h1cn} and   \ref{fig:comb}),  because it does not depend on $\eta$. Second,  as we will show later (in the \emph{Principal Theorem}, \emph{Section \ref{sec:qpam}}), the distances of $w^*_S$ to its immediate left neighbor and immediate right neighbor are either $d^{(\alpha_{opt}, \beta_{opt})}_{\min}$ or $l_{\min}$. As $\eta$ increases, the superimposed symbols that are the left and right neighbors of $w^*_S$ may change, but $d^{(\alpha_{opt}, \beta_{opt})}_{\min}$ can always be found by analyzing the distances of $w^*_S$ to these two neighbors. Thus, although $d^{(\alpha_{opt}, \beta_{opt})}_{\min}$  is a  property of the overall constellation,   $w^*_S$ is a special point in that analyzing its locality allows us to obtain $d^{(\alpha_{opt}, \beta_{opt})}_{\min}$. In particular, the ``global'' optimization problem in \eqref{optab} becomes a ``local'' optimization problem with the aid of $w^*_S$.

Let us now delve into   Figs. \ref{fig:P7h1cn} and  \ref{fig:comb} to interpret the curve of $d^{(\alpha_{opt}, \beta_{opt})}_{\min}$ in Fig. \ref{fig:P713}.  We denote the immediate left and right neighbors of $w^{*}_S$ by $w^{l*}_S$ and $w^{r*}_S$, respectively. The joint symbols associated with $w^{l*}_S$ and $w^{r*}_S$ are denoted by $(w^{l*}_A, w^{l*}_B)$ and $(w^{r*}_A, w^{r*}_B)$, respectively, when they are unique. In the following, we show how $d^{(\alpha_{opt}, \beta_{opt})}_{\min}$ varies with $\eta$ by examining the values of $d^{(\alpha_{opt}, \beta_{opt})}_{\min}$  at a few selected values of $\eta$.

In reading the following paragraphs, the reader is advised to keep a mental picture that the distances between superimposed symbols evolve in a continuous manner in between the selected $\eta$ values below. This continuous evolution leads to the piecewise linear nature of the $d^{(\alpha_{opt}, \beta_{opt})}_{\min}$ versus $\eta$ curve in Fig. \ref{fig:P713}.

1) $\eta=1$

 At $\eta=1$,  $l_{\min}=0$, since  multiple joint symbols have superimposed symbols that overlap with each other. For example,  there are multiple joint symbols $(w_A, w_B)$ satisfying $w_A+w_B=6$. In this paper, joint symbols with the same superimposed symbol are said to overlap.

 We can easily verify that    $(\alpha_{opt}, \beta_{opt})=(1,1)$ can map these overlapped joint symbols to the same NC symbol. Fig. \ref{fig:P7h1cn} shows the constellation of $\mathcal{W}_S$ when $\eta=1$ with NC mapping by $(\alpha_{opt}, \beta_{opt})=(1,1)$. In this case, $d^{(1, 1)}_{\min}=1$. This is the largest $d_{\min}$ that $7$-PAM linear PNC can achieve for any $\eta$.

2) $\eta=11/10$

In  Fig. \ref{fig:comb} (a), we  increase $\eta$ to $11/10$.  Once $\eta$ is slightly larger than $1$,  the overlapped   symbols   in Fig. \ref{fig:P7h1cn} are separated.  The left and right neighbors of $w^*_S$ are   $(w^{l*}_A, w^{l*}_B)=(5,0)$ and $(w^{r*}_A, w^{r*}_B)=(1,5)$, respectively. In this case,    $l_{\min}=|w^{r*}_S-w^*_S|$. With $(\alpha_{opt}, \beta_{opt})=(1,1)$,  joint symbols separated by $l_{\min}$ can be clustered into the same set. Moreover,  $d^{(\alpha_{opt}, \beta_{opt})}_{\min}=|w^*_S-w^{l*}_S|$. Compared with the case of $\eta=1$,  $d^{(\alpha_{opt}, \beta_{opt})}_{\min}$ is decreased while $l_{\min}$ is increased.

3) $\eta=7/6$

  In Fig. \ref{fig:comb} (b),  we increase $\eta$ to $7/6$.   When $\eta=7/6$,   $(w^{l*}_A, w^{l*}_B)$ and $(w^{r*}_A, w^{r*}_B)$ are still the same as in cases 1) and 2). However, $|w^{r*}_S-w^*_S|=|w^{*}_S-w^{l*}_S|=l_{\min}$. Here,  two different $(\Delta_{A}, \Delta_{B})$ are possible---see (34) for the definition of $(\Delta_{A}, \Delta_{B})$. Specifically, the two $(\Delta_{A}, \Delta_{B})$, $(\Delta_{A}, \Delta_{B})  = (1, -1)$ and $(5, -6)$, are given by the  the differences of the reference symbol and the right and left neighbors. In this case, we cannot find a pair $(\alpha_{opt}, \beta_{opt})$ such that $w^{l*}_S$, $w^{*}_S$, and $w^{r*}_S$ are clustered  to the same NC symbol (this will be proved formally in \emph{Lemma \ref{lem:lmin}, Section V}). In this example, we choose to set $(\alpha_{opt}, \beta_{opt})=(1,1)$ so that $w^*_S$ and $w^{r*}_S$ are clustered together. In this case, $d^{(1,1)}_{\min}=l_{\min}$, the distance between $w^*_S$ and $w_S^{l*}$.

4) $\eta=59/50$

If we further  increase $\eta$ a little, the balance $|w^{r*}_S-w^*_S|=|w^{*}_S-w^{l*}_S|$ in case 3) is broken. In particular, $|w^{r*}_S-w^*_S|$ becomes larger than $|w^{*}_S-w^{l*}_S|$. As shown in Fig. \ref{fig:comb} (c), $l_{\min}=|w^*_S-w^{l*}_S|$. Here, we set $(\alpha_{opt}, \beta_{opt})=(4,1)$ to cluster $w^{*}_S$ and $w^{l*}_S$. The left and right neighbors of $w^{*}_S$ are the same as in the previous cases, but we have a different clustering scheme.

At this juncture, let us pause and  introduce the concept of a turning point with respect to the $d^{(\alpha_{opt}, \beta_{opt})}_{\min}$ versus $\eta$ plot in Fig. \ref{fig:P713}.  The case of $d^{(\alpha_{opt}, \beta_{opt})}_{\min}$ with $\eta=7/6$, for example,  is an instance of a turning point in Fig. \ref{fig:P713}. From 1)-4),  note that $d^{(\alpha_{opt}, \beta_{opt})}_{\min}$ decreases as $\eta$ goes from $1$ to $7/6$ and increases as  $\eta$ goes from $7/6$ to $59/50$.   This turning point is a trough.

5) $\eta=6/5$

As shown in Fig. \ref{fig:comb} (d), when $\eta$ is further increased to $6/5$, the previous left neighbor of $w^*_S$ now  overlaps  with $w^*_S$. Thus, $l_{\min}=0$. In this case, the overlapped symbols can be clustered together by $(\alpha_{opt}, \beta_{opt})=(4,1)$. Therefore, $d^{(\alpha_{opt}, \beta_{opt})}_{\min}=|w^{*}_S-w^{l*}_S|=|w^{r*}_S-w^{*}_S|$.

As $\eta$ goes beyond $6/5$ a bit, we see that $d^{(\alpha_{opt}, \beta_{opt})}_{\min}$ decreases while $l_{\min}$  increases. Now,  $d^{(\alpha_{opt}, \beta_{opt})}_{\min}$ goes up from  $\eta=7/6$ to $6/5$. Therefore, we come across a different type of turning point at $\eta=6/5$: the turning point is a peak.

In general, let us define $\eta=1$ as the position of the first turning point. As $\eta$ increases, we come across the second  turning point, the third turning point and so on. Between two adjacent turning points, $d^{(\alpha_{opt}, \beta_{opt})}_{\min}$ is linear.  The \emph{odd turning points} form the bottoms (troughs) and the \emph{even turning points} form the tops  (peaks) of the overall piecewise linear curve. As shown in Fig. \ref{fig:P713}, the point at $\eta=7/6$ (denoted by $1^{o}$) is the first odd turning point, and the the point at $\eta=6/5$ (denoted by $2^{e}$) is the second even turning point. \emph{Section \ref{sec:sys}} will give a systematic way to identify all the odd turning points and all the even turning points.

  As we will show in \emph{Subsidiary  Theorem} of \emph{Section V},  in general,  $d^{(\alpha_{opt}, \beta_{opt})}_{\min}=1$ and   no turning point appears on the curve of $d^{(\alpha_{opt}, \beta_{opt})}_{\min}$ for all $\eta\geq q-1$.


\subsection{Preliminary Results and Problem Statement}

 From  the example in \emph{Part A}, we see that   $d^{(\alpha_{opt}, \beta_{opt})}_{\min}$  changes in a continuous fashion as $\eta$ varies, and there is no abrupt breakdown at rational points. In particular, the example illustrates issues 1), 2), and 4) in \emph{Part E, Section III}, as explained below:
\begin{itemize}
 \item With respect to \emph{Issue 1)}, the 7-PAM example in \emph{Part A} illustrates that each of the odd turning points in  $\{1^o, 2^o, \ldots\}$ corresponds  to a rational $\eta$. At an odd turning point,  $d^{(\alpha_{opt}, \beta_{opt})}_{\min}$ reaches a local minimum. Each of the even turning points $\{1^e, 2^e, \ldots\}$ also corresponds to a rational $\eta$. At an even turning point,  $d^{(\alpha_{opt}, \beta_{opt})}_{\min}$   reaches a local maximum. Between the local minimums and local maximums, there are many other rational $\eta$ that do not form turning points. These rational $\eta$ behave more like irrational $\eta$  in that $d^{(\alpha_{opt}, \beta_{opt})}_{\min}>l_{\min} > 0$.

 \item With respect to \emph{Issue 2)}, the 7-PAM example in \emph{Part A} illustrates that near an odd turning point, $\eta$ is irrational.  In this case, $d^{(\alpha_{opt}, \beta_{opt})}_{\min}$ is very close to $l_{\min}$.

 \item With respect to \emph{Issue 4)}, the 7-PAM example in \emph{Part A} illustrates that $d^{(\alpha_{opt}, \beta_{opt})}_{\min}$ drops from $1$ at $\eta=1$ to a very low value at $\eta=7/6$.  We will see later (in \emph{Section \ref{sec:sen}}) that at larger $q$, a small change in $\eta$ will lead to an even larger drop in $d^{(\alpha_{opt}, \beta_{opt})}_{\min}$.
\end{itemize}

Going beyond illustration by example, we next give a rigorous  treatment for the general $q$-PAM case. In particular, we will answer the following questions:

\begin{itemize}


\item[Q1)]  Why and how  can  the reference symbol  $(w^*_A, w^*_B) = (0, q-1)$ and its associated superimposed symbol $w^*_S$  be used to  track $d^{(\alpha_{opt}, \beta_{opt})}_{\min}$ and $l_{\min}$?

\item[Q2)]  How can the odd and even turning points be determined, and from them the $d^{(\alpha_{opt}, \beta_{opt})}_{\min}$ versus $\eta$ relationship be derived,  in a systematic way (i.e., through an algorithm/equation)?

\item[Q3)] How can the analysis used in the derivation of $d^{(\alpha_{opt}, \beta_{opt})}_{\min}$ versus $\eta$ curve in Q2) be used to determine the SER-optimal $(\alpha, \beta)$?

\item[Q4)] How sensitive is the SER performance to $\eta$, and can the sensitivity be removed?

\end{itemize}

In the following, we address Q1) in \emph{Section \ref{sec:qpam}}, Q2) in \emph{Section \ref{sec:sys}}, Q3) in \emph{Section \ref{sec:sim}}, and  Q4) in \emph{Section \ref{sec:sen}}.

\section{Relationship between reference symbol and its neighbors in $q$-PAM Linear PNC}\label{sec:qpam}

This section  provides the rigorous proof  that the reference symbol $(w^*_A, w^*_B) = (0, q-1)$  and its associated superimposed symbol $w^*_S$  can be used to track $d^{(\alpha_{opt}, \beta_{opt})}_{\min}$  in  general $q$-PAM linear PNC. In particular, we prove that $d^{(\alpha_{opt}, \beta_{opt})}_{\min}$ and $l_{\min}$ must be observable either  as the distance  between $w^*_S$ and its left   neighbor or the distance between $w^*_S$ and its right neighbor; likewise for $l_{\min}$ except that it is also possible for $l_{\min}$ to be equal to $0$ when there is a symbol overlapping with the reference symbol.

\emph{Lemmas \ref{lem2}} to \emph{\ref{lem:lmin}}, leading to the \emph{Principal Theorem},  provide the basis using the reference symbol $(w^*_A, w^*_B)$  to identify $l_{\min}$ and $d^{(\alpha_{opt}, \beta_{opt})}_{\min}$.

For the rest of this paper, for convenience, we will use the word ``symbol'' to refer to either the joint symbol $(w_A, w_B)$ or the superimposed symbol $w_S$. The intended meaning is implied by the context.

\subsection{Observability of $l_{\min}$ and $d^{(\alpha_{opt}, \beta_{opt})}_{\min}$ from Vantage Point of Reference Symbol}

 \begin{lemma}\label{lem2}
Consider two distinct  joint symbols $(w_{A,1}, w_{B,1})$ and $(w_{A,2}, w_{B,2})$ with corresponding superimposed symbols   $w_{S,1}$ and $w_{S,2}$, respectively. Let $(\delta_A, \delta_B)\triangleq (w_{A,1}-w_{A,2}, w_{B,1}-w_{B,2})$. If $|w_{S,1}-w_{S,2}|$ is $l_{\min}$ or $d^{(\alpha_{opt}, \beta_{opt})}_{\min}$, then $\delta_A$ and  $\delta_B$ cannot be both nonzero and of the same sign at the same time (i.e., if $|w_{S,1} - w_{S,2}|$ is $l_{\min}$ or $d^{(\alpha_{opt}, \beta_{opt})}_{\min}$, the only possibility is either $\delta_A\geq 0, \delta_B<0$ or $\delta_A<0, \delta_B\geq0$).
\end{lemma}

\begin{IEEEproof}[Proof of Lemma \ref{lem2}]

Suppose that $\delta_A$ and $\delta_B$ are both nonzero and of the same sign. Let us assume that $0<\delta_A, \delta_B \leq q-1$ (if  $-(q-1)\leq\delta_A, \delta_B<0$, then we just switch the joint symbols $(w_{A,1}, w_{B,1})$ and $(w_{A,2}, w_{B,2})$ to make $0<\delta_A, \delta_B\leq q-1$). Consider a different joint symbol  given by  $(w_{A,3}, w_{B,3})=(w_{A,2}, w_{B,1})$ with corresponding superimposed symbol $w_{S,3}$.

 First, it is easy to verify that $|w_{S,1}-w_{S,3}|<|w_{S,1}-w_{S,2}|$. Thus, $|w_{S,1}-w_{S,2}|$ is not $l_{\min}$.

Second, based on \emph{Proposition \ref{pro:clu}}, we cannot cluster $(w_{A,1}, w_{B,1})$ with $(w_{A,3}, w_{B,3})$
through a linear PNC mapping, since $w_{B,1}=w_{B,3}$. Therefore, $d^{(\alpha_{opt}, \beta_{opt})}_{\min}\leq |w_{S,1}-w_{S,3}|$, and  $|w_{S,1}-w_{S,2}|$ cannot possibly be  $d^{(\alpha_{opt}, \beta_{opt})}_{\min}$.

 Thus,  $|w_{S,1}-w_{S,2}|$ is not equal to $l_{\min}$ or $d^{(\alpha_{opt}, \beta_{opt})}_{\min}$.

\end{IEEEproof}

\begin{lemma}\label{lem3}
Consider the reference  symbols $(w^{*}_A, w^{*}_B)=(0, q-1)$ with corresponding  superimposed symbol  $w^{*}_S$. i) Given any    $(\delta_A, \delta_B)$ such that  $0\leq\delta_A\leq q-1, -(q-1) \leq \delta_B<0$, there is a valid joint symbol given by $(w_A, w_B)=(w^{*}_A, w^{*}_B)+(\delta_A, \delta_B)$.  ii) Given any    $(\delta_A, \delta_B)$ such that   $-(q-1) \leq \delta_A<0, 0\leq \delta_B \leq q-1$, there is a valid joint  symbol given by $(w_A, w_B)=(w^{*}_A, w^{*}_B)-(\delta_A, \delta_B)$.
\end{lemma}

\begin{IEEEproof}[Proof of Lemma \ref{lem3}]

We consider case i). The proof of case ii) is similar. We have $(w_A, w_B)=(0, q-1)+(\delta_A, \delta_B)$. We see that  $0\leq w_A, w_B \leq q-1$.  Thus, $(w_A, w_B)$ is a valid joint symbol.

\end{IEEEproof}

\begin{corollary}[of Lemmas 1 and 2]\label{cor:short}
  $l_{\min}$ and $d^{(\alpha_{opt}, \beta_{opt})}_{\min}$ can be observed by looking at the distances between the reference superimposed symbol $w^*_S$ and other superimposed symbols. That is, $l_{\min} = |w^*_S - w_S|$ for some superimposed symbol $w_S$ induced by joint symbol $(w_A, w_B) \neq (w^*_A, w^*_B)$; $d^{(\alpha_{opt}, \beta_{opt})}_{\min} = |w^*_S - w'_S|$ for some superimposed symbol $w'_S$  induced by joint symbol $(w'_A, w'_B) \neq (w^*_A, w^*_B)$.
\end{corollary}

\begin{IEEEproof}[Proof of Corollary \ref{cor:short}]

The combination of \emph{Lemmas \ref{lem2}} and \emph{\ref{lem3}} gives this corollary. That is, suppose $l_{\min}$ or $d^{(\alpha_{opt}, \beta_{opt})}_{\min}$ is equal to $|w_{S,1}-w_{S,2}|=|\eta\delta_A+\delta_B|$ with respect to the bracketed clause in the statement of  \emph{Lemma \ref{lem2}}. \emph{Lemma \ref{lem3}} says that we can find
$(w_A, w_B)$ such that  $|w^*_S - w_S|=|\eta\delta_A+\delta_B|$.

\end{IEEEproof}

\begin{remark}\label{rem:short}
 Note that $l_{\min}$ and $d^{(\alpha_{opt}, \beta_{opt})}_{\min}$ could also be observed at other places on the constellation. \emph{Corollary \ref{cor:short}} just says that $l_{\min}$ and $d^{(\alpha_{opt}, \beta_{opt})}_{\min}$ must be observable as the distances between $w^*_S$ and some other superimposed symbols. In \emph{Parts B} and \emph{C}, we will prove that these superimposed symbols  are either the immediate left or immediate right neighbor of the reference symbol, or overlap with the reference symbol (in the case where $l_{\min}=0$).
\end{remark}
\rightline{$\blacksquare$}

\subsection{Intermediate Results Leading to Principal Theorem and Algorithm for Identifying $l_{\min}$, $d^{(\alpha_{opt}, \beta_{opt})}_{\min}$, and $(\alpha_{opt}, \beta_{opt})$,}

In this part, we provide some intermediate results that lead to the \emph{Principal Theorem} in \emph{Part C} on how the reference symbol can aid identification of $l_{\min}$ and $d^{(\alpha_{opt}, \beta_{opt})}_{\min}$. These intermediate results are also used to establish our algorithm for systematically identifying $l_{\min}$, $d^{(\alpha_{opt}, \beta_{opt})}_{\min}$, and $(\alpha_{opt}, \beta_{opt})$ as $\eta$ varies presented in \emph{Section VI}.

\begin{definition}\label{def:po}
   Consider the  constellation formed by all  superimposed symbols $w_S$ for a given $\eta$. On this constellation, by the \emph{position} of a joint symbol $(w_A, w_B)$, we mean the value of its superimposed symbol $w_S$. Given a  distinct joint symbol  $(w'_A, w'_B)$ with an associate $w'_S$,  the distance between $(w_A, w_B)$ and $(w'_A, w'_B)$ is $d=|w_S-w'_S|$.  We define the    \emph{orientation} of $(w_A, w_B)$   with respect to another joint symbol $(w'_A, w'_B)$ as follows. The joint symbol $(w_A, w_B)$ is said to
   \begin{itemize}
     \item  reside  on the left of   $(w'_A, w'_B)$ if $w_S-w'_S<0$;
     \item  reside  on the right of   $(w'_A, w'_B)$ if $w_S-w'_S>0$;
     \item   overlap  with  $(w'_A, w'_B)$ when $w_S-w'_S=0$.
   \end{itemize}
   The joint symbol $(w_A, w_B)$ is said to be the left (or right) neighbor of $(w'_A, w'_B)$ if $w_S$ is the superimposed symbol \textbf{closest} to $w'_S$ on the left (or right).
\end{definition}
\rightline{$\blacksquare$}

We remark that the neighbors of a joint symbol $(w_A, w_B)$, as well as the joint symbols overlapping with $(w_A, w_B)$, may be different for different $\eta$. Indeed, as will be explained later, one of our interests is to examine how the neighbors of the reference symbol $(w^{*}_A, w^{*}_B)$ change as $\eta$ varies.

 \begin{definition}
 We define the following notations:
 \begin{itemize}
   \item $(w^{l*}_A, w^{l*}_B)$ is the left neighbor  of the reference joint symbol $(w^{*}_A, w^{*}_B)$,  and $(\delta^{l*}_A, \delta^{l*}_B)\triangleq(w^{l*}_A-w^{*}_A, w^{l*}_B-w^{*}_B)=(w^{l*}_A, w^{l*}_B-(q-1))$;

   \item $(w^{r*}_A, w^{r*}_B)$ is the right neighbor  of the reference joint symbol  $(w^{*}_A, w^{*}_B)$, and $(\delta^{r*}_A, \delta^{r*}_B)\triangleq (w^{r*}_A-w^{*}_A, w^{r*}_B-w^{*}_B)=(w^{r*}_A, w^{r*}_B-(q-1))$;

   \item $(w^{o*}_A, w^{o*}_B)$  overlaps with $(w^{*}_A, w^{*}_B)$, and $(\delta^{o*}_A, \delta^{o*}_B)\triangleq(w^{o*}_A-w^{*}_A, w^{o*}_B-w^{*}_B)=(w^{o*}_A, w^{o*}_B-(q-1))$, if any. Note that $(w^{o*}_A, w^{o*}_B)$ exists   for certain values of $\eta$ only (specifically, at the even turning points).
 \end{itemize}
 \end{definition}
\rightline{$\blacksquare$}

 We emphasize that  $(w^{l*}_A, w^{l*}_B)$, $(w^{r*}_A, w^{r*}_B)$, and $(w^{o*}_A, w^{o*}_B)$ may not be unique. For example,  for a particular $\eta$, we may  have multiple left neighbors of $(w^{*}_A, w^{*}_B)$ that overlap with each other. By $(w^{l*}_A, w^{l*}_B)$, we mean any left neighbor; likewise for $(w^{r*}_A, w^{r*}_B)$  and $(w^{o*}_A, w^{o*}_B)$.

\begin{lemma}\label{lem:ovl}
  Suppose that for a particular $\eta$, $K$ distinct joint symbols, $(w_{A,1}, w_{B,1}), \ldots, (w_{A,K}, w_{B,K})$ overlap with each other. Then, $(w_{A,1}, w_{B,1}), \ldots, (w_{A,K}, w_{B,K})$ can be clustered together and mapped to the same NC symbol  with a specific $(\alpha, \beta)$. Furthermore, for the same $\eta$, if there are $K'$ other distinct joint symbols $(w'_{A,1}, w'_{B,1}), \ldots, (w'_{A,K}, w'_{B,K})$ overlapping with each other, the same $(\alpha, \beta)$ will also cluster them together.
\end{lemma}

\begin{IEEEproof}[Proof of Lemma \ref{lem:ovl}]

  Consider any two joint symbols among $(w_{A,1}, w_{B,1}), \ldots, (w_{A,K}, w_{B,K})$, say $(w_{A,i}, w_{B,i})$ and  $(w_{A,j}, w_{B,j})$. Since $(w_{A,i}, w_{B,i})$ and  $(w_{A,j}, w_{B,j})$ overlap with each other,
  \begin{align}\label{over1}
    \eta w_{A,i}+ w_{B,i}=  \eta w_{A,j}+ w_{B,j}.
  \end{align}
Given that $(w_{A,i}, w_{B,i})$ and  $(w_{A,j}, w_{B,j})$ are distinct---i.e., $(w_{A,i}, w_{B,i}) \neq (w_{A,j}, w_{B,j})$---\eqref{over1} implies that both $w_{A,i}\neq w_{A,j}$ and $w_{B,i}\neq w_{B,j}$ must be true (for example, we cannot have $w_{A,i}\neq w_{A,j}$ but $w_{B,i}=w_{B,j}$). Now, since $w_{A,i}\neq w_{A,j}$, for \eqref{over1} to be true, $\eta$ must be a rational number given by $\frac{w_{B,j}-w_{B,i}}{w_{A,i}-w_{A,j}}$, i.e., overlapping among symbols can only occur under rational $\eta$. Let us write $\eta=m/n$ for some coprime integers $m$ and $n$. Then, \eqref{over1} can be written as
  \begin{align}\label{over2}
    m w_{A,i}+ n w_{B,i}=  m w_{A,j}+ n w_{B,j}.
  \end{align}

  Consider the NC coefficient pair $(\alpha, \beta)={\rm mod}\big((m, n), q\big)$. Taking mod $q$ on both sides of \eqref{over2} gives
    \begin{align}\label{over3}
\alpha \otimes w_{A,i} \oplus \beta\otimes  w_{B,j} =  \alpha\otimes w_{A,j} \oplus \beta\otimes  w_{B,j}.
  \end{align}

  Applying the above argument to all pairs of joint symbols $(w_{A,1}, w_{B,1}), \ldots, (w_{A,K}, w_{B,K})$, we have that
   \begin{align}\label{over4}
\nonumber &\alpha \otimes w_{A,1} \oplus \beta\otimes  w_{B,1}=  \alpha\otimes w_{A,2} \oplus \beta\otimes  w_{B,2}=\ldots\\
&= \alpha\otimes w_{A,K} \oplus \beta\otimes  w_{B,K},\\
&w_{A,1}\neq w_{A,2} \neq \ldots \neq w_{A,K},  \\
& {\rm and} \ w_{B,1}\neq w_{B,2} \neq \ldots \neq w_{B,K}.
  \end{align}

  Thus,  all the $K$ joint symbols can be clustered together and mapped to the same NC symbol through $(\alpha, \beta)={\rm mod}((m,n), q)$.

With respect to the $K'$ other overlapping joint symbols, we note that since $\eta$ is fixed, it is still $m/n$, and  \eqref{over2} simply becomes $ m w'_{A,i}+ n w'_{B,i}=  m w'_{A,j}+ n w'_{B,j}$. And \eqref{over4} becomes
   \begin{align}\label{over5}
\nonumber &\alpha \otimes w'_{A,1} \oplus \beta\otimes  w'_{B,1}=  \alpha\otimes w'_{A,2} \oplus \beta\otimes  w'_{B,2}=\ldots\\
&= \alpha\otimes w'_{A,K} \oplus \beta\otimes  w'_{B,K},\\
&w'_{A,1}\neq w'_{A,2} \neq \ldots \neq w'_{A,K},  \\
& {\rm and} \ w'_{B,1}\neq w'_{B,2} \neq \ldots \neq w'_{B,K}.
  \end{align}
It is easy to note that the same $(\alpha, \beta)$ will cluster these $K'$ joint symbols together (although the NC symbol of these $K'$ symbols may be different from that of the $K$ symbols).

\end{IEEEproof}

Although \emph{Lemma 3} is not about reference symbol \emph{per se}, together with \emph{Corollary 1}, it implies that when $l_{\min}=0$ and the reference symbol overlaps with other joint symbols, the $(\alpha, \beta)$ that clusters the reference symbol with these overlapping joint symbols also clusters other sets of joint symbols overlapping  at other places in the constellation. The implication is that focusing on clustering in the neighborhood of the reference symbol is good enough to ensure $d^{(\alpha_{opt}, \beta_{opt})}_{\min}>0$ when $l_{\min}=0$.

It turns out the $l_{\min}$ and $d^{(\alpha_{opt}, \beta_{opt})}_{\min}$ for $\eta=1$ and $\eta\geq q-1$ can be obtained easily. They will be given in a \emph{Subsidiary  Theorem} later. The derivations of $l_{\min}$ and $d^{(\alpha_{opt}, \beta_{opt})}_{\min}$ for $1< \eta < q-1$ is relatively more difficult. To derive the $l_{\min}$ and $d^{(\alpha_{opt}, \beta_{opt})}_{\min}$ in this range, a careful analysis of how the left and right neighbors of $(w^*_A, w^*_B)$ change as $\eta$ varies is needed. The main results are given in \emph{Lemmas \ref{lem:left}-\ref{lem:lmin}}.

\begin{lemma}\label{lem:left}
For a given $\eta$ in the range  $1< \eta < q-1$, there is at least one joint symbol  residing on the left of $(w^*_A, w^*_B)$ that is closer to $(w^*_A, w^*_B)$ than $(0, q-2)$, separated from $(w^*_A, w^*_B)$ by a distance smaller than $1$. At this $\eta$,  there is also  at least one joint symbol residing on the right of $(w^*_A, w^*_B)$ separated from $(w^*_A, w^*_B)$ by a distance  smaller than or equal to $1$.
\end{lemma}

\begin{IEEEproof}[Proof of Lemma \ref{lem:left}]

 First, let us first consider non-integer $\eta$ in this range (i.e., $\eta \neq 2, 3, \ldots,$ or $q-2$).
 Let $\eta=k+\epsilon$ where $k\in \{1, \ldots, q-2\}$ and $0<\epsilon<1$. That is, $k$ is the integer part and $\epsilon$ is the fractional part of $\eta$.  For any given $k+\epsilon$,  we can verify  that $(w_A,w_B)=(1,q-2-k)$  resides in between $(w^*_A, w^*_B)$ and  $(0, q-2)$. At this $\eta$, we can also verify that another joint symbol $(w_A,w_B)= (1, q-2-(k-1))$ resides on the right of $(w^*_A, w^*_B)$ and $w_S-w^*_S= \epsilon <1$.

Next, we consider $\eta=k$ where $k \in \{2, 3, \ldots, q-2\}$. We can verify that $(w_A, w_B) = (1, q-2-(k-1))$ overlaps with $(w^*_A, w^*_B)$ and hence is closer to $(w^*_A, w^*_B)$ than $(0, q-2)$. At this $\eta$, we can also verify that $(w_A, w_B) = (1, q-2-(k-2))$ resides on the right of of $(w^*_A, w^*_B)$ and $w_S-w^*_S=1$.

\end{IEEEproof}

\begin{remark}\label{rem:left}
  In the first part of the proof of \emph{Lemma \ref{lem:left}}, $(w_A,w_B)=(1,q-2-k)$ may or may not be the left neighbor of the reference symbol. The implication of the proof is that a left neighbor of the reference symbol is closer to the reference symbol than $(0, q-2)$, and that $(0, q-2)$ cannot be a left neighbor for $\eta$ in the range $1< \eta < q-1$. Note that $(0, q-2)$ is a left neighbor of the reference symbol at $\eta=1$ and $\eta\geq q-1$.
\end{remark}
\rightline{$\blacksquare$}

\begin{definition}
  Consider a joint symbol $(w_A,w_B)$. As $\eta$ increases,
  \begin{itemize}
    \item $(w_A,w_B)$ is a \emph{moving} symbol if the value of $w_S$ goes up;
    \item  $(w_A,w_B)$ is a \emph{static} symbol if the value of $w_S$ keeps unchanged.
  \end{itemize}
\end{definition}
\rightline{$\blacksquare$}

Therefore, the joint symbols with $w_A=0$ are the static symbols and the joint symbols with $w_A>0$ are the moving symbols. Note that $w_S$ of a joint symbol is a continuous linear function of $\eta$ with slope $w_A$.

\begin{corollary}\label{cor:left}

 Suppose that $(w^{l*}_A, w^{l*}_B)$ and $(w^{r*}_A, w^{r*}_B)$ are left and right neighbors of $(w^{*}_A, w^{*}_B)$ for  some $\eta$ in the range  $1< \eta < q-1$. Then, $(w^{l*}_A, w^{l*}_B)$  and $(w^{r*}_A, w^{r*}_B)$ must be  two  moving symbols.

\end{corollary}

\begin{IEEEproof} [Proof of Corollary \ref{cor:left}]

When $\eta=1$, $(0, q-2)$, a static symbol,  is a left neighbor of the reference symbol. Thus, any other joint symbol that becomes closer to the reference symbol than $(0, q-2)$ at $\eta>1$  is a moving point. \emph{Lemma \ref{lem:left}} implies a left neighbor of   $(w^{*}_A, w^{*}_B)$, i.e., $(w^{l*}_A, w^{l*}_B)$, is a moving symbol.

Next, we note that the reference symbol $(0, q-1)$ is the ``largest'' static symbol in the sense that we cannot find another static symbol whose superimposed symbol is larger than $w_S^*$. Thus, all joint symbols to the right of the reference symbol, including its right neighbor $(w^{r*}_A, w^{r*}_B)$, must be moving symbols.

\end{IEEEproof}

\begin{definition}
Consider two distinct joint symbols $(w_{A,1},w_{B,1})$ and $(w_{A,2},w_{B,2})$. Suppose that $(w_{A,1},w_{B,1})$ is on the left side of $(w_{A,2},w_{B,2})$ when $\eta=1$, but that $w_{A,1}>w_{A,2}$. As $\eta$ increases, $(w_{A,1},w_{B,1})$ first overlaps with and then overtakes $(w_{A,2},w_{B,2})$. In particular,

  \begin{itemize}
    \item $(w_{A,1},w_{B,1})$ is said to \emph{overlap} with $(w_{A,2},w_{B,2})$ when $\eta w_{A,1}+w_{B,1}=\eta w_{A,2}+w_{B,2}$;
    \item   $(w_{A,1},w_{B,1})$ is said to \emph{overtake}  $(w_{A,2},w_{B,2})$ when $\eta w_{A,1}+w_{B,1}>\eta w_{A,2}+w_{B,2}$.
  \end{itemize}
\end{definition}
\rightline{$\blacksquare$}

\begin{lemma}\label{lem:otake}

  Suppose that at some $\eta$ in the range $1< \eta < q-1$,    $(w^{l*}_A, w^{l*}_B)$ is a unique left neighbor and  $(w^{r*}_A, w^{r*}_B)$ is a unique right neighbor of the reference joint symbol. As we increase $\eta$ further, it is not possible for

  \begin{enumerate}
    \item[i)]  a joint symbol to the left of  $(w^{l*}_A, w^{l*}_B)$ to overtake  $(w^{l*}_A, w^{l*}_B)$ before  $(w^{l*}_A, w^{l*}_B)$ overlaps with the reference symbol $(w^{*}_A, w^{*}_B)$ as $\eta$ increases;
    \item[ii)]  $(w^{r*}_A, w^{r*}_B)$ to overtake  a joint symbol to the right of  $(w^{r*}_A, w^{r*}_B)$ before $(w^{l*}_A, w^{l*}_B)$ overlaps with the reference symbol $(w^{*}_A, w^{*}_B)$ as $\eta$ increases.
  \end{enumerate}

\end{lemma}

\begin{IEEEproof} [Proof of Lemma \ref{lem:otake}]

  We first prove sub-statement i). Suppose that there is a joint symbol $(w_A, w_B)$ to the left of  $(w^{l*}_A, w^{l*}_B)$  overtaking $(w^{l*}_A, w^{l*}_B)$ before  $(w^{l*}_A, w^{l*}_B)$ overlaps with the reference symbol as $\eta$ increases. Before  $(w_A, w_B)$ overtakes $(w^{l*}_A, w^{l*}_B)$, there is a moment (a particular $\eta$) when $(w_A, w_B)$ overlaps with $(w^{l*}_A, w^{l*}_B)$. According to  \emph{Corollary \ref{cor:short}}, at this precise moment (this $\eta$), we should witness a  symbol overlapping with $(w^{*}_A, w^{*}_B)$ as well (i.e., $l_{\min}=0$ must be witnessed in the locality of $(w^{*}_A, w^{*}_B)$). However, according to the statement of \emph{Lemma \ref{lem:otake}},    $(w^{l*}_A, w^{l*}_B)$ is the left neighbor and has not overlapped with it yet. This leads to a contradiction.

 Sub-statement ii) is similar. It is not possible for $(w^{r*}_A, w^{r*}_B)$ to overtake a joint symbol to its right before $(w^{l*}_A, w^{l*}_B)$ overlaps with the reference symbol as  $\eta$ increases, since any overlapping at $(w^{r*}_A, w^{r*}_B)$ implies an overlapping at the reference symbol.

\end{IEEEproof}

The following is a direct corollary of \emph{Lemma \ref{lem:otake}} (it can be treated as a restatement of \emph{Lemma \ref{lem:otake}}):

\begin{corollary}\label{cor:otake}
Suppose that at some $\eta$ in the range $1<\eta<q-1$,   $(w^{l*}_A, w^{l*}_B)$  is a unique left neighbor and $(w^{r*}_A, w^{r*}_B)$ is a unique right neighbor of the reference joint symbol. As we increase $\eta$  further,  the left and right neighbors of the reference symbol remain to be the same symbols until $(w^{l*}_A, w^{l*}_B)$ overlaps with the reference symbol.
\end{corollary}
\rightline{$\blacksquare$}

\begin{lemma}\label{lem:cluo}
  Suppose that at some $\eta$ in the range  $1<\eta<q-1$, there is a joint symbol $(w^{o*}_A, w^{o*}_B)$ overlapping with the reference joint symbol $(w^{*}_A, w^{*}_B)$, and that $(w^{l*}_A, w^{l*}_B)$ is a left neighbor and $(w^{r*}_A, w^{r*}_B)$ is a right neighbor of the reference joint symbol. Then, it is not possible to map all three of $(w^{l*}_A, w^{l*}_B)$, $(w^{o*}_A, w^{o*}_B)$, and $(w^{*}_A, w^{*}_B)$ to the same NC symbol. Neither is it possible to map all three of  $(w^{*}_A, w^{*}_B)$, $(w^{o*}_A, w^{o*}_B)$, and $(w^{r*}_A, w^{r*}_B)$  to the same NC symbol.
\end{lemma}

\begin{IEEEproof}[Proof of Lemma \ref{lem:cluo}]

  According to \emph{Lemma \ref{lem:ovl}}, we can find a $(\alpha, \beta)$ with $\alpha\neq 0$ and $\beta \neq 0$ to cluster $(w^{o*}_A, w^{o*}_B)$ and $(w^{*}_A, w^{*}_B)$. Let $\delta^{o*}_A=w^{o*}_A-w^{*}_A$ and $\delta^{o*}_B=w^{o*}_B-w^{*}_B$.  Then, we have ${\rm mod}(\alpha \delta^{o*}_A+ \beta \delta^{o*}_B, q)=0$ and $\eta\delta^{o*}_A + \delta^{o*}_B=0$, from which we get ${\rm mod}(\alpha\delta^{o*}_A - \beta\eta\delta^{o*}_A, q)=0$. Since $\delta^{o*}_A\neq 0$ and $-(q-1)\leq \delta^{o*}_A\leq q-1$, multiplicative inverse of $\delta^{o*}_A$ exists in $GF(q)$. We then have
  \begin{align}\label{overl1}
    {\rm mod}(\alpha-\eta\beta, q)=0.
  \end{align}

  Consider the left neighbor $(w^{l*}_A, w^{l*}_B)$. Suppose that $(\alpha, \beta)$ maps $(w^{l*}_A, w^{l*}_B)$ to the same cluster. Then ${\rm mod}(\alpha\delta^{l*}_A+\beta\delta^{l*}_B, q)=0$. This and \eqref{overl1} give ${\rm mod}(\eta\delta^{l*}_A+\delta^{l*}_B, q)=0$. Since $(w^{l*}_A, w^{l*}_B)$ does not overlap with the reference symbol,  $\eta\delta^{l*}_A+\delta^{l*}_B\neq 0$. Thus, $|\eta\delta^{l*}_A+\delta^{l*}_B|$ must be an integral multiple of $q$. However, this large distance means $(w^{l*}_A, w^{l*}_B)$ cannot be the left neighbor of $(w^{*}_A, w^{*}_B)$, since the static symbol $(0, q-2)$ is at distance $1$ only from    $(w^{*}_A, w^{*}_B)$. As for the right neighbor $(w^{r*}_A, w^{r*}_B)$, we reason similarly that $|\eta\delta^{r*}_A+\delta^{r*}_B|$ must be an integral multiple of $q$ if it is to be clustered into the same group as $(w^{o*}_A, w^{o*}_B)$ and $(w^{*}_A, w^{*}_B)$. However,  the joint symbol $(1, q-1)$ is at distance less than $q$ to the right of $(w^{*}_A, w^{*}_B)$ for all $\eta$ in the range $1< \eta< q-1$.

\end{IEEEproof}

\begin{lemma}\label{lem:lmin}

 Suppose that there is no joint symbol that overlaps with  $(w^{*}_A, w^{*}_B)$ for some $\eta$ in the range $1<\eta<q-1$. At this $\eta$,
     \begin{enumerate}
     \item[i)] $l_{\min}$   is given by the distance between $(w^{*}_A, w^{*}_B)$ and its left neighbor $(w^{l*}_A, w^{l*}_B)$ or  the distance between $(w^{*}_A, w^{*}_B)$ and its right neighbor $(w^{r*}_A, w^{r*}_B)$;
         \item[ii)] we can cluster  $(w^{*}_A, w^{*}_B)$ with its closer neighbor,  $(w^{l*}_A, w^{l*}_B)$  or $(w^{r*}_A, w^{r*}_B)$, to the same NC symbol;
         \item[iii)] we cannot  cluster all three of $(w^{l*}_A, w^{l*}_B)$, $(w^{*}_A, w^{*}_B)$, and $(w^{r*}_A, w^{r*}_B)$ to the same NC symbol.
     \end{enumerate}
\end{lemma}

\begin{IEEEproof} [Proof of Lemma \ref{lem:lmin}]

  We first prove sub-statement i). By \emph{Corollary \ref{cor:short}}, we can observe $l_{\min}$ in the locality of $(w^{*}_A, w^{*}_B)$. Therefore, $l_{\min}$ must occur between $(w^{*}_A, w^{*}_B)$ and   its left neighbor $(w^{l*}_A, w^{l*}_B)$ or its right neighbor $(w^{r*}_A, w^{r*}_B)$.

  We next prove sub-statement ii).  First, suppose that $l_{\min}$ is the distance between $(w^{*}_A, w^{*}_B)$ and  $(w^{l*}_A, w^{l*}_B)$. According to \emph{Corollary \ref{cor:left}}, $w^{l*}_A> w^{*}_A=0$. Meanwhile, $w^{l*}_B< w^{*}_B=q-1$, since the joint symbol with  $w^{l*}_A>0$ and $w^{l*}_B=q-1$ resides on the right of $(w^{*}_A, w^{*}_B)$. Therefore, we can cluster $(w^{l*}_A, w^{l*}_B)$ and $(w^{*}_A, w^{*}_B)$ together according to \emph{Proposition \ref{pro:wab12}}.

  Now,  suppose that  $l_{\min}$ is the distance between $(w^{*}_A, w^{*}_B)$ and  $(w^{r*}_A, w^{r*}_B)$.  According to \emph{Corollary \ref{cor:left}}, $w^{r*}_A> w^{*}_A=0$. Suppose that $w^{r*}_B= w^{*}_B=q-1$  and we cannot cluster the two joint symbols together.  Then, $w^{r*}_S-w^*_S= \eta w^{r*}_A > 1$. By \emph{Lemma \ref{lem:left}},  we can find a joint symbol residing in between $(w^{*}_A, w^{*}_B)$ and  $(w^{r*}_A, w^{r*}_B)$ at this $\eta$ in the range  $1<\eta<q-1$. This leads to a contradiction. Therefore, $w^{r*}_B< w^{*}_B=q-1$. By \emph{Proposition \ref{pro:wab12}}, we can cluster $(w^{r*}_A, w^{r*}_B)$ and $(w^{*}_A, w^{*}_B)$ together.

  We now prove sub-statement iii). First, we note that at this particular $\eta$, since no joint symbol overlaps with the reference symbol,  the left and right neighbors   $(w^{l*}_A, w^{l*}_B)$ and $(w^{r*}_A, w^{r*}_B)$ are unique according to \emph{Corollary \ref{cor:short}}.  Next, suppose that $(\alpha, \beta)$ clusters $(w^{l*}_A, w^{l*}_B)$, $(w^{*}_A, w^{*}_B)$, and $(w^{r*}_A, w^{r*}_B)$ into the same group. Then ${\rm mod}(\alpha\delta^{r*}_A+\beta\delta^{r*}_B, q)=0$ and $\eta\delta^{r*}_A+\delta^{r*}_B=\eta w^{r*}_A+w^{r*}_B-(q-1)=d^{r*}>0$, where $d^{r*}$ is the distance between $(w^{*}_A, w^{*}_B)$ and $(w^{r*}_A, w^{r*}_B)$; and $|\eta\delta^{l*}_A+\delta^{l*}_B|=|\eta w^{l*}_A+w^{l*}_B-(q-1)|=d^{l*}>0$, where $d^{l*}$ is the distance between $(w^{*}_A, w^{*}_B)$ and $(w^{l*}_A, w^{l*}_B)$. Suppose that we increase $\eta$ until  $(w^{l*}_A, w^{l*}_B)$ overlaps with $(w^{*}_A, w^{*}_B)$ (i.e., until $d^{l*}=0$). Note that as $\eta$   increases, according to \emph{Lemma \ref{lem:otake}}, no joint symbol to the left of $(w^{l*}_A, w^{l*}_B)$ can overtake $(w^{l*}_A, w^{l*}_B)$ to reach $(w^{*}_A, w^{*}_B)$ first. By the same reasoning, we deduce that $(w^{r*}_A, w^{r*}_B)$ will continue to be the right neighbor of $(w^{*}_A, w^{*}_B)$ as $\eta$ is increased at least until $(w^{l*}_A, w^{l*}_B)$ overlaps with $(w^{*}_A, w^{*}_B)$. Let $\eta'$ be the value of $\eta$ when $(w^{l*}_A, w^{l*}_B)$ overlaps with $(w^{*}_A, w^{*}_B)$ (i.e., at $\eta'$, $(w^{l*}_A, w^{l*}_B)$ becomes an overlapped symbol while $(w^{r*}_A, w^{r*}_B)$ remains as a right neighbor). According to \emph{Lemma \ref{lem:cluo}}, it is not possible to cluster all three of $(w^{l*}_A, w^{l*}_B)$, $(w^{*}_A, w^{*}_B)$, and $(w^{r*}_A, w^{r*}_B)$ to the same NC symbol (note: ``clusterabiliy'' does not depend on
    $\eta$; if we cannot cluster the three symbols at $\eta'$, we cannot cluster them at all $\eta$, including the said $\eta$ in the statement of \emph{Lemma \ref{lem:lmin}}).

\end{IEEEproof}

%
%

\subsection{Principal Theorem: Identifying $l_{\min}$, $d^{(\alpha_{opt}, \beta_{opt})}_{\min}$, and $(\alpha_{opt}, \beta_{opt})$ by Examining Symbols Overlapping or Neighboring the Reference Symbol}

In the following, we introduce two theorems. The \emph{Subsidiary  Theorem} covers $l_{\min}$ and $d^{(\alpha_{opt}, \beta_{opt})}_{\min}$ for $\eta=1$ and $\eta\geq q-1$. The \emph{Principal Theorem} describes how $l_{\min}$ and $d^{(\alpha_{opt}, \beta_{opt})}_{\min}$ can be identified for $\eta$ in the range  $1<\eta<q-1$.

\emph{Subsidiary  Theorem}:
When $\eta=1$, $l_{\min}=0$ and $d^{(\alpha_{opt}, \beta_{opt})}_{\min}=1$. As $\eta$ increases from $q-1$ to $q$ in the range $ q-1 \leq \eta\leq q$, $l_{\min}$ increases from $0$ to $1$ and $d^{(\alpha_{opt}, \beta_{opt})}_{\min}=1$. When $\eta> q$, $l_{\min}=d^{(\alpha_{opt}, \beta_{opt})}_{\min}=1$.

\begin{IEEEproof}[Proof of Subsidiary  Theorem]

By \emph{Corollary \ref{cor:short}}, $l_{\min}$ and $d^{(\alpha_{opt}, \beta_{opt})}_{\min}$ must be observable in the locality of $(w^{*}_A, w^{*}_B)$. Therefore, we focus on $(w^{*}_A, w^{*}_B)$ and its left and right neighbors $(w^{l*}_A, w^{l*}_B)$ and $(w^{r*}_A, w^{r*}_B)$.

When $\eta=1$, $l_{\min}=0$ since the joint symbols that satisfy $w_A+w_B=q-1$ overlap  with  $(w^{*}_A, w^{*}_B)$. By \emph{Lemma \ref{lem:ovl}}, we can cluster these overlapped symbols together by $(\alpha_{opt}, \beta_{opt})=\nu\otimes (1,1)$, $\nu\in \{1,\ldots, q-1\}$. At this $\eta$,    $(w^{l*}_A, w^{l*}_B)=(0, q-2)$ and $(w^{r*}_A, w^{r*}_B)=(1, q-1)$. By \emph{Proposition \ref{pro:clu}}, we cannot cluster $(w^{*}_A, w^{*}_B)$ with  $(0, q-2)$ or $(1, q-1)$, since $w^*_A=w^{l*}_A$ and $w^*_B=w^{r*}_B$.
Thus, $d^{(\alpha_{opt}, \beta_{opt})}_{\min}=|w^*_S-w^{l*}_S|= |w^*_S-w^{r*}_S|=1$.

 When $\eta=q-1$, $l_{\min}=0$ since $(1,0)$ overlaps with $(w^{*}_A, w^{*}_B)$. By \emph{Lemma \ref{lem:ovl}}, we can cluster $(w^{*}_A, w^{*}_B)$ and $(1,0)$  together by $(\alpha_{opt}, \beta_{opt})=\nu\otimes (q-1,1)$, $\nu\in \{1,\ldots, q-1\}$.  At this $\eta$, $(w^{l*}_A, w^{l*}_B)=(0, q-2)$ and $(w^{r*}_A, w^{r*}_B)=(1, 1)$, and $|w^*_S-w^{l*}_S|= |w^*_S-w^{r*}_S|=1$.  By  \emph{Proposition \ref{pro:clu}}, we cannot cluster  $(w^{*}_A, w^{*}_B)$ with $(0, q-2)$, since $w^*_A=w^{l*}_A$. Thus, $d^{(\alpha_{opt}, \beta_{opt})}_{\min}=1$.

We note that for $\eta \geq q-1$, there is no moving symbol to the left of $(w^{*}_A, w^{*}_B)$ anymore, and no more symbol will overlap with $(w^{*}_A, w^{*}_B)$ anymore.  Therefore, according to \emph{Corollary \ref{cor:short}}, for $\eta > q-1$, $l_{\min} >0$.  As $\eta$ increases from $q-1$ to $q$, $(1,0)$ moves to the right and becomes the right neighbor of $(w^{*}_A, w^{*}_B)$. The distance between $(1,0)$ and $(w^{*}_A, w^{*}_B)$ is $l_{\min}$.  In particular, $l_{\min}$ increases from $0$ to $1$ as $\eta$ increases from $q-1$ to $q$.   We can cluster $(w^{*}_A, w^{*}_B)$ and $(1,0)$ together by $(\alpha_{opt}, \beta_{opt})=\nu\otimes (q-1,1)$, $\nu\in \{1,\ldots, q-1\}$.  Note that when $\eta\geq q-1$,   $(0, q-2)$ remains   the left neighbor of $(w^{*}_A, w^{*}_B)$. Therefore, $d^{(\alpha_{opt}, \beta_{opt})}_{\min}=1$.

When $\eta> q$,  we have $(w^{l*}_A, w^{l*}_B)=(0,q-2)$  and $(w^{r*}_A, w^{r*}_B)=(1,0)$. In this case,  $|w^*_S-w^{r*}_S|> |w^*_S-w^{l*}_S|=1$.  By \emph{Proposition \ref{pro:clu}}, however,  we cannot cluster $(w^*_A, w^*_B)$   and $(w^{l*}_A, w^{l*}_B)$ together, since $w^*_A=w^{l*}_A$.   Therefore, $l_{\min}=d^{(\alpha_{opt}, \beta_{opt})}_{\min}=1$ for all $\eta> q$.

\end{IEEEproof}

\emph{Principal Theorem}:   Consider  $\eta$ in the range $1< \eta < q-1$. With respect to the reference symbol $(w^{*}_A, w^{*}_B)=(0, q-1)$, when $\eta$ is such that  there is a joint symbol $(w_A^{o*}, w_B^{o*})$   overlapping with the reference symbol, then $l_{\min}=0$ and $d^{(\alpha_{opt}, \beta_{opt})}_{\min}=w^*_S-w^{l*}_S=w^{r*}_S-w^*_S$.

When $\eta$ is such that  an overlapping joint symbol  $(w_A^{o*}, w_B^{o*})$ does not exist, we have a few subcases as follows
\begin{itemize}
  \item[i)] if $w^*_S-w^{l*}_S=w^{r*}_S-w^*_S$, then $l_{\min}=d^{(\alpha_{opt}, \beta_{opt})}_{\min}=w^*_S-w^{l*}_S=w^{r*}_S-w^*_S$;

  \item[ii)] if $w^*_S-w^{l*}_S>w^{r*}_S-w^*_S$, then $l_{\min}=w^{r*}_S-w^*_S$ and $d^{(\alpha_{opt}, \beta_{opt})}_{\min}= w^*_S-w^{l*}_S$;

  \item[iii)]
  if $w^*_S-w^{l*}_S<w^{r*}_S-w^*_S$, then $l_{\min}=w^*_S-w^{l*}_S$ and $d^{(\alpha_{opt}, \beta_{opt})}_{\min}= w^{r*}_S-w^*_S$;
\end{itemize}

 The proof  of the \emph{Principal Theorem}  is given in \emph{Appendix III}.
\rightline{$\blacksquare$}

%

\section{Systematic Derivation of Relationship Between $d^{(\alpha_{opt}, \beta_{opt})}_{\min}$ and $\eta$}\label{sec:sys}

This section presents a ``formula'' for deriving the $d^{(\alpha_{opt}, \beta_{opt})}_{\min}$   versus  $\eta$ curve. In particular, we show that $d^{(\alpha_{opt}, \beta_{opt})}_{\min}$  can be determined by the distances between a subset of joint symbols and the reference symbol. We refer to these joint symbols as characteristic symbols. Each characteristic symbol becomes a $d^{(\alpha_{opt}, \beta_{opt})}_{\min}$  determining symbol for an interval of $\eta$ in that its distance and the reference symbol is $d^{(\alpha_{opt}, \beta_{opt})}_{\min}$. Within that interval, $d^{(\alpha_{opt}, \beta_{opt})}_{\min}$  versus  $\eta$ is linear. Overall,  $d^{(\alpha_{opt}, \beta_{opt})}_{\min}$ is a piecewise linear function of  $\eta$ as different joint symbols take on the role as a $d^{(\alpha_{opt}, \beta_{opt})}_{\min}$  determining. By sorting   the characteristic symbols according to the order in which they become a $d^{(\alpha_{opt}, \beta_{opt})}_{\min}$  determining symbol as $\eta$  increases, we can obtain the overall $d^{(\alpha_{opt}, \beta_{opt})}_{\min}$  versus $\eta$  curve.
 \begin{figure}[t]
 \centering
        \includegraphics[width=0.9\columnwidth]{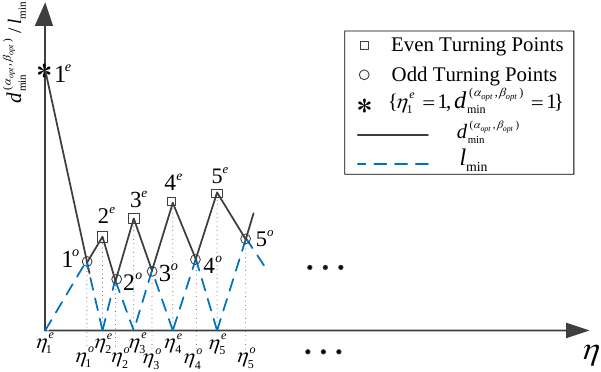}
       \caption{$d^{(\alpha_{opt}, \beta_{opt})}_{\min}$  and $l_{\min}$ versus $\eta$ for $q$-PAM linear PNC.}
        \label{fig:secvdim}
\end{figure}

\subsection{Algorithm for Identifying Characteristic Symbols}

\begin{definition}\label{def:oe}
    Consider  $d^{(\alpha_{opt}, \beta_{opt})}_{\min}$  as a function of $\eta$.    We define a turning point   as a local minimum  or a local maximum  of $d^{(\alpha_{opt}, \beta_{opt})}_{\min}$. In particular,
\begin{itemize}
  \item each peak is an \emph{even turning point} at which   $d^{(\alpha_{opt}, \beta_{opt})}_{\min}$ reaches a local maximum.

  \item each trough is an \emph{odd turning point}  at which   $d^{(\alpha_{opt}, \beta_{opt})}_{\min}$ reaches a local minimum.

\end{itemize}
\end{definition}
\rightline{$\blacksquare$}

Fig. \ref{fig:secvdim} illustrates the definition with the qualitative shapes of the $d^{(\alpha_{opt}, \beta_{opt})}_{\min}$ versus $\eta$, and the  $l_{\min}$ versus $\eta$, curves. As stated in the \emph{Subsidiary Theorem} in the preceding section,    $d^{(\alpha_{opt}, \beta_{opt})}_{\min}=1$ for $\eta\geq q-1$,   thus there are no turning points for $\eta\geq q-1$.

We define a set of moving joint symbols as follows:
 \begin{align}\label{lefts}
 \nonumber \mathcal{W}^{lo}_{(A, B)}=&\big\{(w_{A}, w_{B})\in \mathcal{W}_{(A,B)}| \\ &0<w_{A}+w_{B}\leq q-1, w_{A}> 0\big\}.
  \end{align}
  When $1\leq \eta \leq q-1$, these moving joint symbols are on the
left of or overlap with the reference symbol. When $\eta> q-1$, all these moving symbols are on the right of $(w^*_A, w^*_B)$. All symbols in $\mathcal{W}^{lo}_{(A, B)}$ overlap with the reference symbol at some point as $\eta$ increases in the range $1\leq \eta\leq q-1$.

According to \emph{Corollary \ref{cor:short}},  $d^{(\alpha_{opt}, \beta_{opt})}_{\min}$ and $l_{\min}$  can be determined by the distances between the reference  symbol $(w^*_A, w^*_B)$ and some joint symbols. We will argue shortly that we only need to restrict our attention to the joint symbols in $\mathcal{W}^{lo}_{(A, B)}$. We first put forth three definitions as follows:

\begin{itemize}
  \item  	For a given $\eta$, a joint symbol $(w_A, w_B)\in \mathcal{W}^{lo}_{(A, B)}$ is a  \emph{$d^{(\alpha_{opt}, \beta_{opt})}_{\min}$ determining symbol} if at that  $\eta$ the distance between  $(w_A, w_B)$ and $(w^*_A, w^*_B)$  is $d^{(\alpha_{opt}, \beta_{opt})}_{\min}$.
  \item  	For a given $\eta$, a joint symbol $(w_A, w_B)\in \mathcal{W}^{lo}_{(A, B)}$ is an \emph{$l_{\min}$  determining symbol} if at that $\eta$   the distance between $(w_A, w_B)$  and $(w^*_A, w^*_B)$  is $l_{\min}$.
  \item  	A joint symbol in $(w_A, w_B) \in \mathcal{W}^{lo}_{(A, B)}$  is said to be a \emph{characteristic symbol} if it is a $d^{(\alpha_{opt}, \beta_{opt})}_{\min}$  determining symbol at some $\eta$. We denote the set of characteristic symbols by $\mathcal{W}^{char}_{(A, B)}$.
      \end{itemize}

 We will explain shortly that $\mathcal{W}^{char}_{(A, B)}\subset\mathcal{W}^{lo}_{(A, B)}$. For $q\geq5$,  not all symbols in $\mathcal{W}^{lo}_{(A, B)}$  are characteristic symbols; for $q=3$, $\mathcal{W}^{char}_{(A, B)}=\mathcal{W}^{lo}_{(A, B)}$. The discussion below  applies to general $q$-PAM where $q$ is prime.  Specific examples are given with respect to $7$-PAM  PNC.

For $q$-PAM  PNC, we have two different overlapping cases:
 \begin{itemize}
   \item  Unique overlapping: a unique joint symbol in $\mathcal{W}^{lo}_{(A, B)}$  overlaps with the reference symbol at a particular $\eta$.
   \item  Multiple overlapping: multiple joint symbols in $\mathcal{W}^{lo}_{(A, B)}$  overlap with the reference symbol at a particular $\eta$.
 \end{itemize}

 For $7$-PAM  PNC, at  $\eta=6/5$, the joint symbol $(5, 0)$ overlaps with the reference symbol  uniquely. At  $\eta=3/2$, however, as shown in Fig. \ref{fig:mulo} (b), both $(2, 3)$ and $(4, 0)$ overlap with  the reference symbol.
We have the following general observation:

\begin{figure}[t]
 \centering
        \includegraphics[width=0.9\columnwidth]{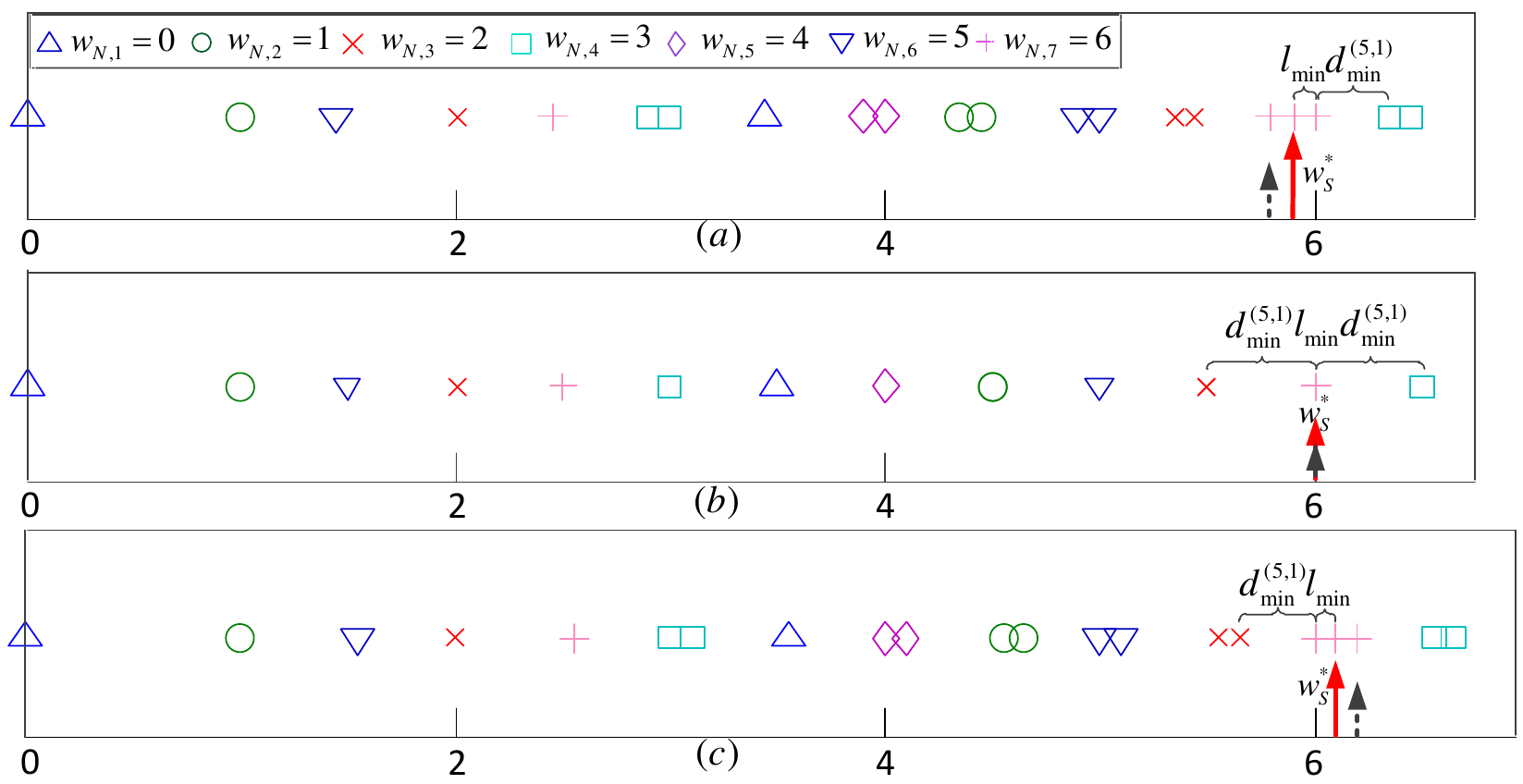}
       \caption{Clustered constellations of $\mathcal{W}_S$ for $7$-PAM linear PNC, when (a) $\eta=29/20$; (b) $\eta=3/2$;  (c) $\eta=31/20$. Constellations beyond $7$ on the real line are omitted to avoid cluttering.}
        \label{fig:mulo}
\end{figure}

\emph{Observation 1}: If a joint symbol $(w_A, w_B)\in\mathcal{W}^{lo}_{(A, B)}$  overlaps with the reference symbol uniquely, then $(w_A, w_B)$  is a characteristic symbol. If multiple joint symbols $(w_{A,1}, w_{B,1}), \ldots, (w_{A,K}, w_{B,K}) \in \mathcal{W}^{lo}_{(A, B)}$  overlap with the reference symbol at the same  $\eta$, then the symbol with the smallest $w_{A,i}$  is a characteristic symbol while the other symbols are not characteristic symbols.

\emph{Explanation}: First, consider the unique overlapping joint symbol  $(w_A, w_B)\in\mathcal{W}^{lo}_{(A, B)}$. Suppose that it overlaps with the reference symbol at  $\eta=\eta_0$.

\begin{itemize}
  \item [i)] 	If we decrease $\eta$ slightly, say, to   $\eta=\eta_0-\Delta\eta$, $\Delta\eta>0$,    $(w_A, w_B)$  will be the left neighbor of the reference symbol for a small  $\Delta\eta>0$. In particular, symbol $(w_A, w_B)$  is an $l_{\min}$  determining symbol that accords with subcase iii) in the \emph{Principal Theorem}. If we continue to increase  $\Delta\eta$, at some  $\Delta\eta$, we will have the situation that accords with subcase i) in the \emph{Principal Theorem}, at which point $(w_A, w_B)$  becomes a $d^{(\alpha_{opt}, \beta_{opt})}_{\min}$  determining symbol\footnote{Note that according to \emph{Corollary  \ref{cor:otake}}, it is not possible for us to bypass subcase i) in the \emph{Principal Theorem} as  $\Delta\eta$ increases. This is because in order to bypass subcase i),  $(w_A, w_B)$ must first overlap with and then bypass another symbol on its left (so that  $(w_A, w_B)$  ceases to be the left neighbor of the reference symbol) before any overlapping occurs at the reference symbol as $\Delta \eta$  increases; however, \emph{Corollary \ref{cor:otake}} says this is not possible.}. Hence,  $(w_A, w_B)$ is a characteristic symbol according to our definition.

  \item [ii)]  Similarly, if we increase $\eta$  to $\eta=\eta_0+\Delta\eta$,  $\Delta\eta>0$, symbol  $(w_A, w_B)$ will first become an  $l_{\min}$ determining symbol  that accords with subcase (ii) in the \emph{Principal Theorem} and then a  $d^{(\alpha_{opt}, \beta_{opt})}_{\min}$ determining symbol that accords with subcase (i) in the \emph{Principal Theorem} (on the right side of the reference symbol), as  $\Delta\eta$ increases.
\end{itemize}

 Next, consider the multiple overlapping case  (which occurs when $q\geq5$). We note that as we decrease or increase $\eta$  as in the above, only the overlapping symbol with the smallest $w_{A,i}$  will ever become the left and right neighbors of the reference symbol and follow the patterns as per i) and ii) above. In particular, the other overlapping symbols with larger $w_{A,i}$  will never become a $d^{(\alpha_{opt}, \beta_{opt})}_{\min}$  determining symbol; also, they are  $l_{\min}$ determining symbols only at the singular $\eta=\eta_0$  when they overlap with the reference symbol  (i.e., they are not $l_{\min}$ determining at any other $\eta$, whereas the symbol with the smallest $w_{A,i}$ is $l_{\min}$ determining for a range of $\eta$ and $d^{(\alpha_{opt}, \beta_{opt})}_{\min}$ determining for two ranges of $\eta$, once when $(w_{A,i}, w_{B,i})$ is on the left of the reference symbol,  and once when $(w_{A,i}, w_{B,i})$ is on the right of the reference symbol).  Lastly, we note that the symbol with the smallest $w_{A,i}$  is unique because it is not possible for two different joint symbols with the same $w_{A,i}$  but different $w_{B,i}$  to overlap with the reference  symbol simultaneously.

Based on \emph{Observation 1}, \emph{Appendix IV} outlines an algorithm for identifying all characteristic symbols and orders them in a sequence according to the $\eta$  at which they overlap with the reference symbol.

 \subsection{Identifying Turning Points and $d^{(\alpha_{opt}, \beta_{opt})}_{\min}$  Versus  $\eta$ Curve}

Given a sequence of ordered characteristic symbols  $(w^{char}_{A,i}, w^{char}_{B,i})_{i=1,2,\ldots, I}$,  we can systematically identify the turning points and derive the $d^{(\alpha_{opt}, \beta_{opt})}_{\min}$  versus  $\eta$ curve, as follows:

  \textbf{Identifying turning points}

 Consider any three characteristic symbols $(w^{char}_{A,i-1}, w^{char}_{B, i-1})$, $(w^{char}_{A,i}, w^{char}_{B, i})$, and $(w^{char}_{A,i+1}, w^{char}_{B, i+1})$  in $\mathcal{W}^{char}_{(A,B)}$  that overlap with the reference symbol $(w^*_A, w^*_B)$ at $\eta^e_{i-1}, \eta^e_{i}$, and  $\eta^e_{i+1}$, respectively.

\begin{figure}[t]
 \centering
        \includegraphics[width=0.7\columnwidth]{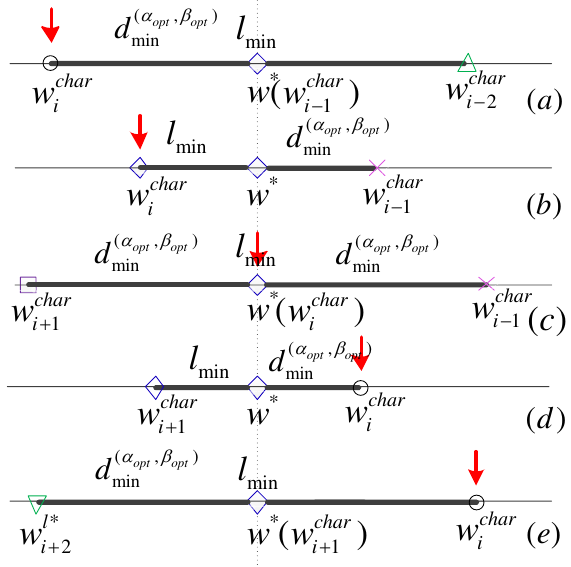}
       \caption{Movements of   characteristic symbols in  $\mathcal{W}^{char}_{(A, B)}$.  $w^{char}$ and $w^*$ denote  $(w^{char}_{A}, w^{char}_{B})$ and $(w^*_A, w^*_B)$,  respectively.}
        \label{fig:role}
\end{figure}

 This paragraph draws on the results from the \emph{Principal Theorem}. As depicted in Fig.  \ref{fig:role} (a), at  $\eta^e_{i-1}$,  $(w^{char}_{A,i-1}, w^{char}_{B, i-1})$ overlaps with $(w^*_A, w^*_B)$ and is an  $l_{\min}$ determining symbol, while $(w^{char}_{A,i}, w^{char}_{B, i})$ is a $d^{(\alpha_{opt}, \beta_{opt})}_{\min}$ determining symbol on the left. As  $\eta$ increases,  $(w^{char}_{A,i}, w^{char}_{B, i})$  continues to be a $d^{(\alpha_{opt}, \beta_{opt})}_{\min}$ determining symbol until $(w^*_A, w^*_B)$ is in the middle of $(w^{char}_{A,i-1}, w^{char}_{B, i-1})$  and  $(w^{char}_{A,i}, w^{char}_{B, i})$  at  $\eta=\eta^o_{i-1}$, as shown  in Fig.  \ref{fig:role} (b). Then, $(w^{char}_{A,i-1}, w^{char}_{B, i-1})$  and $(w^{char}_{A,i}, w^{char}_{B, i})$ are both  $l_{\min}$ determining symbols and $d^{(\alpha_{opt}, \beta_{opt})}_{\min}$ determining symbols since $l_{\min}=d^{(\alpha_{opt}, \beta_{opt})}_{\min}$ at this   $\eta=\eta^o_{i}$. However, for $\eta\in (\eta^o_{i-1}, \eta^e_{i}]$, $(w^{char}_{A,i}, w^{char}_{B, i})$  is only an  $l_{\min}$ determining symbol and $(w^{char}_{A,i-1}, w^{char}_{B, i-1})$ is only a $d^{(\alpha_{opt}, \beta_{opt})}_{\min}$ determining symbol.   At  $\eta=\eta^e_{i}$,  $(w^{char}_{A,i}, w^{char}_{B, i})$  overlaps with $(w^*_A, w^*_B)$  in Fig.  \ref{fig:role} (c) and continues to be an  $l_{\min}$ determining symbol; at the same time  $(w^{char}_{A,i+1}, w^{char}_{B, i+1})$ on the left and  $(w^{char}_{A,i-1}, w^{char}_{B, i-1})$ on the right are both  $d^{(\alpha_{opt}, \beta_{opt})}_{\min}$  determining symbols. Symbol $(w^{char}_{A,i-1}, w^{char}_{B, i-1})$  ceases to be a  $d^{(\alpha_{opt}, \beta_{opt})}_{\min}$ or $l_{\min}$  determining symbol when $\eta>\eta^e_{i}$.

Based on the above, we have the following observations:

\emph{Observation 2}:

(1) An even turning point occurs when a characteristic symbol overlaps with the reference symbol in the range  $1\leq \eta <q-1$. Thus, the  $i$th even turning point occurs at  $\eta^e_i=(q-1-w^{char}_{B,i})/ w^{char}_{A,i}$  for  $i=1,2, \ldots, I-1$. Since $d^{(\alpha_{opt}, \beta_{opt})}_{\min}=1$ for $\eta\geq q-1$, no turning point occurs when $(w^{char}_{A, I}, w^{char}_{B, I})$ overlaps with the reference symbol at $\eta=q-1$. Therefore, the $I-1$th even turning point is the last even turning point when $(w^{char}_{A, I-1}, w^{char}_{B, I-1})$ overlaps with the reference symbol.

(2) The  $i$th odd turning point occurs when the reference symbol resides in the middle of characteristic symbols  $(w^{char}_{A,i}, w^{char}_{B, i})$ and $(w^{char}_{A,i+1}, w^{char}_{B, i+1})$. This gives $q-1-(\eta^o_{i} w^{char}_{A,i+1}+ w^{char}_{B,i+1})=\eta^o_i w^{char}_{A,i}+ w^{char}_{B,i}-(q-1)$. Thus, the $i$th odd turning point occurs at $\eta^o_i=\frac{2(q-1)-(w^{char}_{B,i+1}+w^{char}_{B,i})}{w^{char}_{A,i+1}+ w^{char}_{A,i}}$ for $i=1,2,\ldots,I-1$.

\emph{Observation 3}: As $\eta$  increases in the range  $1\leq \eta<q-1$, a characteristic symbol $(w^{char}_{A,i}, w^{char}_{B, i})$ becomes a $d^{(\alpha_{opt}, \beta_{opt})}_{\min}$  determining symbol at the  $i-1$th even turning point; then becomes an  $l_{\min}$  determining symbol at the  $i-1$th odd turning point; continues to be an $l_{\min}$ determining symbol at the  $i$th even turning point; then becomes a $d^{(\alpha_{opt}, \beta_{opt})}_{\min}$ determining symbol again at the  $i$th odd turning point;  then ceases to be an $l_{\min}$  or $d^{(\alpha_{opt}, \beta_{opt})}_{\min}$  determining symbol.

\textbf{Linear $d^{(\alpha_{opt}, \beta_{opt})}_{\min}$ versus $\eta$ curve between two consecutive turning points}

Consider $\eta=\eta^e_i$. At $\eta^e_i$, characteristic symbol  $(w^{char}_{A,i}, w^{char}_{B, i})$ overlaps with the reference symbol  and is an  $l_{\min}$ determining symbol; and  the next characteristic symbol $(w^{char}_{A,i+1}, w^{char}_{B,i+1})$ is a $d^{(\alpha_{opt}, \beta_{opt})}_{\min}$  determining symbol. In particular, $(w^{char}_{A,i}, w^{char}_{B,i})$  and $(w^{char}_{A,i+1}, w^{char}_{B,i+1})$  will respectively remain   $l_{\min}$  and $d^{(\alpha_{opt}, \beta_{opt})}_{\min}$  determining symbols, for $\eta$ in the internal $[\eta^e_i, \eta^o_i)$---the movements of $(w^{char}_{A,i}, w^{char}_{B,i})$ and $(w^{char}_{A,i+1}, w^{char}_{B,i+1})$ correspond to those in between Fig.  \ref{fig:role} (c) and Fig.  \ref{fig:role} (d).  Thus,  $d^{(\alpha_{opt}, \beta_{opt})}_{\min}=-(\eta w^{char}_{A,i+1}+w^{char}_{B,i+1})+q-1$  and $l_{\min}=(\eta w^{char}_{A,i}+w^{char}_{B,i})-q-1$. At  $\eta=\eta^o_i$, the reference symbol is sandwiched in the middle of $(w^{char}_{A,i}, w^{char}_{B,i})$ and  $(w^{char}_{A,i+1}, w^{char}_{B,i+1})$. As shown  in Fig.  \ref{fig:role} (d) and Fig.  \ref{fig:role} (e), the roles of  $(w^{char}_{A,i+1}, w^{char}_{B,i+1})$ and $(w^{char}_{A,i}, w^{char}_{B,i})$ are changed to the  $l_{\min}$ and $d^{(\alpha_{opt}, \beta_{opt})}_{\min}$   determining symbols, respectively, within the interval $[\eta^o_i, \eta^e_{i+1})$. In this interval,  $d^{(\alpha_{opt}, \beta_{opt})}_{\min}=\eta w^{char}_{A,i}+w^{char}_{B,i}-(q-1)$  and $l_{\min}=-(\eta w^{char}_{A,i+1}+w^{char}_{B,i+1})+q-1$. The above pattern repeats between any two turning points according to two successive characteristic symbols. The $d^{(\alpha_{opt}, \beta_{opt})}_{\min}$  versus $\eta$ and $l_{\min}$  versus  $\eta$  curves can be derived by examining successive pairs of characteristic symbols.

\subsection{$d^{(\alpha_{opt}, \beta_{opt})}_{\min}$ at Turning Points}

It turns out that $d^{(\alpha_{opt}, \beta_{opt})}_{\min}$  at turning points can be derived in a simple closed form, as described in this subsection.

\begin{lemma}\label{lem:raeta}
At turning points, $\eta$ can be expressed as a rational number $m/n$, where $\gcd(m,n)=1$. Furthermore, at even turning points, $1\leq m, n\leq q-1$.
\end{lemma}

\begin{IEEEproof}[Proof of Lemma \ref{lem:raeta}]

Let us first consider the $i$th even turning point, with $(w^{char}_{A,i}, w^{char}_{B,i})$  overlapping with the reference symbol. We have $\eta\delta^{o*}_A+\delta^{o*}_B=0$  with $\delta^{o*}_A=w^{char}_{A,i}$ and  $\delta^{o*}_B=w^{char}_{B,i}-(q-1)$. Since $\delta^{o*}_A$ and $\delta^{o*}_B$ are integers,  $\eta=-\delta^{o*}_B/\delta^{o*}_A$  must be rational. Furthermore, since $1\leq |\delta^{o*}_A|, |\delta^{o*}_B|\leq q-1$, we can express  $\eta$ in the form of  $\eta={m}/{n}$, where $\gcd(m,n) =1$ and  $1\leq m, n\leq q-1$.

Next consider the $i$th odd turning point. The reference symbol is sandwiched in the middle of two characteristic symbols $(w^{char}_{A,i}, w^{char}_{B,i})$   and $(w^{char}_{A,i+1}, w^{char}_{B,i+1})$. Thus, $-\eta w^{char}_{A,i+1}+(q-1-w^{char}_{B,i+1})=\eta w^{char}_{A,i}+[w^{char}_{B,i}-(q-1)]$. This gives $\eta=\frac{2(q-1)-(w^{char}_{B, i+1}+w^{char}_{B,i})}{w^{char}_{A,i+1}+ w^{char}_{A,i}}$.  Since $w^{char}_{A,i}, w^{char}_{B,i}, w^{char}_{A,i+1}$, and $w^{char}_{B,i+1}$  are all integers, we can express $\eta$  in the form of  $\eta={m}/{n}$, where $\gcd(m,n) =1$.

\end{IEEEproof}

\emph{Lemma \ref{lem:bez}} below  is the well-known result of \emph{B\'{e}zout's Identity}  and a proof will not be given here.

\begin{lemma}[B\'{e}zout's Identity \cite{bez}] \label{lem:bez}
 Let $a$ and $b$ be integers, not both zero, and $d=\gcd(a,b)$. There exist integers $x$ and $y$ such that $ax+by=d$. Furthermore,

\begin{enumerate}
  \item[i)] $d$ is the smallest positive integer that can be written as $ax+by$;
  \item[ii)]  if both $a$ and $b$ are nonzero, there are two pais of $(x,y)$ such that $|x|<|b/d|$ and $y<|a/d|$;
  \item[iii)]  there are an infinite number of solutions, and given any solution $(x,y)$, all other solutions can be obtained by $(x+kb/d, y-ka/d)$, where $k$ is an arbitrary integer.
\end{enumerate}

\end{lemma}
\rightline{$\blacksquare$}

\begin{corollary}\label{cor:dmin}
For a turning point at $\eta=m/n$, $\gcd(m,n)=1$, we have $d^{(\alpha_{opt}, \beta_{opt})}_{\min}=1/n$. Furthermore, $d^{e_i}_{\min}=1/w^{char}_{A,i}$  and  $d^{o_i}_{\min}=1/(w^{char}_{A,i}+w^{char}_{A,i+1})$, where $d^{e_i}_{\min}$  and $d^{o_i}_{\min}$ are the $d^{(\alpha_{opt}, \beta_{opt})}_{\min}$  at the  $i$th even turning point and the  $i$th odd turning point respectively.
\end{corollary}

\begin{IEEEproof}[Proof of Corollary \ref{cor:dmin}]

The statement is trivially true for the first even turning point, where  $\eta=1$, $(w^{char}_{A,1}, w^{char}_{B, 1})=(1, q-2)$  and  $d^{(\alpha_{opt}, \beta_{opt})}_{\min} =d^{e_1}_{\min}=1$ . In the following, we focus on the turning points in the range  $1<\eta< q-1$.

Let us first consider the $i$th even turning point,  $i\geq 2$. Let the corresponding $\eta$  be  $\eta^e_i=m^e_i/n^e_i$, $\gcd(m^e_i, n^e_i)=1$, $1\leq m^e_i, n^e_i\leq q-1$  as per \emph{Lemma \ref{lem:raeta}}.

According to the \emph{Principal Theorem} and \emph{Observation 3},    $d^{e_i}_{\min}$ can be found by the distance between the reference symbol $(0, q-1)$ and  $(w^{char}_{A,i-1}, w^{char}_{B,i-1})$, which is a right neighbor of the reference symbol. Thus, we have $\eta^e_i w^{char}_{A, i-1}+[w^{char}_{B, i-1}-(q-1)]=d^{e_i}_{\min}$, giving
\begin{eqnarray}
m^e_i w^{char}_{A, i-1}+n^e_i[w^{char}_{B, i-1}-(q-1)]=n^e_i d^{e_i}_{\min}.
\end{eqnarray}
Since $m^e_i, n^e_i, w^{char}_{A, i-1},$ and $w^{char}_{B, i-1}-(q-1)$ are all integers, and $d^{e_i}_{\min}$ is nonzero by the \emph{Principal Theorem}, we have that
\begin{eqnarray}\label{dmin1}
n^{e_i} d^{e_i}_{\min}\geq 1.
\end{eqnarray}

By \emph{Lemma \ref{lem:bez}}, there exist integers $x$ and $y$ such that $m^e_ix+n^e_i y=1$. By statement ii) of \emph{Lemma \ref{lem:bez}}, there are two pairs $(x, y)$ such that  $|x|<n^e_i\leq q-1$ and $|y|<m^e_i\leq q-1$. Given that $1\leq m^e_i, n^e_i \leq q-1$ and $m^e_i\neq n^e_i$ (since we consider $\eta>1$) in order that $m^e_ix+n^e_i y=1$, $x$ and $y$ must have opposite signs and neither can be zero.

Between the two pairs $(x, y)$, there is a pair in which  $0<x<q-1, -(q-1)<y<0$. To see this, suppose that we have a pair $(x, y)$ with  $x<0, y>0$. We can apply statement iii) of \emph{Lemma \ref{lem:bez}} repeatedly for $k=1,2,\ldots$  until we find a pair $(x, y)$  such that  $0<x<q-1, -(q-1)<y<0$. Now, given the pair $(x, y)$, the duple $(w_A, w_B)=(x, y+(q-1))$ is a valid symbol because  $1\leq w_A, w_B\leq q-1$. The distance between  $(w_A, w_B)$  and the reference symbol $(0, q-1)$ is
\begin{eqnarray}
d^{*}=\eta^{e_i}w_A+ [w_B-(q-1)]=\frac{m^e_i}{n^e_i}x+y=\frac{1}{n^e_i}.
\end{eqnarray}
According to the \emph{Principal Theorem}, $d^{e_i}_{\min}$ cannot be larger than  $d^{*}$   (i.e., $d^{e_i}_{\min}\leq \frac{1}{n^e_i}$) . Together with \eqref{dmin1}, we have that
\begin{eqnarray}
d^{e_i}_{\min}= \frac{1}{n^{e_i}}.
\end{eqnarray}

Now, at this even turning point,  $(w^{char}_{A,i}, w^{char}_{B,i})$ overlaps with the reference symbol, and $(w^{char}_{A,i-1}, w^{char}_{B,i-1})$  is at a distance $d^{e_i}_{\min}= 1/n^{e_i}$  from the reference symbol. This gives two equations:
\begin{eqnarray}\label{dminev3}
\nonumber &\frac{m^e_i}{n^e_i} w^{char}_{A, i}+[w^{char}_{B, i}-(q-1)]=0,\\
&\frac{m^e_i}{n^e_i} w^{char}_{A, i-1}+[w^{char}_{B, i-1}-(q-1)]=\frac{1}{n^{e_i}}.
\end{eqnarray}

We can  see that $(w^{char}_{A,i}, w^{char}_{B,i})=(n^e_i, q-1-m^e_i)$  clearly satisfies the first equation in \eqref{dminev3}. Thus, if the overlapping is unique, this is the $i$th characteristic symbol. If the overlapping is not unique, we can also see that this is the overlapping symbol with the smallest $w_{A}$  that can satisfy the first equation because $\gcd(m^{e}_i, n^{e}_i) = 1$ in our rational number representation (i.e.,  $\frac{m^e_i}{n^e_i} w_{A}+[w_{B}-(q-1)]=0$ means $\frac{m^e_i}{n^e_i} w_{A}$ must be an integer, and therefore all overlapping symbols must have $w_A$  being a multiple of  $n^{e}_i$). Thus,
\begin{eqnarray}
d^{e_i}_{\min}= \frac{1}{w^{char}_{A,i}}.
\end{eqnarray}

Let us now consider the $(i-1)$th odd turning point, $i\geq 2$. At this turning point, the reference symbol is sandwiched in the middle of the $i$th characteristic symbol (on the left) and the $(i-1)$th characteristic symbol (on the right).  Let the corresponding  $\eta$ be  $\eta^o_{i-1}=\frac{m^o_{i-1}}{n^o_{i-1}}$, $\gcd(m^o_{i-1}, n^o_{i-1}) =1$. We can write
\begin{eqnarray}\label{dminev2}
 \nonumber &-\frac{m^o_{i-1}}{n^o_{i-1}} w^{char}_{A, i}+[q-1-w^{char}_{B, i}]=d^{o_{i-1}}_{\min},\\
&\frac{m^o_{i-1}}{n^o_{i-1}} w^{char}_{A, i-1}+[w^{char}_{B, i-1}-(q-1)]=d^{o_{i-1}}_{\min}.
\end{eqnarray}

Add the first equation of \eqref{dminev2} to the first equation of \eqref{dminev3}, and subtract the second equation of \eqref{dminev2} from the second equation of \eqref{dminev3}, we get
\begin{eqnarray}\label{dminev}
 \nonumber &\big(\frac{m^e_i}{n^e_i}-\frac{m^o_{i-1}}{n^o_{i-1}}\big)w^{char}_{A,i}=d^{o_{i-1}}_{\min},\\
 &\big(\frac{m^e_i}{n^e_i}-\frac{m^o_{i-1}}{n^o_{i-1}}\big)w^{char}_{A,i-1}=\frac{1}{n^e_i}-d^{o_{i-1}}_{\min}.
\end{eqnarray}

By eliminating $\big(\frac{m^e_i}{n^e_i}-\frac{m^o_{i-1}}{n^o_{i-1}}\big)$ from \eqref{dminev}  (after substituting $w_{A, i}^{char} = n^e_i$ in the second equation in \eqref{dminev}), we get
\begin{eqnarray}
 d^{o_{i-1}}_{\min}=\frac{1}{w^{char}_{A,i}+w^{char}_{A,i-1}}.
\end{eqnarray}

It remains to show that $w^{char}_{A,i}+w^{char}_{A,i-1}=n^o_{i-1}$  where $n^o_{i-1}$   is the denominator in
\begin{eqnarray}
 \eta^o_{i-1}=\frac{m^o_{i-1}}{n^o_{i-1}}, ~\gcd(m^o_{i-1},n^o_{i-1})=1.
\end{eqnarray}

Subtract the first equation of \eqref{dminev2} from the second equation of \eqref{dminev2}, we have
 \begin{eqnarray}\label{dminev1}
 \eta^o_{i-1}=\frac{m^o_{i-1}}{n^o_{i-1}}=\frac{2(q-1)-(w^{char}_{B, i-1}+w^{char}_{B, i})}{w^{char}_{A, i-1}+w^{char}_{A, i}}.
\end{eqnarray}

Using \eqref{dminev3}, we can express  the numerator in \eqref{dminev1} as
 \begin{align}
\nonumber & 2(q-1)-(w^{char}_{B, i-1}+w^{char}_{B, i})\\
&=\frac{m^e_{i}}{n^e_{i}} (w^{char}_{A, i}+w^{char}_{A, i-1})-\frac{1}{n^e_{i}}.
\end{align}

Thus,
 \begin{eqnarray}
\eta^o_{i-1}=\frac{m^o_{i-1}}{n^o_{i-1}}=\frac{\frac{1}{n^e_{i}}[m^e_i(w^{char}_{A, i}+w^{char}_{A, i-1})-1]}{(w^{char}_{A, i}+w^{char}_{A, i-1})}.
\end{eqnarray}

By definition, there is no common factor between $m^o_{i-1}$ and $n^o_{i-1}$. In order that  $w^{char}_{A, i}+w^{char}_{A, i-1}\neq n^o_{i-1}$ (i.e,  $w^{char}_{A, i}+w^{char}_{A, i-1}> n^o_{i-1}$), there must be a common factor between  $w^{char}_{A, i}+w^{char}_{A, i-1}$ and  $\frac{1}{n^e_{i}}[m^e_i(w^{char}_{A, i}+w^{char}_{A, i-1})-1]$. However, this is not possible, because an integer of the form $ab -1$, where $a$ and $b$ are integers, is not divisible by $b$. Thus,  $w^{char}_{A, i}+w^{char}_{A, i-1}= n^o_{i-1}$.

\end{IEEEproof}

\section{Sensitivity and Robustness Studies}\label{sec:sen}
Based on the results in \emph{Section VI}, this section first points out a sensitivity problem that causes $q$-PAM linear PNC systems to be non-robust. After that, a tentative solution is given to achieve robust SER performance.

\subsection{Sensitivity Problem}

For $q$-PAM linear PNC, the best SER performance is achieved when $\eta=1$  and  $\eta\geq q-1$, where $d^{(\alpha_{opt}, \beta_{opt})}_{\min}=1$. In terms of power efficiency, the operating point $\eta=1$ is the most efficient. In particular, for $\eta>1$, where the power of node $A$ increases while the power of node $B$ is kept constant, better performance cannot be achieved, as can be inferred from the $d^{(\alpha_{opt}, \beta_{opt})}_{\min}$ versus $\eta$ curve.

To maintain $\eta=1$, a straightforward solution is to employ power control at the transmitters to ensure receive powers are balanced at the relay. In real communication systems, however, perfect power balance is probably not realizable due to imperfect CSIT (i.e., the channel state information at the transmitters may not be perfect). Imperfect CSIT can be due to channel estimation error or simply due to changing channel gains that cause outdated channel estimates. A slight imperfect CSIT may cause a slight deviation from perfect power control, leading to a slight deviation from the ideal case of  $\eta=1$, i.e., $\eta$  may be close to $1$ but not exactly $1$.

In the following, we explain that slight deviation from perfect power balance (i.e.,  $\eta=1$) may cause catastrophic SER degradations in $q$-PAM PNC, particularly for higher-order modulations.

From the $d^{(\alpha_{opt}, \beta_{opt})}_{\min}$ versus $\eta$  curve in Fig.  \ref{fig:secvdim}, we expect the SER performance of  $q$-PAM linear PNC to be poor at the odd turning points. In particular, for $\eta$  close to 1, $d^{(\alpha_{opt}, \beta_{opt})}_{\min}$ drops drastically  at the first and second odd turning points. A slight deviation from an even turning point will cause a large drop in  $d^{(\alpha_{opt}, \beta_{opt})}_{\min}$. To see how sensitive the SER performance to  $\eta$ is, let us focus on the two odd turning points closest to the ideal  $\eta=1$  case.


By \emph{Observation 2}, the first odd turning point occurs when the reference symbol resides in the middle of the first and second characteristic symbols, i.e. $(w^{char}_{A,1}, w^{char}_{B,1})=(1, q-2)$ and $(w^{char}_{A,2}, w^{char}_{B,2})=(q-2, 0)$,   at  $\eta^o_1=\frac{q}{q-1}$. By \emph{Corollary \ref{cor:dmin}}, at this  $\eta^o_1$,  $d^{(\alpha_{opt}, \beta_{opt})}_{\min}=\frac{1}{q-1}$.

\begin{corollary}\label{cor:12odd}
  The second odd turning point occurs at $\eta^o_2=\frac{2q-3}{2q-5}$  and it has the smallest   $d^{(\alpha_{opt}, \beta_{opt})}_{\min}=\frac{1}{2q-5}$ among all odd turning points.
\end{corollary}

\begin{IEEEproof}[Proof of Corollary \ref{cor:12odd}]

  In this proof, we focus on the odd turning points, since the global minimum must be found among all the local minimums. By \emph{Corollary \ref{cor:dmin}}, $d^{(\alpha_{opt}, \beta_{opt})}_{\min}=1/(w^{char}_{A,i}+w^{char}_{A,i+1})$  when the  $i$th odd turning point occurs. Therefore, to find the smallest $d^{(\alpha_{opt}, \beta_{opt})}_{\min}$ among all odd turning points, it is sufficient to maximize $(w^{char}_{A,i}+w^{char}_{A,i+1})$.

 By definition, the characteristic symbol in  $\mathcal{W}^{char}_{(A, B)}$ satisfies that $w^{char}_A<q-1$, since none of the symbols in $\mathcal{W}^{char}_{(A, B)}$  has $w^{char}_A=q-1$  (i.e., the joint symbols of the form  $(w_A, w_B)=(q-1, w_B)$, where  $w_B>0$, are all to the right of the reference symbol $(0, q-1)$  for  $\eta\geq 1$). Therefore, in the range    $0<w^{char}_A\leq q-2$, $w^{char}_A=q-2$   and $q-3$  are possible solutions to maximize $(w^{char}_{A,i}+w^{char}_{A,i+1})$.

  Next, we show that $(q-2, 0)$  and $(q-3, 1)$  are the second and third characteristic symbols (i.e., $w^{char}_{A,2}=q-2$  and $w^{char}_{A,3}=q-3$).  First, we note that $(w^{char}_{A, 1}, w^{char}_{B, 1})=(1, q-2)$  is the first characteristic symbol overlapping with the reference symbol at  $\eta=1$. For all other characteristic symbols  $(w^{char}_A, w^{char}_B)$, the equality $w^{char}_A+ w^{char}_B\leq q-2$  must be satisfied. Among all these characteristic symbols,  $(q-2, 0)$ must overlap with the reference symbol next because it is one of the closest symbols to the reference symbol on the left at $\eta=1$ (the other are $(q-3, 1), (q-4, 2), \ldots$ ) and among these closest symbols,  $(q-2, 0)$ is the fastest moving symbol as $\eta$  increases. Thus, $(q-2, 0)$  must the second characteristic symbol.  Now, among these closest symbols at  $\eta=1$, after  $(q-2, 0)$, symbol $(q-3, 1)$ is the next fastest moving symbol. Thus, $(q-3, 1)$  is a candidate for the third characteristic symbol. The only way it is not the third characteristic symbol is that a symbol to the left of $(q-3, 1)$  at $\eta=1$  overtakes $(q-3, 1)$ as before $(q-3, 1)$   overlaps with the reference symbol as $\eta$  increases. However, all the symbols to the left of $(q-3, 1)$  at $\eta=1$  cannot overtake it because their  $w^{char}_A\leq q-3$ (i.e., they all must satisfy $w^{char}_A+w^{char}_B\leq q-3$  in order to be to the left of $(q-3, 1)$  at  $\eta=1$). Therefore, $(q-3, 1)$  is the third characteristic symbol. Lastly, by \emph{Observation 2}, the second odd turning point occurs at $\eta^o_{2}=\frac{2(q-1)-(w^{char}_{B, 1}+w^{char}_{B, 2})}{w^{char}_{A, 1}+w^{char}_{A, 2}}=\frac{2q-3}{2q-5}$.

\end{IEEEproof}

\begin{figure}[t]
 \centering
        \includegraphics[width=0.98\columnwidth]{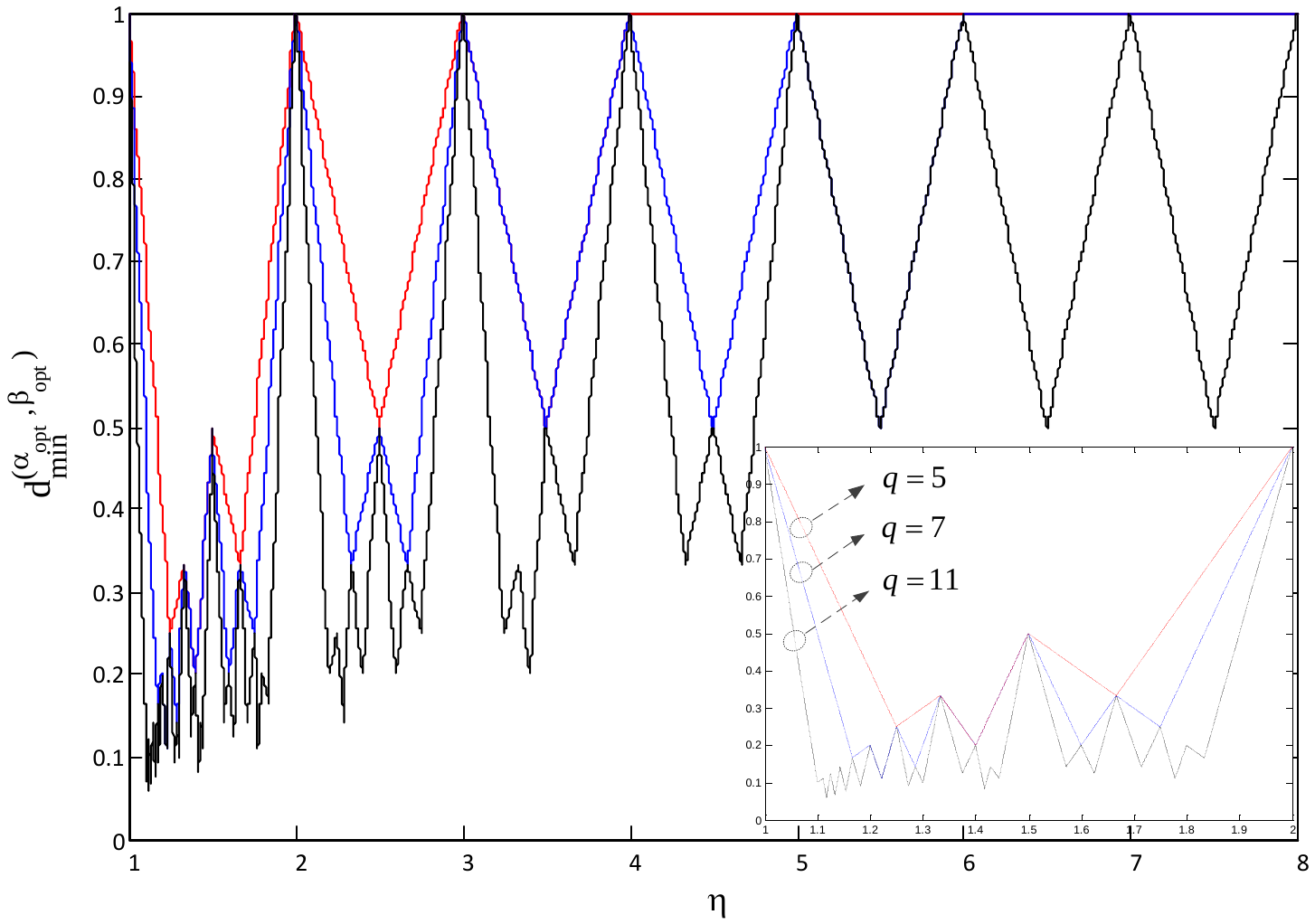}
       \caption{$d^{(\alpha_{opt}, \beta_{opt})}_{\min}$ versus $\eta$ for    $q=5, 7,$ and $11$.}
        \label{fig:sim1}
\end{figure}

From \emph{Corollary \ref{cor:12odd}}, we can see the higher the order of modulation (i.e., the larger the $q$), the smaller the $d^{(\alpha_{opt}, \beta_{opt})}_{\min}$   (i.e., $d^{(\alpha_{opt}, \beta_{opt})}_{\min}$ approaches $0$ as $q$ increases). Furthermore, this occurs with  $\eta$   progressively closer to $1$ as $q$ increases. To illustrate the above observations,   we plot the $d^{(\alpha_{opt}, \beta_{opt})}_{\min}$ versus $\eta$  curves for $q=5, 7$, and $11$ in Fig. \ref{fig:sim1}.  Due to the drastic drop in $d^{(\alpha_{opt}, \beta_{opt})}_{\min}$ near $\eta=1$, the SER is likely to degrade significantly with even a tiny deviation from the perfect power-balanced case  $\eta=1$. Such deviations are unavoidable in real communications systems due to slight wireless channel gain variation, slight imperfection in power control, and slight quantization error at the receiver.  The simulation results shown in  Figs. \ref{fig:sim2} and \ref{fig:sim3} confirm the SER degradation of $5$-PAM and $7$-PAM PNC (highlighted with the dashed lines). Further discussions about these figures together with a solution to the sensitivity problem  will be given later.

We next propose an asynchronized $q$-PAM linear PNC system to restore the SER performance under tiny channel variations, hence allowing robust operation in practice. In this scheme, we deliberately introduce symbol misalignment between the received signals of the two nodes at the relay. At the receiver (relay), we use a belief propagation (BP) decoder to obtain ML estimates of the NC symbols.

\subsection{System Model of Asynchronized q-PAM Linear PNC}

In asynchronized PNC, we control the timing of transmissions at the transmitters to deliberately introduce a symbol misalignment  $\mathcal{D}\in [0, T)$ between the symbols of the two nodes at the relay, where $T$ is a symbol duration. The received signal with symbol misalignment  $\mathcal{D}$  at baseband is
 \begin{align}\label{gyr}
 \nonumber y_R(t) = &\sum^N_{n=1}\big\{ h_A \sqrt{P} x_A[n] p(t-nT) + \\
  &h_B \sqrt{P} x_B[n]p(t-\mathcal{D}-nT) \big\} +z(t),
    \end{align}
where  the signal arrival time of $B$ lags behind that of $A$ by $\mathcal{D}$. For simple exposition, we assume that $\mathcal{D}$ is within one symbol period   and $p(\cdot)$  is a rectangular pulse. After matched filtering, we oversample the signal to obtain  $2N+1$ samples \cite{lulu,lutec}
 \begin{align}\label{ayr1}
 \nonumber y_R[2n-1] &=  h_A \sqrt{P} x_A[n] + h_B \sqrt{P} x_B[n-1] +z[2n-1],\\
  \nonumber y_R[2n]& =  h_A \sqrt{P} x_A[n] + h_B \sqrt{P} x_B[n] + z[2n], \\
y_R[2N+1] &= h_A \sqrt{P} x_B[N]  + z[2N+1],
    \end{align}
where $n=1, 2, \ldots, N$, and $z[2n-1]$ and $z[2n]$ are zero-mean Gaussian noise with variances $N_0/(2\mathcal{D})$ and  $N_0/[2(1-\mathcal{D})]$, respectively. By definition, $x_B[0]=0$, since the signal of node $B$ has not arrived yet when the signal of node $A$ first arrives.

\begin{figure}[t]
 \centering
        \includegraphics[width=0.82\columnwidth]{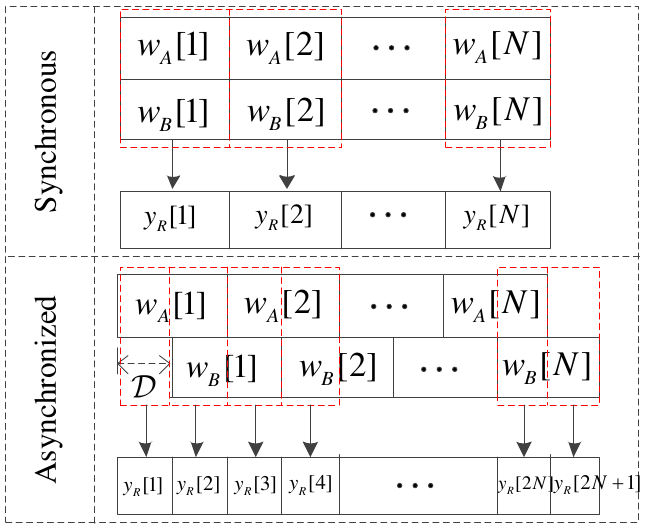}
       \caption{Sampling schemes in synchronous and asynchronized linear PNC systems.}
        \label{fig:samp}
\end{figure}

Fig. \ref{fig:samp} illustrates the sampling schemes of synchronous PNC in \eqref{syr2} and  asynchronized PNC in \eqref{ayr1},  where synchronous PNC is simply the PNC discussed in the preceding sections.   With symbol offset and  oversampling in asynchronized  PNC, we not only have the sample containing information on $(w_A[n], w_B[n])$, but also samples containing information on $(w_A[n+1], w_B[n])$ and $(w_A[n], w_B[n-1])$. The $2N+1$ samples are correlated with each other.

\subsection{Asynchronized $q$-PAM Linear PNC with the BP Decoder}

In asynchronized PNC, the samples in $\mathbf{y}_R=(y_R[n])_{n=1,2,\ldots, 2N+1}$ are not independent and information on $(w_A[n], w_B[n])$ for a particular $n$ is contained in all samples $(y_R[n])_{n=1,2,\ldots, 2N+1}$ through correlations among the samples. Here we consider an ML decoding rule for the  NC symbols that make use of all samples  $(y_R[n])_{n=1,2,\ldots, 2N+1}$. Recall that  \eqref{map} describes the ML decoding rule for  synchronous PNC. For asynchronized PNC, since all samples $(y_R[n])_{n=1,2,\ldots, 2N+1}$ contain  information on $(w_A[n], w_B[n])$, in place of $Pr((w_A[n], w_B[n])|y_R[n])$   in which $(w_A[n], w_B[n])$ depends only on one sample, we have  $Pr((w_A[n], w_B[n])|\mathbf{y}_R)$ for asynchronized system,   in which $(w_A[n], w_B[n])$ depends on all samples.

A decoder based on the belief propagation (BP) algorithm can be used to compute  $Pr((w_A[n], w_B[n])|\mathbf{y}_R)$. The BP decoder, also known as sum-product decoder, makes use of the Bayes' rule to compute  $Pr((w_A[n], w_B[n])|\mathbf{y}_R)$ for all $n$ via a message passing algorithm \cite{yedbp,ksc}. Details of the BP decoder can be found in \cite{lulu, lutec}, where oversampling was also used, but for a different problem (specifically, the problem being tackled in \cite{lulu, lutec} was penalty caused  by the phase offset between the two end nodes in the PNC system). We omit the details here and refer the interested reader to \cite{lulu, lutec},  since the algorithm there can be easily extrapolated and adapted  for our application here.


\subsection{Intuitive Explanation of the Advantages of Asynchronization}

We now explain intuitively why asynchronized PNC with the BP decoder can outperform synchronous PNC. Suppose that $\eta$ is such that the system is at an odd turning point, where $d^{(\alpha_{opt}, \beta_{opt})}_{\min}$ is small. Furthermore, suppose that nodes $A$ and $B$ transmit a joint symbol $(w_A[n], w_B[n])$ whose superimposed symbol is at distance $d^{(\alpha_{opt}, \beta_{opt})}_{\min}$  from the superimposed symbol of another joint symbol  $(w'_A[n], w'_B[n])$, and that $(w_A[n], w_B[n])$  and  $(w'_A[n], w'_B[n])$ are mapped to different NC symbols via $(\alpha_{opt}, \beta_{opt})$. With the small  $d^{(\alpha_{opt}, \beta_{opt})}_{\min}$, it is easy to make mistake in the decoding of the NC symbol if we have only one sample $y_R$ to base our decision on, giving rise to high SER. This is the case for the synchronous system.

For the asynchronized system with the BP decoder,  two effects   come into play to reduce SER. First, symbol misalignment introduces diversity. Although $(w_A[n], w_B[n])$ is at $d^{(\alpha_{opt}, \beta_{opt})}_{\min}$ from another joint symbol $(w'_A[n], w'_B[n])$, the joint symbols $(w_A[n], w_B[n-1])$ and/or $(w_A[n+1], w_B[n])$ associated with the adjacent samples may have neighbors that are further away than $d^{(\alpha_{opt}, \beta_{opt})}_{\min}$. If so, this allows us to decode $(w_A[n], w_B[n-1])$  and/or $(w_A[n+1], w_B[n])$  with high certainty. Say, we manage to decode  $(w_A[n], w_B[n-1])$, then $w_A[n]$  is known. This in turn allow us to  make decision on $(w_A[n], w_B[n])$ by selecting a joint symbol among joint symbols of the form $(w_A[n], \cdot)$ rather than among all joint symbols (i.e., only $w_B[n]$ is unknown and   $w_A[n]$ is already known). And among joint symbols of the form  $(w_A[n], \cdot)$, the neighbors of $(w_A[n], w_B[n])$  may be much further from it than $(w'_A[n], w'_B[n])$.  Thus, it is more likely for us to detect the correct joint symbol $(w_A[n], w_B[n])$.

The second effect that reduces the SER is certainty propagation. The previous paragraph explained how the decoding of  $(w_A[n], w_B[n])$ is assisted by the decoding of $(w_A[n], w_B[n-1])$ and $(w_A[n+1], w_B[n])$  when $(w_A[n], w_B[n-1])$ and $(w_A[n+1], w_B[n])$ have neighbors that are far from them. Now, even if both $(w_A[n], w_B[n-1])$  and $(w_A[n+1], w_B[n])$  have neighbors close to them (e.g., at $d^{(\alpha_{opt}, \beta_{opt})}_{\min}$  from them),  $(w_A[n-1], w_B[n-1])$ and/or $(w_A[n+1], w_B[n+1])$  associated with the samples even further away may not. Once $(w_A[n-1], w_B[n-1])$ or $(w_A[n+1], w_B[n+1])$  can be decoded with certainty, so can $(w_A[n], w_B[n-1])$  or $(w_A[n+1], w_B[n])$, and from   $(w_A[n], w_B[n-1])$  or $(w_A[n+1], w_B[n])$, the certainty is propagated to  $(w_A[n], w_B[n])$. The BP algorithm allows certainty to propagate from sample to sample in a chain-like manner, drastically reducing the SER. The underlying fundamental of such certainty propagation is Bayes' rule. Rather than absolute certainty as expounded in the above intuitive explanation, the degree of certainty is expressed in terms of probability in a rigorous manner under the BP framework. The interested reader is referred to \cite{lulu, lutec} for further details.

 \section{SER Performance} \label{sec:sim}

Prior to this section, we have focused on  $d^{(\alpha_{opt}, \beta_{opt})}_{\min}$, assuming that maximizing $d^{(\alpha_{opt}, \beta_{opt})}_{\min}$ will lead to lower SER. While this is true most of the time, it is not always so: as will be explained shortly, we need to be careful at odd turning points. \emph{Part A} is devoted to clarifying the relationship between $d^{(\alpha_{opt}, \beta_{opt})}_{\min}$ and SER, focusing on the synchronous PNC. \emph{Part B} then presents and compares the SER performance of synchronous and asynchronized PNC.

\subsection{Clarifying the Relationship between $d^{(\alpha_{opt}, \beta_{opt})}_{\min}$ and SER }

Let us first reexamine the different detection rules as presented in \emph{Part C, Section III}. Recall that the decoding error probabilities of MD and ML rules are almost the same in the high SNR regime. As indicated in \eqref{ser}, for the MD rule under high SNR, the decoding error probability is dominated by  $d^{(\alpha_{opt}, \beta_{opt})}_{\min}$ with secondary effects in $\mathcal{A}^{(\alpha, \beta)}_{\min}$.

In the preceding sections, we focused on deriving $(\alpha_{opt}, \beta_{opt})$, i.e.,  $d_{\min}$-optimal $(\alpha,\beta)$  that maximizes $d_{\min}$. From our analytical results in \emph{Section VI}, we know that at odd turning points, the reference joint symbol can either cluster with its left neighbor or its right neighbor, both of which are at equal distance to the reference symbol. Although the  $(\alpha,\beta)$ are different in these two cases, they are both $(\alpha_{opt}, \beta_{opt})$. The resulting $d^{(\alpha, \beta)}_{\min}$  are also the same. This is where the effect of the multiplicity  $\mathcal{A}^{(\alpha, \beta)}_{\min}$  comes in. Specifically, $\mathcal{A}^{(\alpha, \beta)}_{\min}$  refers to the number of pairs of NC symbols that are at distance $d^{(\alpha, \beta)}_{\min}$  apart under the NC mapping induced by $(\alpha, \beta)$. Clustering with the left neighbor and the right neighbor can result in different $\mathcal{A}^{(\alpha, \beta)}_{\min}$.

\begin{figure}[t]
 \centering
        \includegraphics[width=0.9\columnwidth]{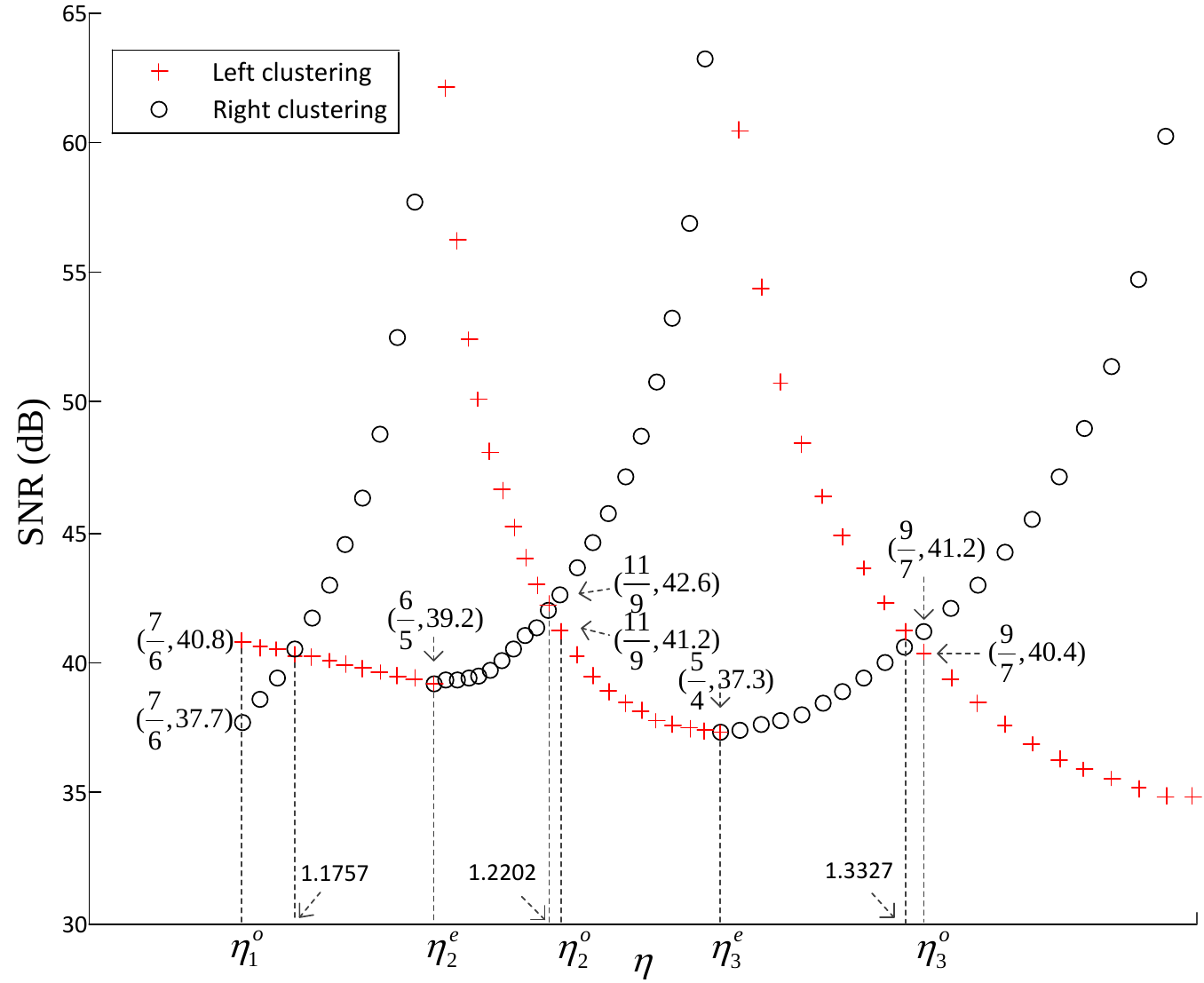}
       \caption{SNR required  to meet ${\rm SER}=10^{-3}$ in 7-PAM linear PNC for different $\eta$. }
        \label{fig:mdex}
\end{figure}

In Fig. \ref{fig:mdex}, we plot the required SNR to meet ${\rm SER}=10^{-3}$ for different $\eta$ in 7-PAM PNC. At a particular  $\eta$, we use cross and circle  to denote the required SNR when the reference symbol is clustered with its left neighbor (i.e., left clustering) and its right neighbor (i.e., right clustering), respectively. As shown in Fig. \ref{fig:mdex}, left and right clustering can have different SER performance at and in the neighborhood of odd turning points, especially at the first and second odd turning points.

\begin{figure}[t]
 \centering
        \includegraphics[width=1\columnwidth]{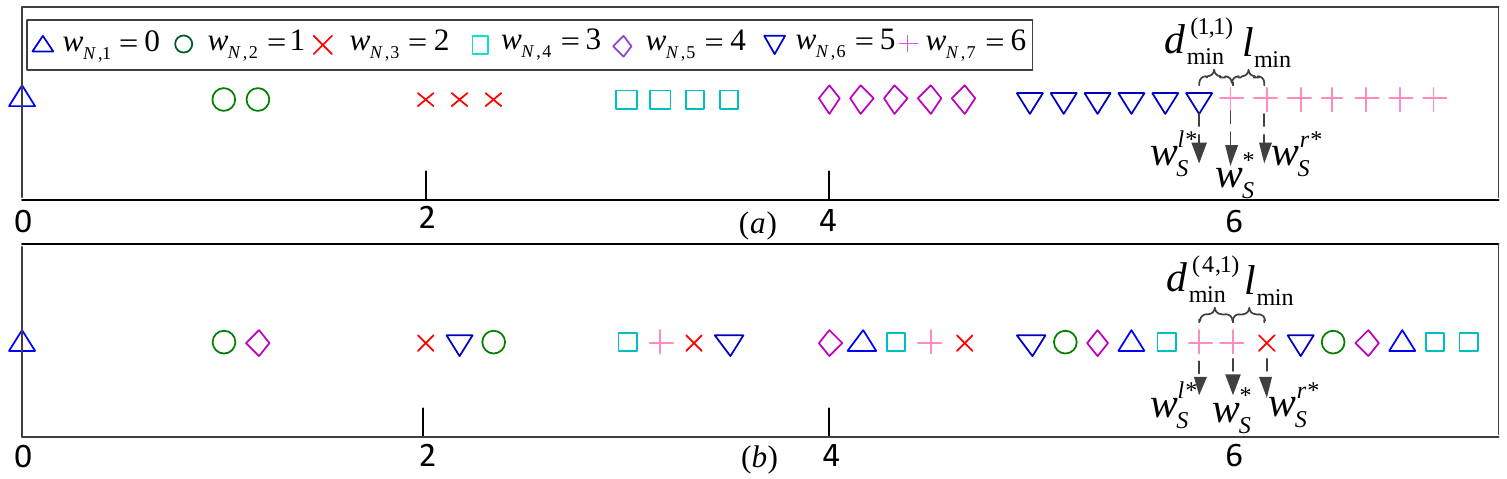}
       \caption{Clustered constellations of $\mathcal{W}_S$ for $7$-PAM linear PNC at  $\eta^o_1=7/6$, where the reference symbol is clustered with the left neighbor by $(\alpha_{opt}, \beta_{opt})=(1,1)$  in (a), and clustered with the right neighbor by $(\alpha_{opt}, \beta_{opt})=(4,1)$   in (b). Constellations beyond $7$ on the real line are omitted to avoid cluttering.}
        \label{fig:leftclu}
\end{figure}

At the first odd turning point, the right clustering needs a smaller SNR to achieve ${\rm SER}=10^{-3}$  than the left clustering and the SNR gap is over $3$dB. To shed light on the SNR gap, Fig. \ref{fig:leftclu} (a) and Fig. \ref{fig:leftclu} (b) show the clustered constellations for left and right clustering, respectively, at the first odd turning point. On the right of the reference symbol, we have a sequence of adjacent symbols separated by the same difference $(\delta^{r*}_A, \delta^{r*}_B)=(1, 5)-(0, 6)=(1,1)$  and the same distance $l_{\min}=\eta^o_1\delta^{r*}_A+\delta^{r*}_B$. In Fig. \ref{fig:leftclu} (a), we use $(\alpha, \beta)=(1,1)$ so that these symbols separated by $(\delta^{r*}_A, \delta^{r*}_B)$ are clustered to the same NC symbol. In Fig. \ref{fig:leftclu} (b), however, the left clustering does not cluster these adjacent symbols on the right of the reference symbol. In other words, the NC symbols of these adjacent symbols are different and the distances between each pair of the adjacent symbols is $d_{\min}$. As a result, the multiplicity $\mathcal{A}^{(\alpha, \beta)}_{\min}$  is larger for left clustering, resulting in worse SER performance.  Specifically,  $\mathcal{A}^{(\alpha, \beta)}_{\min}=36$ for left clustering and  $\mathcal{A}^{(\alpha, \beta)}_{\min}=2$ for right clustering.

At the second and third odd turning points, left clustering is better than right clustering,   since left clustering has a lower $\mathcal{A}^{(\alpha, \beta)}_{\min}$ at these two points. The SNR gap between these two clustering becomes less significant than that between the first odd turning point, since the ratio of $\mathcal{A}^{(\alpha, \beta)}_{\min}$ between these two clustering becomes smaller.

\begin{figure}[t]
 \centering
        \includegraphics[width=0.9\columnwidth]{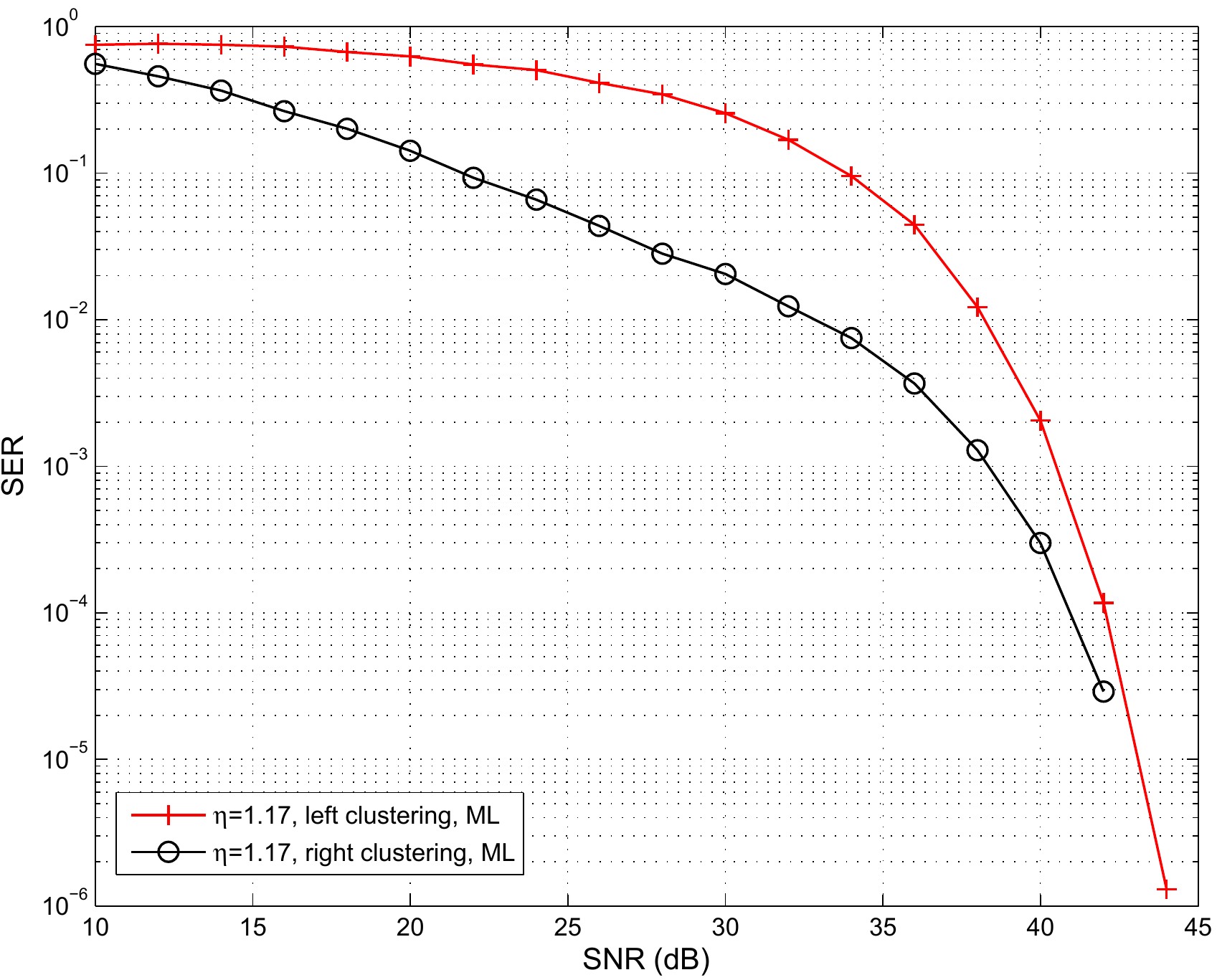}
       \caption{SER performance of $7$-PAM linear PNC at $\eta=1.17$ with left and right clustering.}
        \label{fig:a1sto}
\end{figure}

With respect to Fig. \ref{fig:leftclu}, let us focus on $\eta$ slightly to the right of $\eta^o_1$. Here, the left neighbor is closer to the reference symbol than the right neighbor. Therefore, left clustering will maximize $d_{\min}$. However, as shown in Fig. \ref{fig:leftclu}, it is the right clustering that has a lower SNR requirement (up to several dB in difference). Thus, around this odd turning point, the effect of $\mathcal{A}^{(\alpha, \beta)}_{\min}$  persists even after we depart from the odd turning point, so much so that its effect dominates over the effect of  $d_{\min}$. In Fig. \ref{fig:a1sto}, at   $\eta=1.17$, which is slightly larger than  $\eta^o_1=7/6$, right clustering with a higher $\mathcal{A}^{(\alpha, \beta)}_{\min}$  needs around $3$dB less of the required SNR than left clustering to meet ${\rm SER}=10^{-3}$, consistent with the SNR gap at this $\eta$ in Fig. \ref{fig:mdex}.

Lastly, we note that at even turning points, we have no choice but to cluster the reference symbol with the symbols overlapping with it if we are to avoid  $d_{\min}=0$. Thus, the issues of left clustering versus right clustering and the associated SNR gap do  not arise.

Overall, our conclusion is that we need to pay attention to $\mathcal{A}^{(\alpha, \beta)}_{\min}$  at odd turning points and choose left or right clustering based on  $\mathcal{A}^{(\alpha, \beta)}_{\min}$ to break the tie. At  $\eta$ in the neighborhood of odd turning points, we also have to be careful because the effect of $\mathcal{A}^{(\alpha, \beta)}_{\min}$  will persist for a while. At other  $\eta$, we can simply focus on maximizing $d_{\min}$  to minimize SER.

\subsection{Robustness of Asynchronized PNC}


We now look at the SER performance of synchronous and asynchronized PNC at odd turning points. For asynchronized PNC, we assume a symbol offset of half symbol duration is introduced (i.e., $\mathcal{D}=T/2$ in \eqref{gyr}).

\begin{figure}[t]
 \centering
        \includegraphics[height=0.75\columnwidth]{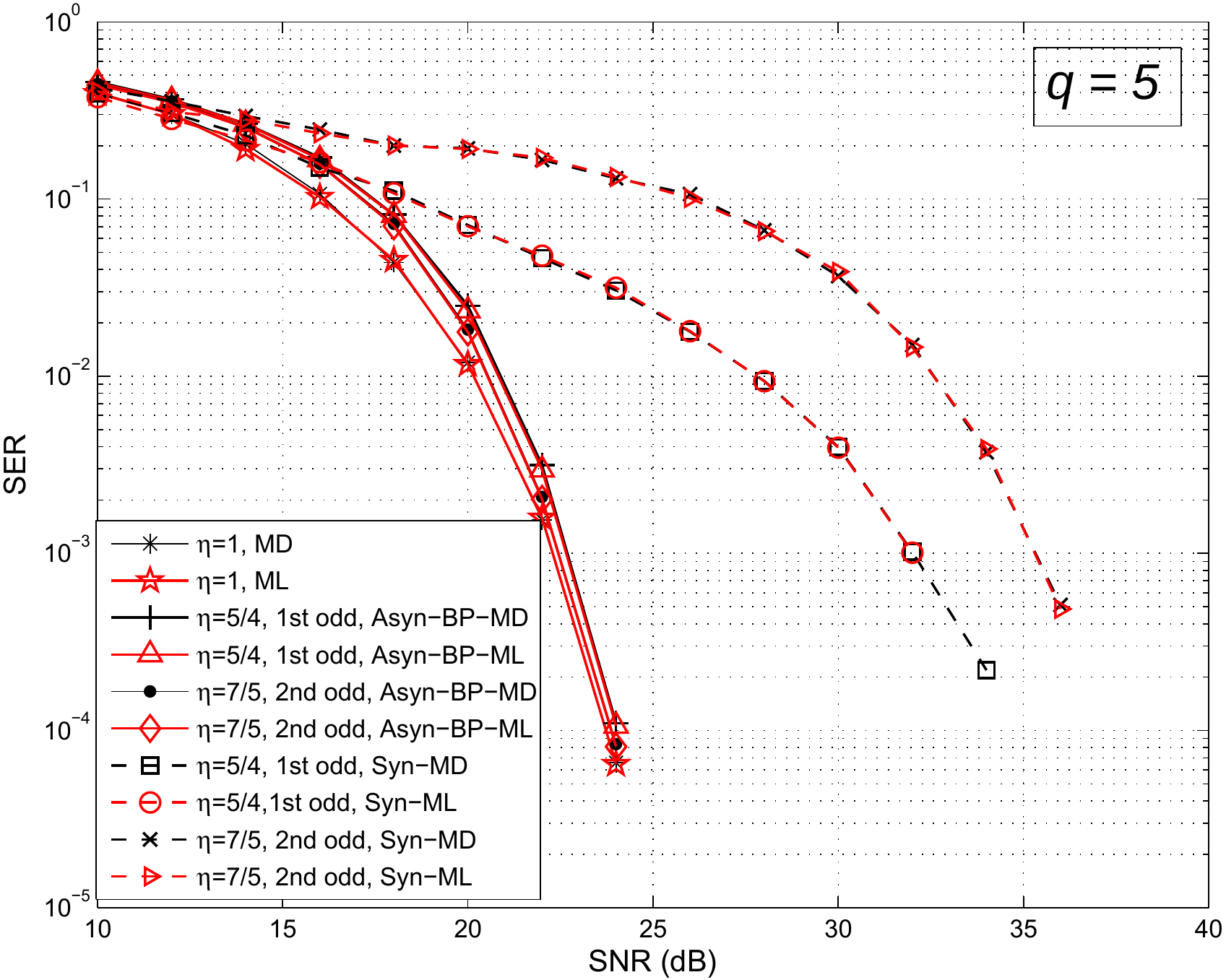}
       \caption{SER performance of synchronous and asynchronized $5$-PAM linear PNC.}
        \label{fig:sim2}
\end{figure}

Fig. \ref{fig:sim2}  presents the SER of synchronous and asynchronized PNC when $q=5$, under $\eta=1$,  $\eta=5/4$ (first odd turning point), and $\eta=7/5$ (second odd turning point).  The dashed and solid lines correspond to synchronous and asynchronized PNC, respectively. For synchronous PNC, we observe that the SER degrades significantly at the first and second odd turning points. In particular, the second odd turning point has the worst SER due to the minimum  $d^{(\alpha_{opt}, \beta_{opt})}_{\min}$, and the best SER is obtained at $\eta=1$ with the maximum $d^{(\alpha_{opt}, \beta_{opt})}_{\min}=1$. For asynchronized PNC, symbol misalignment with the BP decoder can significantly improve the SER performance in the high SNR regime. This agrees with our analytical results in \emph{Part C}, \emph{Section \ref{sec:sen}}.  Moreover, we see that the SER gap between ML and MD decoders is negligible, consistent with our analysis in \emph{Part C}, \emph{Section \ref{syn}}.

\begin{figure}[t]
  \centering
        \includegraphics[height=0.75\columnwidth]{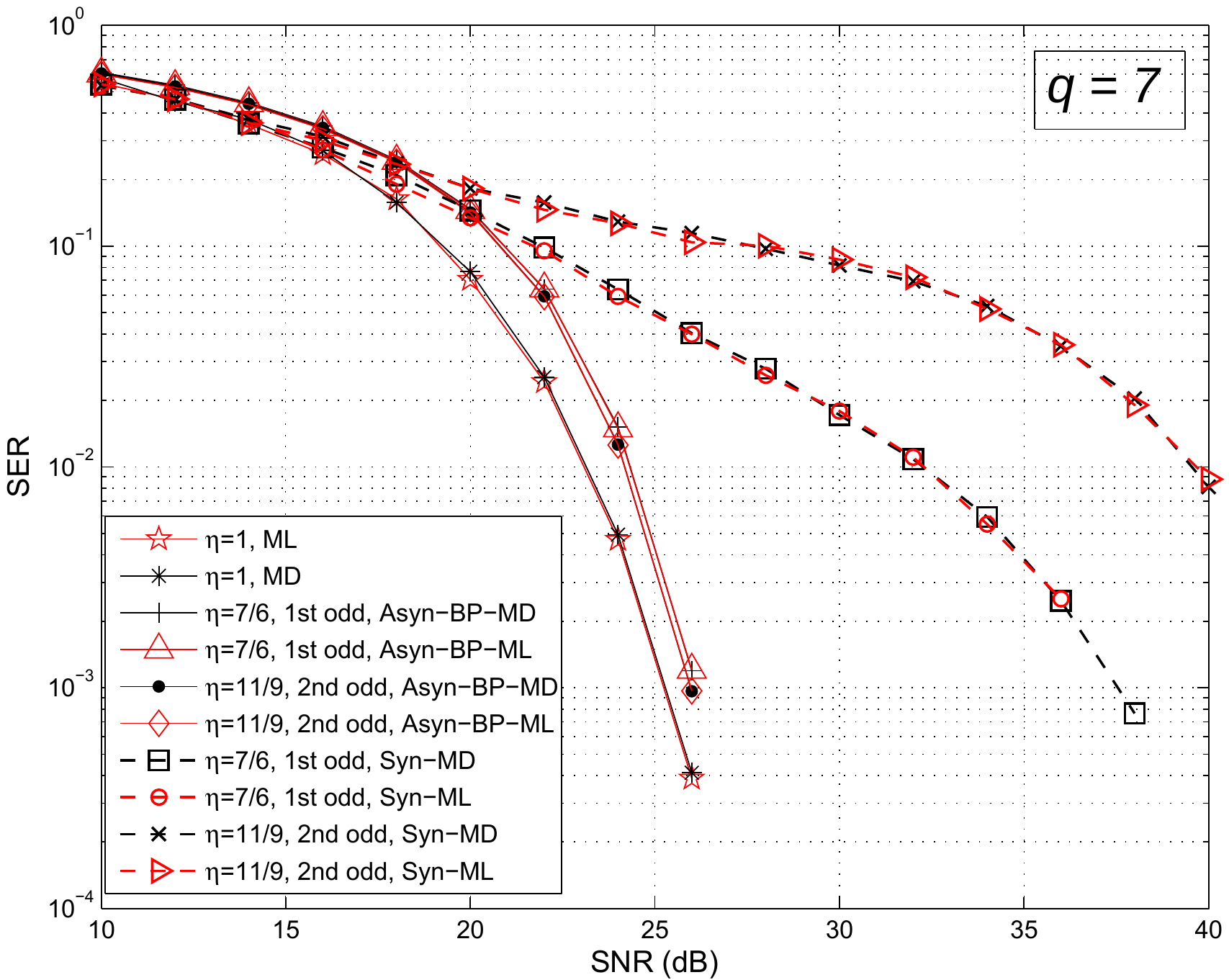}
       \caption{SER performance of synchronous and asynchronized $7$-PAM linear PNC.}
        \label{fig:sim3}
\end{figure}

Fig. \ref{fig:sim3} presents the SER of synchronous and asynchronized PNC when  $q=7$, under $\eta=1$,  $\eta=7/6$ (first odd turning point), and $\eta=11/9$ (second turning point). Observations similar to those of $q=5$  apply here also, except that higher SNR is needed  to obtain the same SER, due to the higher order modulation.

\section{Conclusion}

In this paper, we have thoroughly investigated the subtleties of applying $q$-PAM linear PNC in TWR channels. Going beyond \cite{yangtwc}, we derived the analytical dependence of minimum distance between superimposed symbols (a key performance determining factor) on the relative channel gains of the two users. In particular, we gave a systematic way to obtain the analytical relationship between minimum distance and channel-gain ratio for all $q$, allowing us to examine the exact dependence of minimum distance on $q$ and channel-gain ratio. An insight obtained is that the performance of $q$-PAM linear PNC systems is extremely sensitive to the slight imbalance in the received powers from two users at the relay, particularly when $q$ is large. Thus, a negative conclusion is that high-order $q$-PAM PNC is not robust.

We proposed a solution---the introduction of symbol asynchrony and the use of a BP decoder---to overcome the sensitivity problem. We showed that such asynchronized $q$-PAM PNC can significantly recover the SER performance loss caused by the sensitivity problem, making the system robust against power imbalance.

Going forward, several areas of research deserve further investigation. The analytical relationship between minimum distance and channel-gain ratio under $q^2$-QAM is yet to be derived. As shown in this paper, establishing this analytical relationship is already non-trivial for $q$-PAM, the use of $q^2$-QAM introduces an additional dimension, the phase difference between the two users, that will make the derivation of such a relationship even more challenging. In addition, we have not considered the use of channel coding in this paper. The investigation of the robustness of channel-coded linear PNC deserves further attention.

\section*{Appendix I: Relationships among Propositions, Lemmas, Corollaries and Theorems}

\begin{figure}[t]
  \centering
        \includegraphics[height=0.8\columnwidth]{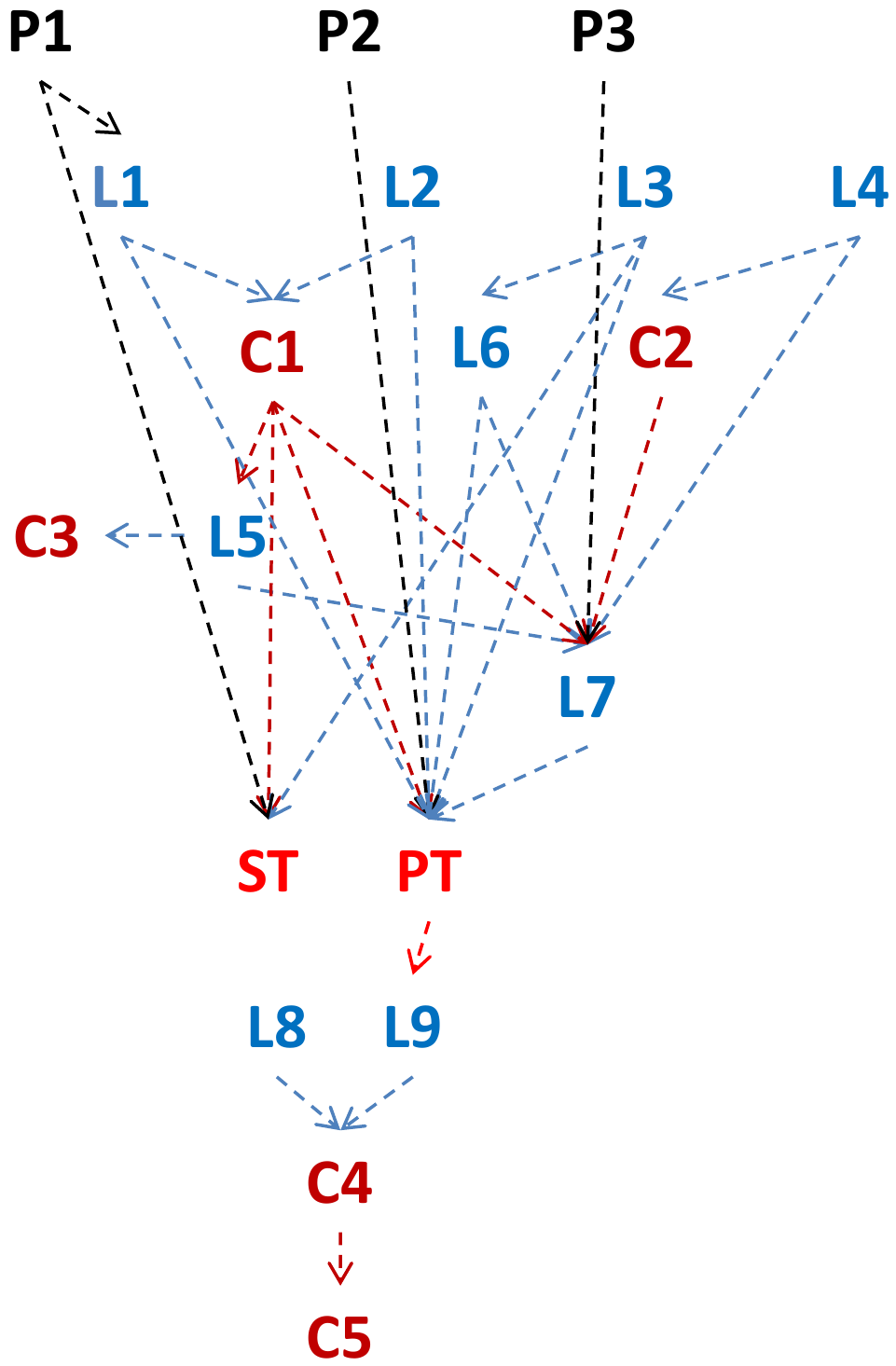}
       \caption{Relationships among Propositions (P), Lemmas (L), Corollaries (C), and Theorems (T).}
        \label{fig:graph}
\end{figure}

For ease of understanding, Fig. \ref{fig:graph} illustrates relationships among Propositions (P), Lemmas (L), Corollaries (C), and Theorems (T) in this paper, where we use the dashed line with arrow to link two related objectives.

\section*{Appendix II: No loss in generality in assuming positive $h_A$ and $h_B$ }

Suppose that $h_A$ is negative and $h_B$ is positive. Define $h'_A = - h_A$ and $x'_A = - x_A$. We could ``pretend'' that $h'_A$ is the channel gain and that $x'_A$ is the modulated signal from node $A$. Here, $x'_A$  corresponds to $w'_A=q-1-w_A$.   Then, \eqref{yr1} can be written as
 \begin{align}\label{yrph}
\nonumber   y_R  =  &  \frac{\sqrt{P}}{\mu}( {h'_A (q-1-w_A) + h_B w_B} )+z\\
&- \frac{\sqrt{P}(q-1)}{2\mu}(h'_A+h_B).
    \end{align}

As a result of this transformation, the NC symbol will be
\begin{align}\label{wnph}
w^{(\alpha', \beta)}_N=\alpha'\otimes w'_A \oplus \beta \otimes w_B.
\end{align}

At the relay, we define $\alpha = -\alpha'$ (in $GF(q)$), and $w_A = - w'_A$ (in $GF(q)$). Then, \eqref{wnph} can be written as
\begin{align}
w^{(\alpha', \beta)}_N=w^{(\alpha, \beta)}_N=\alpha\otimes w_A \oplus \beta \otimes w_B
\end{align}

Thus, after computing $w^{(\alpha', \beta)}_N$ in \eqref{wnph}, instead of sending $(\alpha', \beta)$, the relay sends $(\alpha, \beta)$ as the NC coefficients to the end nodes for decoding purposes. The NC symbols sent by the relay are not changed; only the NC coefficients are.

Similar treatment applies when both $h_A$ and $h_B$ are negative. The relay  pretends the channel coefficients are positive when performing PNC mapping. The relay then negates the NC coefficients sent to the end nodes for decoding purposes over there.

\section*{Appendix III: Proof of Principal Theorem}

\emph{(Part 1)}: In Part 1, we consider the case where $(w^{o*}_A, w^{o*}_B)$ exists. If $(w^{o*}_A, w^{o*}_B)$ exists, then $l_{\min}=0$. By \emph{Lemma \ref{lem:ovl}},   we can cluster  overlapping  symbols. Thus, $d^{(\alpha_{opt}, \beta_{opt})}_{\min}>l_{\min}=0$.   By \emph{Lemma \ref{lem:cluo}},  we cannot cluster  $(w^{l*}_A, w^{l*}_B)$   into the same NC symbol as $(w^{*}_A, w^{*}_B)$  and $(w^{o*}_A, w^{o*}_B)$; neither can we cluster $(w^{r*}_A, w^{r*}_B)$ into the same NC symbol as $(w^{*}_A, w^{*}_B)$ and $(w^{o*}_A, w^{o*}_B)$. By \emph{Corollary \ref{cor:short}}, $d^{(\alpha_{opt}, \beta_{opt})}_{\min}$ must be observable in the locality of $(w^{*}_A, w^{*}_B)$.  Therefore,  the distance from $(w^*_A, w^*_B)$ to its left or right neighbor is a candidate for $d^{(\alpha_{opt}, \beta_{opt})}_{\min}$.

It remains to be shown the distances between $(w^{*}_A, w^{*}_B)$ and its left and right neighbors are
equal. Without loss of generality, we consider the case where $d^{r*}\triangleq w^{r*}_S-w^{*}_S=d^{(\alpha_{opt}, \beta_{opt})}_{\min}$.  The proof for the case where $d^{l*}\triangleq w^{*}_S-w^{l*}_S=d^{(\alpha_{opt}, \beta_{opt})}_{\min}$ is similar. Define $(\delta^{ro}_A, \delta^{ro}_B)\triangleq (w^{r*}_A-w^{o*}_A, w^{r*}_B-w^{o*}_B)$. By \emph{Lemma \ref{lem2}}, $(\delta^{ro}_A, \delta^{ro}_B)$ cannot be   both nonzero and of the same sign. By \emph{Lemma \ref{lem3}}, we can find a $(w_A, w_B)$ such that one of the following is true:

1) $-(q-1)\leq\delta^{ro}_A<0, 0\leq \delta^{ro}_B\leq q-1$ and $(w_A, w_B)=(w^{*}_A, w^{*}_B)-(\delta^{ro}_A, \delta^{ro}_B)$;

or 2) $0 \leq \delta^{ro}_A\leq q-1, -(q-1)\leq\delta^{ro}_B<0$ and $(w_A, w_B)=(w^{*}_A, w^{*}_B)+(\delta^{ro}_A, \delta^{ro}_B)$.

We first consider case 1), where we have a valid symbol $(w_A, w_B)=(w^{*}_A, w^{*}_B)-(\delta^{ro}_A, \delta^{ro}_B)$. Now, $w^*_S-w_S=\eta\delta^{ro}_A+\delta^{ro}_B=d^{r*}>0$. Since $w^*_S-w_S > 0$,  $(w_A, w_B)$ resides on the left side of $(w^{*}_A, w^{*}_B)$. By definition, however, $(w^{l*}_A, w^{l*}_B)$  is a left neighbor of $(w^{*}_A, w^{*}_B)$, and $d^{l*}$ must be no larger than the distance between $w^{*}_S$ and any other superimposed symbol to the left of $w^*_S$, including $w_S$. Thus, $d^{l*}\leq w^{*}_S-w_S=d^{r*}$. But $d^{r*}=d^{(\alpha_{opt}, \beta_{opt})}_{\min}$ by our earlier supposition, and this means $d^{l*} \geq d^{r*}$. Therefore, $d^{l*}=d^{r*}=d^{(\alpha_{opt}, \beta_{opt})}_{\min}$.

Next, we consider case 2), where we have a valid symbol $(w^{r'*}_A, w^{r'*}_B)=(w^{*}_A, w^{*}_B)+(\delta^{ro}_A, \delta^{ro}_B)=(\delta^{ro}_A, (q-1)+\delta^{ro}_B)$. Unfortunately, $(w^{r'*}_A, w^{r'*}_B)$ resides on the right side of $(w^{*}_A, w^{*}_B)$, the same side as $(w^{r*}_A, w^{r*}_B)$. However, we note that $(w^{r'*}_A, w^{r'*}_B)$ overlaps with $(w^{r*}_A, w^{r*}_B)= (w^{o*}_A, w^{o*}_B)+(\delta^{ro}_A, \delta^{ro}_B)$ and is distinct from it. Thus, $(w^{r'*}_A, w^{r'*}_B)$ is another right neighbor of $(w^{*}_A, w^{*}_B)$. Let us replace $(w^{r*}_A, w^{r*}_B)$ with $(w^{r'*}_A, w^{r'*}_B)$ in our choice as the right neighbor of $(w^{*}_A, w^{*}_B)$ to be considered. Define $(\delta^{r'o}_A, \delta^{r'o}_B)\triangleq (w^{r'*}_A-w^{o*}_A, w^{r'*}_B-w^{o*}_B)$. Note that
\begin{align}\label{ptc2}
  w^{r'*}_S-w^{o*}_S&=\eta\delta^{r'o}_A+\delta^{r'o}_B=d^{r*}=d^{(\alpha_{opt}, \beta_{opt})}_{\min}.\\
 \nonumber  (\delta^{r'o}_A, \delta^{r'o}_B) &= (w^{r'*}_A-w^{o*}_A, w^{r'*}_B-w^{o*}_B)\\ \label{ptc3} &= (\delta^{ro}_A-w^{o*}_A, (q-1)+\delta^{ro}_B-w^{o*}_B).
\end{align}
Note also that since $(w^{o*}_A, w^{o*}_B) \neq (w^*_A, w^*_B)=(0,q-1)$,  we must have $w^{o*}_A>0$ and $w^{o*}_B<q-1$ in order for them to overlap. From \eqref{ptc3}, we have that $\delta^{r'o}_A<\delta^{ro}_A$, $\delta^{r'o}_B>\delta^{ro}_B$. If $\delta^{r'o}_A<0$, then the condition for case 1) is satisfied and we can then apply the previous argument to prove that $d^{l*}=d^{r*}=d^{(\alpha_{opt}, \beta_{opt})}_{\min}$. If not (i.e., $\delta^{r'o}_A\geq 0, \delta^{r'o}_B<0$), we repeat the above procedure (and drawing on the results of \emph{Lemmas \ref{lem2}} and \emph{\ref{lem3}} repeatedly for the existence of an additional right neighbor that has not been considered thus far) until we find a particular right neighbor of $(w^{*}_A, w^{*}_B)$ for which the condition for case 1) is satisfied. In particular,   $\delta^{r' o}_A$ of the progressive right neighbor $(w^{r'*}_A, w^{r'*}_B)$ thus found keeps decreasing until we find one that is negative in value.  The above procedure implies that there is always one right neighbor for which the condition for case 1) is satisfied.

\emph{(Part 2)}: In Part 2, we consider the case where $(w^{o*}_A, w^{o*}_B)$ does not exist  (thus, $l_{\min} > 0$ by \emph{Corollary \ref{cor:short}}). By \emph{Lemma \ref{lem:lmin}}, we cannot  map all three of  $(w^{*}_A, w^{*}_B)$, $(w^{l*}_A, w^{l*}_B)$, and $(w^{r*}_A, w^{r*}_B)$  to the same NC symbol. However, $(w^{*}_A, w^{*}_B)$ can be clustered with $(w^{l*}_A, w^{l*}_B)$  or $(w^{r*}_A, w^{r*}_B)$, depending on which is the closer neighbor. In case the two distances are the same, subcase i) in the \emph{Principal
Theorem} then follows immediately.

Subcases ii) and iii) are similar and can be proved similarly, since ii) corresponds to the situation in which the right neighbor is closer to $(w^{*}_A, w^{*}_B)$ and iii) corresponds to the situation in which the left neighbor is closer. We focus on the proof for ii) here.

Since $w^{r*}_S-w^{*}_S=l_{\min}$, we cluster $(w^{*}_A, w^{*}_B)$ and $(w^{r*}_A, w^{r*}_B)$ together. Let us label the superscripts of the symbols to the right of $(w^{*}_A, w^{*}_B)$ the following way: $r(0)$ is the right neighbor of $*$; $r(1)$ is the right neighbor of $r(0)$;  $r(2)$ is the right neighbor of $r(1)$; and so on. Let $d^{r(i)}$ be the distance between the two symbols labeled by $r(i)$ and $r(i-1)$. We have
\begin{align}\label{pri2}
 d^{r(i)} \triangleq w^{r(i)}_S-w^{r(i-1)}_S\geq l_{\min}, i=1,2, \ldots.
\end{align}
By \emph{Corollary 1},  $d^{(\alpha_{opt}, \beta_{opt})}_{\min}$ is either $w_S^* - w_S^{l*}$ or $w_S^{r(i)} - w_S^*$ for some $i\geq 1$ and $w^{r(i)}_S$ that is not clustered with the reference symbol $w^*_S$. To prove that $d^{(\alpha_{opt}, \beta_{opt})}_{\min}=w^*_S-w^{l*}_S$, we   use an ``algorithmic'' argument  as follows:

\vspace{0.1in}
{\noindent \emph{\textbf{Algorithmic Argument}}}
\vspace{0.1in}

{\noindent \textbf{{Initialization}} $i=1$;}

{\noindent\textbf{{Start}};}
\footnotesize
(at this point, $(w^{*}_A, w^{*}_B), (w^{r(0)}_A, w^{r(0)}_B), \ldots,   (w^{r(i-1)}_A, w^{r(i-1)}_B)$ are known to be mapped to the same NC symbol.  So the distance between $w^*_S$ and $w_S^{r(j)}, j = 0, ..., i-1$ cannot be $d^{(\alpha_{opt}, \beta_{opt})}_{\min}$. )
\normalsize

Stop if $r(i)$ does not exist (i.e., there is no more joint symbols to the right of the reference symbol). The proof is complete since all symbols to the right of the reference symbol are clustered with the reference symbol.  If $r(i)$ exists, one of the following two cases applies:

 \textbf{Case (1)}: $d^{r(i)}>l_{\min}$

 If  $d^{l*}\leq d^{r(i)}$, then $d^{l*}$ must be $d^{(\alpha_{opt}, \beta_{opt})}_{\min}$ since $d^{l*}<l_{\min}+d^{r(i)}\leq w^{r(i)}_S-w^{*}_S<w^{r(k)}_S-w^{*}_S$, $\forall \ k>i$, and our proof is complete.

  If  $d^{l*}>d^{r(i)}$ and $(w^{r(i)}_A, w^{r(i)}_B)$ is mapped to the same NC symbol as $(w^{*}_A, w^{*}_B), (w^{r(0)}_A, w^{r(0)}_B), \ldots, (w^{r(i-1)}_A, w^{r(i-1)}_B)$ then $d^{r(i)}$ is not a candidate for $d^{(\alpha_{opt}, \beta_{opt})}_{\min}$. We increment $i$ and go to \textbf{Start};

   At this point, we have $d^{l*}> d^{r(i)}$ and $(w^{r(i)}_A, w^{r(i)}_B)$ is not mapped to the same NC symbol as $(w^{*}_A, w^{*}_B), (w^{r(0)}_A, w^{r(0)}_B), \ldots, (w^{r(i-1)}_A, w^{r(i-1)}_B)$. Thus, $d^{r(i)}$ is a potential candidate for $d^{(\alpha_{opt}, \beta_{opt})}_{\min}$. If $d^{r(i)}$ were equal to $d^{(\alpha_{opt}, \beta_{opt})}_{\min}$, then $w^{r(i)}_S-w^{*}_S=d^{r(i)}+d^{r(i-1)}+\ldots+d^{r(1)}+l_{\min}>d^{(\alpha_{opt}, \beta_{opt})}_{\min}$.  Note that $w_S^{r(i)} - w_S^* > d^{(\alpha_{opt}, \beta_{opt})}_{\min}$ and the clustering of joint symbols $(w^{*}_A, w^{*}_B), (w^{r(0)}_A, w^{r(0)}_B), \ldots, (w^{r(i-1)}_A, w^{r(i-1)}_B)$  imply that $d^{(\alpha_{opt}, \beta_{opt})}_{\min}$ cannot be observed at the locality of the reference symbol by looking to the right, contradicting \emph{Corollary \ref{cor:short}}. Neither can $d^{(\alpha_{opt}, \beta_{opt})}_{\min}$ be observed at the locality by looking to the left because here we have $d^{l*}>d^{r(i)}=d^{(\alpha_{opt}, \beta_{opt})}_{\min}$. This leads to a contradiction and our proof is complete.

 \textbf{Case (2)}: $d^{r(i)}=l_{\min}$

Here, we argue that $(w^{r(i)}_A, w^{r(i)}_B)$ is mapped to the same NC symbol as $(w^{*}_A, w^{*}_B), (w^{r(0)}_A, w^{r(0)}_B), \ldots, (w^{r(i-1)}_A, w^{r(i-1)}_B)$ and therefore we will  increment $i$ and go to \textbf{Start} in our proof.

Suppose that $(w^{r(i)}_A, w^{r(i)}_B)$ is not mapped to the same NC symbol as $(w^{*}_A, w^{*}_B), (w^{r(0)}_A, w^{r(0)}_B), \ldots, (w^{r(i-1)}_A, w^{r(i-1)}_B)$. Define $(\delta^{r(i)}_A, \delta^{r(i)}_B)\triangleq(w^{r(i)}_A-w^{r(i-1)}_A, w^{r(i)}_B-w^{r(i-1)}_B)$ and $(\delta^{r(0)}_A, \delta^{r(0)}_B)\triangleq(w^{r(0)}_A-w^{*}_A, w^{r(0)}_B-w^{*}_B)$. Since $(w^{r(i)}_A, w^{r(i)}_B)$ cannot be clustered with  $(w^{r(i-1)}_A, w^{r(i-1)}_B)$, we have that ${\rm mod}((\delta^{r(i)}_A, \delta^{r(i)}_B), q)\neq \nu \otimes {\rm mod}((\delta^{r(0)}_A, \delta^{r(0)}_B), q)$,  $\nu\in\{1, \ldots, q-1\}$  by \emph{Proposition \ref{pro:twop}}.  Since the inequality applies for $\nu = 1$, we have  $(\delta^{r(i)}_A, \delta^{r(i)}_B)\neq (\delta^{r(0)}_A, \delta^{r(0)}_B)$.

 By \emph{Lemmas \ref{lem2}} and \emph{\ref{lem3}}, since $d^{r(i)} = l_{\min}$,  we can observe a joint symbol $(w_A, w_B)$ separated from $(w^{*}_A, w^{*}_B)$ by the difference $(\delta^{r(i)}_A, \delta^{r(i)}_B)$ with distance from $(w^{*}_A, w^{*}_B)$ equal to  $l_{\min}$. First, we consider $\delta^{r(i)}_A\geq0$ and $\delta^{r(i)}_B<0$. By case (i) in the statement of  \emph{Lemma \ref{lem3}}, we can find a valid joint symbol $(w_A, w_B)=(0, q-1)+(\delta^{r(i)}_A, \delta^{r(i)}_B)=(\delta^{r(i)}_A, q-1+\delta^{r(i)}_B)$. Now, $w_S-w^*_S=\eta \delta^{r(i)}_A+ \delta^{r(i)}_B=d^{r(i)}=l_{\min}$. Thus, this $(w_A, w_B)$ is on the right side of  $(w^{*}_A, w^{*}_B)$. Since  $(\delta^{r(i)}_A, \delta^{r(i)}_B)\neq (\delta^{r(0)}_A, \delta^{r(0)}_B)$, we have $(w_A, w_B)\neq (w^{r(0)}_A, w^{r(0)}_B)$ and $(w_A, w_B)$ must be a distinct joint symbol overlapping with  $(w^{r(0)}_A, w^{r(0)}_B)$.   This cannot be true because any such overlap would mean that $l_{\min}$ is actually zero, and by \emph{Corollary 1}, there must also be a symbol overlapping with the reference symbol (but we are considering the situation where such overlapping does not occur). Second, we consider $\delta^{r(i)}_A<0$ and $\delta^{r(i)}_B\geq 0$. By case (ii) in the statement of \emph{Lemma \ref{lem3}}, we can find a valid joint symbol $(w_A, w_B)=(0, q-1)-(\delta^{r(i)}_A, \delta^{r(i)}_B)=(-\delta^{r(i)}_A, q-1-\delta^{r(i)}_B)$.  Now, $w_S-w^*_S=-(\eta \delta^{r(i)}_A+ \delta^{r(i)}_B)=-d^{r(i)}=-l_{\min}$.  Thus, this $(w_A, w_B)$ is on the left side of  $(w^{*}_A, w^{*}_B)$. However, the left neighbor $(w^{l*}_A, w^{l*}_B)$ is at a distance larger than $l_{\min}$ from $(w^{*}_A, w^{*}_B)$.  This leads to a contradiction and our proof is complete.

\section*{Appendix IV: Algorithm for Identifying Characteristic Symbols  }

In the following, $J$  is the cardinality of  $\mathcal{W}^{lo}_{(A, B)}$, and the characteristic symbols form a subset of  $\mathcal{W}^{lo}_{(A, B)}$.

\begin{enumerate}
  \item[Step 1:] For each $(w_{A,j}, w_{B,j})\in \mathcal{W}^{lo}_{(A, B)}$, $\forall j\in \{1,2,\ldots, J\}$, find $\eta_j$ at which $(w_{A,j}, w_{B,j})$ overlaps with the reference symbol. Specifically, $\eta_jw_{A,j}+w_{B,j}=q-1$.
  \item [Step 2:] Sort $\eta_1, \eta_2, \ldots, \eta_J$  in ascending order. Let $(\tilde{\eta}_1, \tilde{\eta}_2, \ldots, \tilde{\eta}_J)$  denote the ordered sequence, with the associated sequence of joint symbols denoted by  $\big((\tilde{w}_{A,1}, \tilde{w}_{B,1}), (\tilde{w}_{A,2}, \tilde{w}_{B,2}), \ldots, (\tilde{w}_{A,J}, \tilde{w}_{B,J})\big)$, $(\tilde{w}_{A,j}, \tilde{w}_{B,j}) \in \mathcal{W}^{lo}_{(A, B)}$.
  \item [Step 3:] Eliminate non-characteristic symbols as follows: Suppose that $\tilde{\eta}_v=\tilde{\eta}_{v+1}=\ldots=\tilde{\eta}_{v+U}$ for some $v\geq 1$ and $v+U\leq J$, with  $(\tilde{w}_{A,v}, \tilde{w}_{B,v}), \ldots, (\tilde{w}_{A,v+U}, \tilde{w}_{B,v+U})$ overlapping with the reference symbol together. Retain only the joint symbol with the smallest $w_A$  among $(\tilde{w}_{A,v}, \tilde{w}_{B,v}), \ldots, (\tilde{w}_{A,v+U}, \tilde{w}_{B,v+U})$ as a characteristic symbol. Also, retain only one of $\tilde{\eta}_v, \tilde{\eta}_{v+1}, \ldots, \tilde{\eta}_{v+U}$  in the sequence  $(\tilde{\eta}_1, \tilde{\eta}_2, \ldots, \tilde{\eta}_J)$. Go through the above procedure for all other multiple overlappings.
\end{enumerate}

At the conclusion of Step 3, we have a sorted sequence  $({\eta}^e_1, {\eta}^e_2, \ldots, {\eta}^e_I)$, where $\eta^e_i\neq \eta^e_j$ for all $i\neq j$, and the corresponding sequence of characteristic symbols  $\big(({w}^{char}_{A,1}, {w}^{char}_{B,1}), ({w}^{char}_{A,2},  {w}^{char}_{B,2}), \ldots, ({w}^{char}_{A,I},  {w}^{char}_{B,I})\big)$.


\end{document}